\documentclass[fleqn,usenatbib]{mnras}
\usepackage[T1]{fontenc}
\usepackage{newtxtext,newtxmath}

\usepackage{graphicx}	
\usepackage{amsmath}	
\usepackage{float}  

\usepackage{caption}
\usepackage{enumitem}

\usepackage{hyperref}
\usepackage{url}

\title[Beyond the Merger - Quasar - Quench Paradigm]{Beyond the Merger - Quasar - Quench Paradigm I: Mergers are neither necessary nor sufficient to quench central galaxies in IllustrisTNG}  

\author[Casimiro et al.]{Camilo A. Casimiro$^{1}$\thanks{E-mail: ccasimir@fiu.edu},
Asa F. L. Bluck$^{1}$\thanks{E-mail: abluck@fiu.edu},
Paul Goubert$^{1}$,
Thomas Pinto Franco$^{1}$,
Joanna M. Piotrowska$^{2}$
\\\\
$^{1}$Stocker AstroScience Center, Dept. of Physics, Florida International University,11200 SW 8th Street, Miami, 33199, Florida, USA\\
$^{2}$Cahill Center for Astronomy and Astrophysics, California Institute of Technology, Pasadena, CA, USA\\
}

\date{Accepted 2026 March 10. Received 2026 March 09; in original form 2025 November 26.}
\pubyear{2025}

\newcommand{\window}{Merger-Quenching Association Window }
\newcommand{\wind}{MQAW}

\begin{document}
\label{firstpage}
\pagerange{\pageref{firstpage}--\pageref{lastpage}}
\maketitle

\begin{abstract}
The cessation of star formation in galaxies, known as `quenching', is a complex, multi-scale process, which has been theorized to be linked to galaxy mergers. In this paper, we investigate the potential role of mergers in quenching galaxies in the IllustrisTNG cosmological hydrodynamical simulation. We track the evolution of over 11,000 central galaxies in the simulation with $M_\star~\ge~10^9\,M_\odot$ at $z=0$ throughout the entirety of cosmic history. We compare their star formation and merger histories to test whether mergers are necessary or sufficient for inducing quenching in the simulation. Only a very small fraction of mergers ($\sim $3~per~cent of major mergers and $\sim$12~per~cent of all mergers) lead to quenching within 1~Gyr, indicating that mergers are not sufficient by themselves to cause quenching. Furthermore, the vast majority of quenching events are not preceded by a merger within 1~Gyr. Once random coincidences are accounted for and a stellar mass–matched control sample is applied, no merger excess is observed. Hence, mergers are clearly not necessary for quenching to occur in the simulation. Finally, we perform a series of random forest classification and regression analyses to assess the integrated role of mergers in galaxy quenching, and supermassive black hole growth, in IllustrisTNG. We determine that secular processes dominate the growth of supermassive black holes and the quenching of central galaxies in this simulation, in stark contrast to prior theoretical expectations from idealized hydrodynamical simulations. 
\end{abstract}


\section{Introduction}

The population of galaxies in both the local and high-redshift Universe exhibits strong bimodality in a variety of fundamental properties.  In rest-frame optical colors, specific star formation rates (sSFR~$\equiv~\text{SFR}/M_*$), stellar ages, and structural parameters there are two broad classes of systems: actively star-forming, blue, disk-dominated galaxies and quiescent, red, bulge-dominated galaxies \citep[e.g.,][]{Strateva2001, Brinchmann2004, Driver2006, Bluck2014}.  These bimodal distributions persist out to at least $z\sim$ 2 - 3 \citep[e.g.][]{Williams2009, Santini2009, Whitaker2011, Bluck2022, Bluck2024b} and are accompanied by a steep decline in the cosmic star formation rate density from its peak at $z\sim2$ to the present epoch \citep[e.g.,][]{Lilly1996, Madau2014}.  

A further observational clue is the `downsizing' trend, whereby more massive galaxies exhaust their gas reservoirs and cease star formation earlier in cosmic history than their lower-mass counterparts \citep[e.g.,][]{Cowie1996, Heavens2004, Thomas2005}.  Any viable theory of galaxy evolution must account for this diversity in galaxy properties and explain the physical processes that drive galaxies from the star-forming `blue cloud' onto the quiescent `red sequence'. In this paper we refer to this process as `quenching'.

Many mechanisms have been proposed to suppress star formation in galaxies.  External, or environmental, processes operate primarily on satellite galaxies in dense environments and include ram pressure stripping, strangulation, tidal harassment and galaxy–cluster interactions \citep[e.g.,][]{Gunn1972, Larson1980, vanDenBosch2008}.  Such mechanisms can remove or heat gas in satellites, leading to rapid or gradual quenching.  Although these processes are crucial for understanding galaxy populations in clusters and groups, our focus here is on \emph{central} galaxies, which are the most massive galaxies in their dark matter haloes and are much less susceptible to environmental effects \citep[e.g.,][]{Bluck2016, Bluck2020a, Bluck2020b, Goubert2024, Goubert2025}. 

Internal quenching mechanisms include feedback from supernovae and stellar winds, which can drive outflows and heat the interstellar medium in low-mass systems \citep[e.g.,][]{Muratov2015}, and morphological stabilization, in which the growth of a massive bulge may tidally suppress the formation of gravitational instabilities that would otherwise collapse gas into stars \citep[e.g.][]{Martig2009, Gensior2020}.  A particularly important internal mechanism invokes feedback from supermassive black holes at galaxy centers.  Accretion onto black holes releases vast amounts of energy in radiative or mechanical form, which can heat, expel, or redistribute gas on galactic (and even halo) scales, thereby regulating star formation \citep[e.g.,][]{Croton2006, Maiolino2012, Silk2012, Fabian2012, Weinberger2017}.

Modern hydrodynamical simulations and semi-analytic models distinguish two main modes of active galactic nucleus (AGN) feedback.  In the \emph{radiative} or \emph{quasar} mode, accretion at high Eddington ratios produce multi-phase gas outflows \citep[e.g.][]{DiMatteo2005, Hopkins2006, Hopkins2008, Maiolino2012}.  In the \emph{kinetic} or \emph{radio} mode, low-Eddington accretion powers collimated jets that form expanding bubbles in the circum-galactic medium (CGM), depositing vast quantities of energy, increasing the entropy within the hot gaseous halo \citep[e.g.,][]{Croton2006, Sijacki2007, Fabian2012, Weinberger2017, Zinger2020}.  Both of these modes of AGN feedback are implemented in contemporary cosmological hydrodynamical simulations such as Illustris and IllustrisTNG \citep[e.g.,][]{Vogelsberger2014a, Vogelsberger2014b, Weinberger2017, Weinberger2018, Pillepich2018, Nelson2018}.

Galaxy mergers, especially gas-rich major mergers, have long been posited as a key route to quenching.  In the canonical picture, merging disks undergo violent tidal perturbations that drive gas into the nuclear regions, triggering intense starbursts and fueling rapid black hole growth.  The ensuing quasar activity, and associated outflows, can expel or heat the remaining gas, leading to a rapid shut-down of star formation \citep[e.g.][]{Mihos1996, Springel2005, DiMatteo2005, Hopkins2006, Hopkins2008}.  

Mergers may also rearrange the stellar distribution in galaxies, transforming high angular momentum disks into low-angular momentum spheroids, and potentially lead to dynamical stabilization of the remnant against further star formation \citep[e.g.,][]{Toomre1972, Barnes1992, Naab2003, Lotz2010}.  Many theoretical and observational studies have therefore explored the `merger–quasar–quench' paradigm, in which major mergers are both a trigger for luminous AGN and a catalyst for quenching \citep[e.g.,][]{Wild2010, Ellison2013, Pawlik2018, Wilkinson2022, Ellison2024}.

However, this paradigm was developed largely from idealized hydrodynamical simulations of isolated galaxy pairs on $\lesssim 1$\,cMpc scales and over $\sim$1\,Gyr timescales.  Cosmological hydrodynamical simulations, such as IllustrisTNG, provide a vastly broader and more self-consistent laboratory to test these ideas.  

IllustrisTNG100 follows the evolution of $\sim$10$^4$ central galaxies with $M_\star \gtrsim 10^9\,M_\odot$ over 14\,Gyr in a $\sim(110\,\mathrm{cMpc})^3$ volume. Moreover, this simulation suite includes detailed subgrid physical models for star formation, stellar feedback, black hole accretion, and both radiative and kinetic AGN feedback \citep[see,][]{Weinberger2017, Nelson2018, Pillepich2018}.  Merger trees from the simulation enable the reconstruction of each galaxy’s full star formation and merger/ interaction history, while its relatively high time resolution allows for the measurement of quenching onset times and durations.  These features make IllustrisTNG ideally suited for quantifying whether mergers are necessary or sufficient for central galaxy quenching within the contemporary theoretical paradigm of cosmological simulations.

A number of prior studies have begun to explore the possible link between mergers and quenching in IllustrisTNG (see \citealt{Weinberger2017, Weinberger2018} for discussions on the model itself). Of particular note, \citealt{Quai2021} examined star-forming post-merger galaxies at $z\lesssim1$, finding that only a few per cent quench within $0.5$\,Gyr post coalescence, with little difference from a control sample by $1.5$\,Gyr thereafter. Expanding upon this, \citealt{Quai2023} compared the quenching of post-mergers in Illustris, EAGLE and IllustrisTNG and also concluded that rapid post-merger quenching is rare. These works considered only the \emph{sufficiency} of mergers for quenching (i.e., how often post-mergers quench), and focused on star-forming progenitors with $M_\star\ge10^{10}\,M_\odot$.  Their results suggest that mergers are not the dominant driver of quenching, but they leave open several important questions, explicitly: (i)~whether mergers are \emph{necessary} for quenching (i.e., how often quenching events are preceded by mergers); and (ii)~whether the cumulative merger history impacts quenching in a more complex manner (e.g., via growth of the supermassive black holes, with subsequent quenching via AGN feedback).

In this work, we analyze all central galaxies in IllustrisTNG100 that reach stellar masses of $M_\star \geq 10^9\,M_\odot$ by $z=0$, and trace their major progenitors along the merger tree. This stellar-mass-limited selection, unrestricted in redshift, enables us to probe quenching across all of cosmic time (unlike in previous works). We reconstruct both star formation and merger histories for all galaxies in our sample, measuring the onset and duration of quenching using a bespoke redshift-dependent sSFR classification approach. We then test whether mergers are \emph{sufficient} for quenching (i.e., do mergers frequently lead to quenching, or not?) and whether mergers are \emph{necessary} for quenching (i.e., are mergers frequently involved in quenching events, or not?). We achieve this by comparing merger frequencies in quenched systems to those in a stellar mass-matched star-forming control sample. To capture both rapid and delayed pathways, we adopt wide merger association windows around quenching events. These choices are motivated by both theoretical modeling and observational constraints (see Section~\ref{sec:Merger_Quenchin_Window}). 

Finally, we employ sophisticated random forest classifiers and regressors to assess the predictive power of cumulative merger metrics relative to intrinsic properties (including, stellar mass, halo mass, and central black hole mass). This machine learning approach enables us to disentangle merger-driven and secular quenching channels, in addition to quantifying the role of mergers in supermassive black hole growth, and any possible indirect impact of mergers on central galaxy quenching. The random forest technique is ideal for problems of this type due to its proven capacity to expose nuisance/ spurious correlators in complex inter-correlated astrophysical data (see \citealt{Bluck2022} for many tests and examples of this technique).

This paper is structured as follows. In Section~\ref{sec:data} we describe the IllustrisTNG100 simulation, our galaxy sample selection, and the construction of merger trees.  Section~\ref{sec:methods} outlines our methods for identifying mergers, classifying galaxies by star formation rate, and defining quenching onset and duration times.  Section~\ref{sec:results} presents the results of our sufficiency and necessity tests, along with the machine-learning analysis of cumulative merger effects.  In Section~\ref{sec:discussions} we discuss the implications of our findings for quenching models and compare to previous work. Section~\ref{sec:summary} summarizes our contributions. In the appendix we provide additional examples of star formation and merger histories for IllustrisTNG galaxies, as well as giving full details on our machine learning set-up for reproducibility.  Throughout the paper we assume a spatially flat $\Lambda$CDM cosmology consistent with \citealt{Planck2016} and express all logarithmic quantities in base ten.

\section{Data}
\label{sec:data}

Our analysis is based on the IllustrisTNG100-1 cosmological hydrodynamical simulation \citep{Pillepich2018, Nelson2018, Springel2018, Naiman2018, Marinacci2018}. For brevity, we refer to this simulation simply as `TNG' from this point onward. This simulation follows the evolution of dark matter, gas, stars, and black holes within a periodic cube of side length $L=75\,h^{-1}\,\mathrm{cMpc}$, corresponding to $\approx110.7\,\mathrm{cMpc}$ for $h=0.6774$. The simulation contains $2\times1820^3$ resolution elements (dark matter particles and gas cells), yielding a baryonic mass resolution of $1.4\times10^6\,\mathrm{M}_\odot$ and a Plummer-equivalent gravitational softening length of $0.74\,\mathrm{kpc}$ (physical) for stellar particles at $z=0$.

TNG evolves from $z=127$ to $z=0$ over 100 snapshots using the moving-mesh code \textsc{AREPO} \citep{Springel2010}, which solves the coupled equations of gravity and magneto-hydrodynamics on a dynamic Voronoi mesh. The typical snapshot spacing at $z\lesssim1$ is $\sim150\,\mathrm{Myr}$, allowing reasonably precise tracking of merger events and quenching timescales. The galaxy formation model includes the following crucial subgrid physical recipes: star formation and stellar feedback from supernovae (both Type I \& II); black hole seeding at $M_{\rm seed}=8\times10^5\,h^{-1}M_\odot$ and growth via Eddington-limited, Bondi-Hoyle accretion (\citealt{Hoyle1939, Bondi1944}); and AGN feedback in quasar (thermal) and radio (kinetic) modes \citep[see,][]{Weinberger2017, Weinberger2018}.

The initial conditions assume a flat $\Lambda$CDM cosmology consistent with Planck 2015 results \citep{Planck2016}, with parameters $(\Omega_{\mathrm{m}}, \Omega_\Lambda, \Omega_{\mathrm{b}}, \sigma_8, n_s, H_0) = (0.3089,\,0.6911,\,0.0486,\,0.8159,\,0.9667,\,67.74~\mathrm{km\,s^{-1}\,Mpc^{-1}})$.

In this work, we define stellar mass as the total bound stellar mass of each subhalo and halo mass via the total mass associated with the friends-of-friends (FoF) group. We select \emph{central} galaxies (the most massive subhalo in each FoF group) at $z=0$ with $M_\star\ge10^9\,M_\odot$, yielding a final sample of 11\,724 galaxies. This mass cut ensures completeness well above the simulation’s resolution limit ($M_* \sim 10^{7}\,M_{\odot}$), as well as being amenable to observational follow-up via various extant galaxy surveys. 

Evolutionary histories are reconstructed using the \textsc{SubLink} merger trees \citep{RodriguezGomez2015}, and we define the onset of reliable stellar tracking as the earliest snapshot where $M_\star~\geq~10^8~\,~M_\odot$. Galaxies below $M_\star~=~10^7~\,~M_\odot$ are poorly resolved in TNG, hence this threshold enables a minimum merger ratio assessment of 1:10, which rises to $\sim$1:10$^5$ for the most massive galaxies in TNG. Moreover, virtually no low-mass centrals are quenched in TNG (or in SDSS observations; see, \citealt{Piotrowska2022}; \citealt{Bluck2022}), so imposing a lower mass limit would likely not exclude any quenched systems in any case.

\section{Methods}
\label{sec:methods}
\subsection{Merger Identification and Classification}
\label{sec:Merger_Id}

We reconstruct each galaxy's main progenitor branch as the most massive stellar progenitor in each preceding snapshot, from $z=0$ back to its first resolved phase (defined here as the first snapshot where $M_\star \geq 10^8\,M_\odot$). Along this branch, we record the stellar mass ($M_\star$), halo mass ($M_{\rm halo}$), central black hole mass ($M_{\rm BH}$), and instantaneous star formation rate (SFR).

Merger events are identified by following secondary progenitor branches in the SubLink merger trees \citep{RodriguezGomez2015}, using the publicly available illustris\_python merger-tree tools provided with the Illustris(TNG) data release \citep{Nelson2015}. When a secondary progenitor ceases to appear as a distinct object in the catalog (i.e., it coalesces into the main branch), we record the corresponding merger event and time. 

For each merger event, we compute the stellar mass ratio $r \equiv M_{\star,\mathrm{sec}}/M_{\star,\mathrm{main}}$, where $M_{\star,\mathrm{main}}$ is the stellar mass of the main progenitor (main-branch galaxy) and $M_{\star,\mathrm{sec}}$ is the stellar mass of the secondary progenitor (the merging galaxy). Due to the impact of stellar stripping and issues with robust subhalo reconstruction of merging systems in SubFind, we do not simply evaluate these masses at the last snapshot available (see \citealt{Patton2020}). Instead we evaluate the merger ratio at the snapshot selected by our novel robust ratio procedure, which is described fully in Appendix~\ref{sec:App_AA}. 

Briefly, this approach filters the stellar-mass histories of both galaxies to mitigate transient particle-misclassification artifacts and identifies a set of jointly reliable snapshots from which $r$ is measured. Crucially, we also test the stability of our results against alternative methods, including the last snapshot, the peak of the secondary galaxy's stellar mass history, and the ratio of the medians of merging systems stellar masses. All of the results and conclusions of this paper are completely stable to these methodological choices (see Appendix~\ref{sec:App_AA}).

We classify mergers using both a resolution threshold and a major--minor division. Specifically, we define \textit{resolved mergers} as events for which the secondary progenitor has $M_\star \geq 10^7\,M_\odot$ at the snapshot used to measure the stellar mass ratio $r$ (that is, the snapshot selected by the procedure described in Appendix~\ref{sec:App_AA}). We refer to this inclusive set as \emph{all mergers} or \emph{any mergers}, since it includes every well-resolved merger experienced by any galaxy over its history. We further define \textit{major mergers} as the subset of resolved mergers with $r \ge 1/4$ (that is, 4:1 or more equal).

\begin{figure*}
    \centering
    \includegraphics[width=0.88\textwidth]{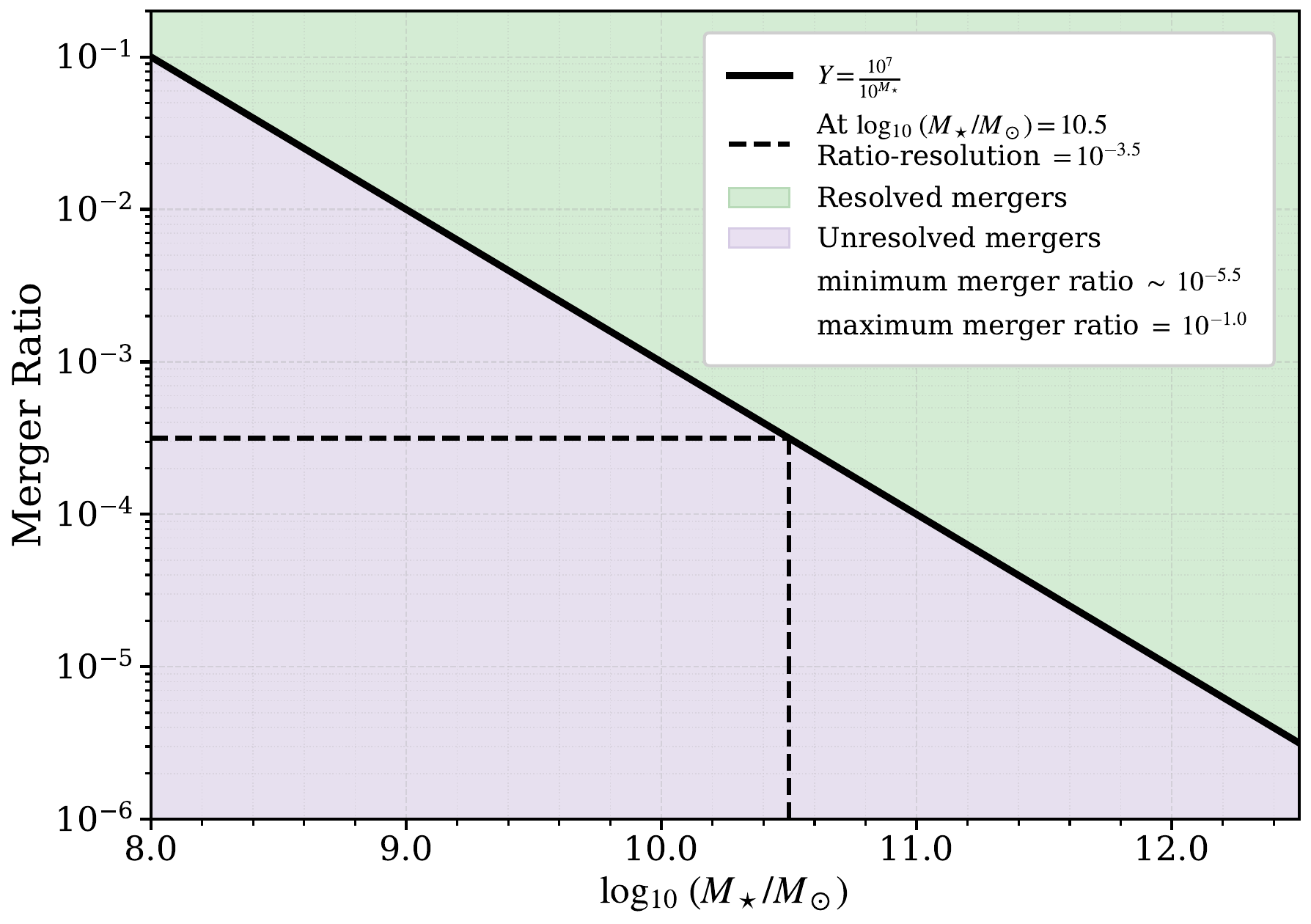}
\caption{
    Effective merger-ratio completeness implied by our resolved-merger criterion.
    Requiring $M_{\star,\mathrm{sec}}\ge 10^7\,M_\odot$ sets a host-mass-dependent minimum ratio $r_{\rm min}(M_{\star,\mathrm{main}})=10^7/M_{\star,\mathrm{main}}$ (solid black line). Mergers above this limit are included (green shaded region), while those below are excluded (purple shaded region). The vertical dashed line marks $\log_{10}(M_{\star,\mathrm{main}}/M_\odot)=10.5$, near the stellar-mass scale where central galaxies typically begin quenching, illustrating completeness in merger ratio down to $\sim$1:3000 at that mass.
    }
    \label{fig:MergerRatioCompleteness}
\end{figure*}

Because the inclusive sample is defined by a fixed secondary-mass resolution threshold (rather than a fixed mass-ratio cut), the minimum stellar mass ratio we can probe depends explicitly on the host stellar mass. For a host of stellar mass $M_{\star,\mathrm{main}}$, the corresponding limit is
$r_{\min}(M_{\star,\mathrm{main}})=10^{7}\,M_\odot/M_{\star,\mathrm{main}}$.
This mapping is illustrated in Fig.~\ref{fig:MergerRatioCompleteness}. The key implication is that our mass-ratio sensitivity improves substantially for more massive hosts. Near the stellar-mass scale where central galaxies typically begin quenching, $M_{\star,\mathrm{main}} \sim 10^{10.5}\,M_\odot$, the \emph{all mergers} sample is complete down to $r \sim 10^{-3.5}$ (about 1:3000). In contrast, at the low-mass end of our host sample ($M_{\star,\mathrm{main}} \gtrsim 10^{8}\,M_\odot$), the same resolution requirement corresponds to $r_{\min}\sim 0.1$ (about 1:10).

This is why imposing a stricter universal ratio cut (for example, $r>0.1$) is not well matched to our goals. For massive galaxies, mergers with $r<0.1$ can still involve secondaries that are substantial in absolute stellar mass, and they would be excluded despite being well resolved.

From the resulting merger catalog we compute for each galaxy:

\begin{itemize}
  \item $M_{\rm m} = \sum_i M_{\star,\mathrm{sec},i}$, the total stellar mass accreted in resolved mergers, where $M_{\star,\mathrm{sec},i}$ is the stellar mass of the secondary (merging) galaxy evaluated at the snapshot selected for measuring $r$ by Appendix~\ref{sec:App_AA}; \\

  \item  $M_{\rm M} = \sum_{j:\,r_j\ge 1/4} M_{\star,\mathrm{sec},j}$, the stellar mass accreted in major mergers only (same definition of $M_{\star,\mathrm{sec}}$ as above).
\end{itemize}

\noindent This procedure yields a sequence of resolved merger events and main-branch galaxy properties throughout the full cosmic history of central galaxies within TNG.

\subsection{Quenching classification}
\label{sec:Q-SF_Class}

\begin{figure*}
    \centering
    \includegraphics[width=0.95\textwidth]{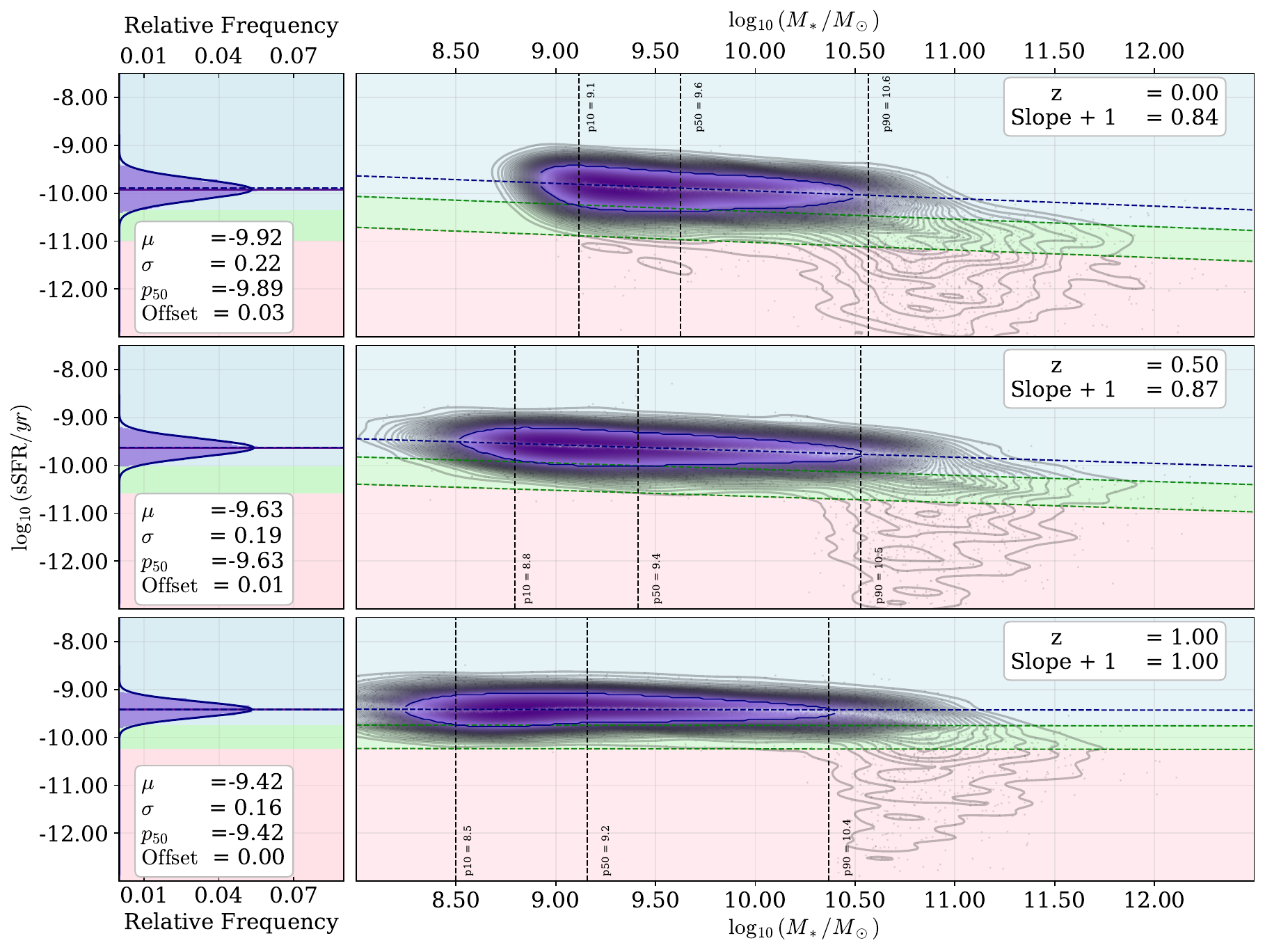}
    \caption{
    Examples of the KDE-based procedure used to fit the main sequence (MS) and estimate its scatter at three snapshots (top to bottom: $z=0,\,0.5,\,1$, as labeled).
    Right: KDE contours in the $\log_{10}(\mathrm{sSFR}/\mathrm{yr}) - \log_{10}(M_\bigstar/M_\odot)$ plane with the adopted HDR boundary (dark blue), the linear MS fit (dashed blue; the reported \textit{Slope+1} corresponds to $a+1$ for a fit of the form $y = a\,x + b$), and the population boundaries defined relative to the MS as $y - y_{\rm MS} = -2\sigma$ and $y - y_{\rm MS} = -5\sigma$ (dashed green), with shaded star-forming (blue), green-valley (green), and quenched (pink) regions.
    Vertical dashed black lines mark the stellar-mass percentiles $p_{10}$, $p_{50}$, and $p_{90}$ of $x=\log_{10}(M_\bigstar/M_\odot)$ for the sample in that snapshot.
    Left: the collapsed distribution of $y\equiv\log_{10}(\mathrm{sSFR}/\mathrm{yr})$ for galaxies inside the HDR boundary (relative frequency) and the Gaussian fit used to define $\mu$ and $\sigma$; the inset also reports $p_{50}$ evaluated on the MS (i.e., $y_{\rm MS}$ at the median stellar mass) and the offset $|\mu - p_{50}|$.}
    \label{sSFR_Peaks}
\end{figure*}

We classify galaxies as star-forming, transitioning, or quenched using their offset from the star-forming main sequence (MS) in the $\log_{10}(M_\bigstar/M_\odot) - \log_{10}(\mathrm{sSFR}/\mathrm{yr})$ plane. 
Rather than defining thresholds from the one-dimensional distribution of $\log_{10}(\mathrm{sSFR})$ alone, we first determine a mass-dependent MS relation at each snapshot and then measure the intrinsic vertical scatter of the star-forming population around this relation. See Fig.~\ref{sSFR_Peaks} for a representation of this relationship and our classification scheme at $z = 0, 0.5, 1$.

For each snapshot, we estimate the joint density $\hat{p}(x,y)$ of
$x \equiv \log_{10}(M_\bigstar/M_\odot)$ and $y \equiv \log_{10}(\mathrm{sSFR}/\mathrm{yr})$ using a two-dimensional kernel density estimate (KDE). 
We then define a highest-density region (HDR) as the set of points enclosed by a density level $\lambda(\alpha)$ such that a fraction $\alpha$ of the total integrated KDE mass lies within the region,
\begin{equation}
\mathcal{R}_\alpha \equiv \{(x,y): \hat{p}(x,y)\ge \lambda(\alpha)\},
\end{equation}
where $\lambda(\alpha)$ is chosen so that $\iint_{\mathcal{R}_\alpha} \hat{p}(x,y)\,dx\,dy = \alpha$.
HDRs are a standard way to isolate the high-density core of a distribution in a non-parametric and reproducible manner (e.g., \citealt{Hyndman1996}; \citealt{SamworthWand2010}). 
In practice, we retain the connected component of $\mathcal{R}_\alpha$ associated with the dominant high-density locus in the $(x,y)$ plane (see Fig.~\ref{sSFR_Peaks}, dark-blue boundary).

Within this HDR-selected region, we trace the density ridge (the locus of maximum density in $y$ at fixed $x$) and fit a linear MS relation, of the form:

\begin{equation}
y_{\rm MS}(x,z) = a(z)\,x + b(z),
\end{equation}

\noindent where $a(z)$ and $b(z)$ are the best-fitting slope and intercept at snapshot redshift $z$ (Fig.~\ref{sSFR_Peaks}, dashed blue line).

Next, we estimate the vertical scatter of the star-forming population by collapsing the HDR-selected points onto the $y$-axis and fitting a Gaussian to this one-dimensional distribution (Fig.~\ref{sSFR_Peaks}, left panels). 
The fitted dispersion defines $\sigma(z)$, our snapshot-dependent estimate of the characteristic MS width in $\log_{10}(\mathrm{sSFR}/\mathrm{yr})$. 
We verify that the Gaussian peak is close to the MS value at the median stellar mass, i.e. the reported offset $|\mu - p_{50}|$ in Fig.~\ref{sSFR_Peaks} is small, supporting the interpretation of $\sigma(z)$ as the characteristic vertical scatter about the MS.

We adopt a fiducial HDR fraction of $\alpha_0 = 0.70$. 
This choice provides a conservative definition of the high-density core while avoiding sensitivity to low-density tails. 
To ensure robustness, we repeated the analysis over a range of $\alpha$ values and find that the inferred MS parameters vary only weakly for $\alpha \in [0.60, 0.80]$ (e.g., the $z=0$ slope is nearly constant across this interval), indicating that our results are not sensitive to the precise HDR fraction.

Using the fitted MS and scatter, we define a normalized MS offset for each galaxy $i$ at redshift $z$ and stellar mass $M_{\bigstar}$:
\begin{equation}
r(\mathrm{sSFR}_i(z,M_{\bigstar})) \equiv 
\frac{\log_{10}(\mathrm{sSFR}_i(z,M_{\bigstar})) - \log_{10}(\mathrm{sSFR}_0(z,M_{\bigstar}))}{\sigma(z)}
\label{eq:r_sSFR}
\end{equation}
where the MS reference value is
\begin{equation}
\log_{10}(\mathrm{sSFR}_0(z,M_{\bigstar})) \equiv y_{\rm MS}(x_i,z) = a(z)\,x_i + b(z),
\end{equation}
and $x_i \equiv \log_{10}(M_{\bigstar}/M_\odot)$. 
Note that both $\mathrm{sSFR}_i$ and the reference $\mathrm{sSFR}_0$ depend on stellar mass, while the normalization uses the snapshot-dependent scatter $\sigma(z)$.

Finally, we classify galaxies using their MS offset in units of $\sigma(z)$:
\begin{itemize}
\item \textit{Star-forming:} $r \ge -2$ \hfill (blue region)
\item \textit{Transitioning (green valley):} $-5 \le r < -2$ \hfill (green region)
\item \textit{Quenched:} $r < -5$ \hfill (pink/red region)
\end{itemize}

\noindent This methodology provides a redshift-dependent and mass-aware definition of quenching tied directly to the evolving star-forming locus, while remaining straightforward to interpret in terms of a galaxy's displacement from the MS.

In Fig.~\ref{sSFR_Peaks}, we present the distribution of $\log_{10}(\mathrm{sSFR})$ at several representative redshifts. The corresponding Gaussian fits, their parameters, and the shaded color regions (corresponding to the bespoke thresholds for each galaxy type as a function of redshift) are displayed on each panel. Clearly, the peak sSFR moves to higher values at earlier cosmic times (as is well known observationally; e.g., \citealt{Madau2014}). Additionally, the width of the distribution in log-space also narrows systematically with increasing redshift \citep[as noted in][]{Goubert2025}. We account for both of these effects in the classification of galaxies into star forming and quenched categories, based on what is typical at each point in its evolution.

\begin{figure*}
    \centering
    \includegraphics[width=1.0\textwidth]{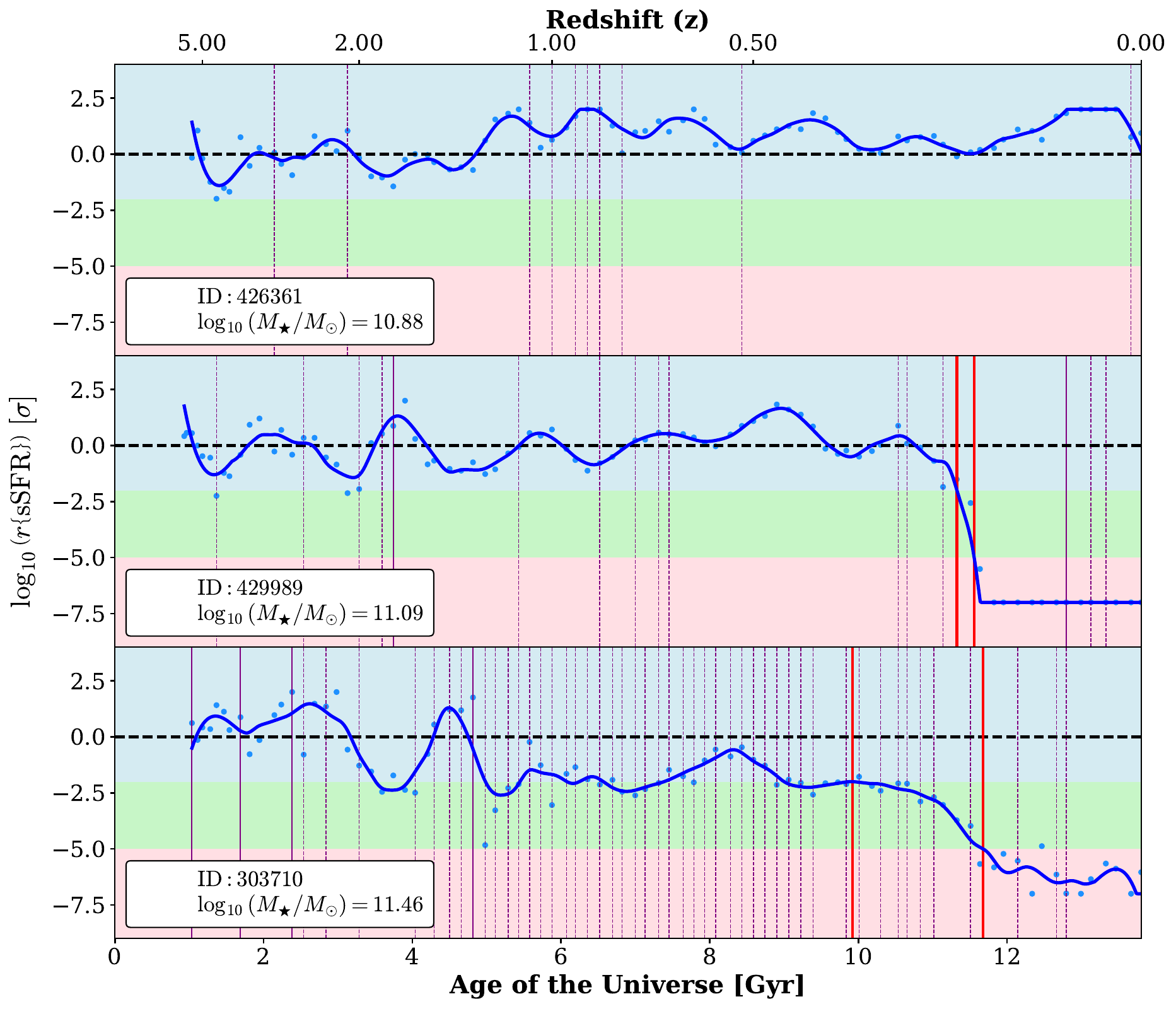}
    \caption{Star formation histories for three example galaxy types: a star forming galaxy (top panel), a fast-quenched galaxy (center panel), and a slow-quenched galaxy (bottom panel). On each panel, the $Y$-axis shows the evolution of ${\mathrm{r(sSFR)}}$ as a function of the age of the Universe (bottom $X$-axis) and redshift (top $X$-axis). The ${\mathrm{r(sSFR)}}$ values are measured as offsets from the main sequence at each epoch, and are normalized by the standard deviation ($\sigma$) obtained from a Gaussian fit to the sSFR distributions (see eq. 3). The shaded regions represent different states of star formation (as defined in Figure~\ref{sSFR_Peaks}). The blue region corresponds to star forming systems, the green region to quenching systems, and the red region to fully quenched systems. The vertical dashed and solid lines indicate times when minor and major mergers occur, respectively. The blue points represent the individual ${\mathrm{r(sSFR)}}$ values at each epoch, while the smoothed solid curve (displayed as a solid blue line on each panel) is obtained using a Savitzky-Golay filter to highlight the overall trend, averaging out stochasticity due to burstiness in star formation. The two solid red vertical lines indicate the times when each galaxy begins and completes quenching, defined by entering and departing the green valley region.}
    \label{RsSFR_Hist}
\end{figure*}

To analyze individual galaxies via our classification scheme, we track the evolution of ${\mathrm{r(sSFR)}}$ for all systems within our sample. By plotting ${\mathrm{r(sSFR)}}$ as a function of cosmic time, we can visualize the movement between the star-forming, transitioning, and quenched states (defined above). Moreover, we can also associate this movement with the timing and frequency of merger events, which are potentially associated with these transitions. 

In Fig.~\ref{RsSFR_Hist}, we present evolutionary tracks for examples of star-forming and quenched galaxies (see Appendix~\ref{sec:App_B} for several more examples). Here, the shaded regions correspond to the classification boundaries established above, with vertical purple lines marking the occurrence of major (solid lines) and minor (dotted lines) mergers. The top panel shows an example of a star forming galaxy (i.e., a system which remains star forming throughout its entire evolution), the middle panel shows an example of a galaxy which undergoes fast quenching, and the bottom panel shows an example of a galaxy which undergoes slow quenching (the exact definitions for fast and slow quenching are defined later in Section 4.1).

The dark blue lines in Figure~\ref{RsSFR_Hist} show the smoothed evolution of ${\mathrm{r(sSFR)}}$, obtained by first applying cubic interpolation to the original data and then filtering the result with a Savitzky-Golay filter (window size 100, polynomial order 3). The Savitzky-Golay filter is a local polynomial regression method that preserves local features and trends while reducing noise and short-term fluctuations. This approach is particularly well-suited for our analysis, as it smooths out stochastic variability in star formation while retaining genuine, physically-motivated rapid drops, such as those associated with fast quenching events, without artificially broadening or dampening these transitions \citep{Savitzky1964,Steinier1972}. As a result, the filter allows us to more robustly identify the onset, duration, and pace of both rapid and gradual quenching processes in simulated galaxies. 

Although some galaxies may fluctuate near the boundaries of adjacent regions, we restrict attention to complete transitions from the star-forming (SF) region to the quenched (Q) region. Temporary excursions across regional boundaries are not counted as genuine quenching events. We define quenching to begin at the cosmic time \(t_{Q}\) when a galaxy first enters the green valley from the SF region, and to end when it leaves the green valley and enters the Q region, during a full transition. The quenching duration, \(\Delta\tau_{Q}\), is the time interval between these two epochs.

Conversely, complete transitions in the opposite direction, from Q to SF, are identified as `rejuvenation' events. The red solid vertical lines indicate the epochs at which each galaxy begins quenching and completes a full transition from the star-forming to the quenched region, as defined by our classification scheme. Note that, in the examples shown in Figure~\ref{RsSFR_Hist}, we do not display any rejuvenation events (since these are very rare in TNG). Hence, all highlighted transitions correspond to quenching episodes only. 

Although rejuvenation is a genuine physical process that galaxies can experience, the classification of rejuvenated systems using our approach is highly sensitive to uncertainties and limitations in the model. We observe that some galaxies are classified as rejuvenated or not depending on the specific smoothing filter applied, or on the choice to omit smoothing altogether. This sensitivity introduces significant ambiguity in identifying robust rejuvenation events. Combined with the fact that rejuvenated galaxies consistently represent only a very small fraction of our sample (typically 1–3\%), these considerations motivate us to exclude such cases from the present analysis.

We define "green valley galaxies" (GVG) as those systems that, at $z = 0$, have ${\mathrm{r(sSFR)}}$ values lying within the green valley region of our classification scheme. The evolutionary status of these galaxies is inherently ambiguous, as it is not possible to determine from a single snapshot whether they are in the process of transitioning from the star-forming region to the quenched region, or vice versa. As such, GVGs may represent systems in the midst of quenching, galaxies experiencing rejuvenation, or those with long-term intermediate star formation rates. In this study, we treat green valley galaxies as a distinct category to reflect this uncertainty in their current evolutionary phase. 

With this classification scheme, we identify three principal categories of galaxies in our sample for further analysis: star-forming (SF), quenched (Q), and green valley (GV) systems.

\subsection{Defining the \window\ (\wind)}
\label{sec:Merger_Quenchin_Window}

A critical aspect of our methodology is defining the time window within which a merger can plausibly be causally linked to galaxy quenching, which we refer to as the \window\ (\wind). This choice must be physically motivated to ensure robust conclusions. Both simulations and observations show that the morphological and star-forming impacts of mergers are transient. Controlled merger simulations and cosmological hydrodynamical models establish that visible tidal features and structural disturbances largely dissipate within roughly 1~Gyr after coalescence, with merger remnants subsequently becoming morphologically indistinguishable from non-merging galaxies of similar mass and type \citep[e.g.][]{Lotz2008}. For example, \citealt{Lotz2008} find that the visual postmerger phase ends about 1~Gyr after final coalescence under typical imaging depths. This indicates that the immediate impact of a merger on a galaxy ought to be observable within this timescale. Moreover, in idealized hydrodynamical simulations, the triggering of quenching via mergers is frequently found to be $<$1\,Gyr (e.g., \citealt{Hopkins2006, Hopkins2008}).

Ultimately, merger induced gas inflows and central star formation can persist into the postmerger phase for several 100's of Myr, as shown by simulations of sustained inflow, black-hole growth, and feedback \citep[e.g.,][]{Springel2005,DiMatteo2005,DiMatteo2008}, and by observations that post-mergers exhibit elevated star-formation rates and enhanced AGN incidence relative to controls \citep[e.g.,][]{Wild2010,Ellison2013}.  However, observations and theory agree that these effects do not extend much beyond 1\,Gyr.

However, we also need to allow for the possibility that a merger occurring during the green valley phase (i.e., the transitional stage between our star-forming and quenched classes defined in Section~\ref{sec:Q-SF_Class}) could plausibly aid in the shutdown of star formation. For instance, a merger-driven perturbation could convert a transient excursion into the green valley into a complete transition to quiescence.

Based on the brief discussion above, we define the Merger-Quenching Association Window (MQAW[$\Delta t$]) as:

\begin{equation}
    \mathrm{MQAW}[\Delta t]  \equiv \big[\,T_{\mathrm{Q}}-\Delta t,\; T_{\mathrm{Q}} + \Delta\tau_Q\,\big]
\end{equation}

\noindent with $\Delta t =$ 1 or 2 Gyr  (\wind[1] and \wind[2]). This window spans $\Delta t$ before the onset of quenching and continues through to quenching completion (no matter how long that may take). It is important to stress that the 1\,Gyr window is already long enough for one to reasonably expect a merger event to at least trigger quenching (if it ever does). We include the extremely wide association window of 2\,Gyr primarily as a test. 

Additionally, beyond the MQAW approach, we also consider the full integrated impact of mergers on quenching later in our analyses via our random forest method, where no time limits are employed. Hence, our complete analysis is robust to potential variability in the timescales in merger - quenching association.

\subsection{Random Forest}
\label{RF}
\subsubsection{Random Forest Classification}
\label{Random_Forest_Methods_Class}

We utilize a Random Forest (RF) classifier \citep{Breiman2001}, as implemented in \textsc{scikit-learn} \citep{Pedregosa2011}, to quantify which galaxy properties are most predictive of quenching in our sample (for similar approaches, see \citealt{Piotrowska2022}; \citealt{Bluck2022}; \citealt{Brownson2022}; \citealt{Goubert2024}). RFs are ensemble methods that construct a large number of decision trees, each trained on bootstrap-resampled data, with the final classification based on majority voting across all trees. This approach is well suited for astrophysical data sets because it captures non-linear relationships and is capable of breaking spurious correlations between inter-correlated parameters (see \citealt{Bluck2022} for a discussion).

Each decision tree recursively splits the data at nodes so as to maximize class purity, as quantified in our implementation by the Gini impurity:

\begin{equation}
{G(n) = 1 - \sum_{k=1}^{C} p_k(n)^2}
\end{equation}

\noindent where $p_k(n)$ is the fraction of samples belonging to class $k$ at node $n$, and $C$ is the total number of classes (here, quenched or star-forming). At each split, the feature and threshold are chosen to yield the greatest reduction in Gini impurity in a fully deterministic manner within a given tree. Randomness is introduced in the forest via bootstrapped random sampling of input data and regeneration of star forming control samples. This process is repeated until a stopping criterion is met, such as a minimum number of samples per leaf node, used in our implementation to prevent over-fitting.

In this work, the RF is trained on a balanced set of quenched and star-forming galaxies, using all available features at each split (\texttt{max\_features=None}; see \citealt{Bluck2022} Appendix B for justification). All input features are standardized to zero mean and unit variance prior to training. We adopt 1000 trees to ensure convergence of the feature importance rankings. The data is randomly divided into training and testing subsets (each containing 50\% of the full data), with model performance evaluated using the area under the receiver operating characteristic curve (AUC), as in \citealt{Teimoorinia2016, Bluck2022, Goubert2024}.

The relative importance of each feature is quantified by the mean decrease in Gini impurity resultant from that feature across the entire random forest, calculated as:

\begin{equation}
{\mathrm{I}_C (k) = \frac{1}{N_{\mathrm{trees}}} \sum_{\mathrm{trees}} \bigg\{ \frac{\sum_{n\in \mathcal{N}_k} N(n) \, \Delta G(n)}{\sum_n N(m) \, \Delta G(n)}} \bigg\}
\end{equation}

\noindent where $N_{\mathrm{trees}}$ is the number of trees in the forest; $n$ indexes each split; $N(n)$ is the number of training samples reaching a fork, $n$; $\Delta G(n)$ is the reduction in Gini impurity at a given fork, $n$; and the denominator sums over all forks in the tree. Crucially, the numerator sums over only those forks which utilize variable, $k$. This approach naturally accounts for inter-correlations among features, providing a robust ranking of the predictors most relevant for distinguishing between quenched and star-forming galaxies.

This RF framework enables an objective, quantitative assessment of the physical parameters most strongly associated with galaxy quenching in our sample. In Appendix~\ref{sec:App_C} we provide a full list of hyper-parameters used in each RF analysis to enable reproducibility by the community.

\subsubsection{Random Forest Regression}
\label{Random_Forest_Methods_Reg}

Since supermassive black hole mass is known to be the key predictor of quenching in TNG (see \citealt{Piotrowska2022, Bluck2023, Bluck2024b}), we also explore the role of mergers in growing supermassive black holes. This represents an indirect manner in which mergers may lead to quenching in the simulation.

To identify which galaxy parameters best predict supermassive black hole mass, $M_{\rm BH}$, we employ a Random Forest regressor (\citealt{Breiman2001}; \citealt{Pedregosa2011}).  This regressor is an ensemble of decision trees, each trained on a bootstrap‑resampled subset of the data, with splits chosen to minimize the mean squared error (MSE).  The target variable in this case is $\log(M_{\rm BH})$, and (as in classification) all input features are standardized to zero mean and unit variance.

Model performance is evaluated on the training set using the root mean square error, which is defined as:

\begin{equation}
\mathrm{RMSE} = \sqrt{\frac{1}{N}\sum_{i=1}^{N}\bigl(y_i - \hat{y}_i\bigr)^2}\,,
\end{equation}

\noindent and the coefficient of determination,

\begin{equation}
R^2 = 1 - \frac{\sum_{i=1}^{N}(y_i - \hat{y}_i)^2}{\sum_{i=1}^{N}(y_i - \bar{y})^2}\,,
\end{equation}

\noindent where $y_i$ and $\hat{y}_i$ are the true and predicted values of $\log(M_{\rm BH})$, respectively, and $\bar{y}$ is the mean of the true values.

Feature importance is quantified by the mean decrease in MSE across the random forest for a given feature, as follows:

\begin{equation}
I_R (k) = \frac{1}{N_{\rm trees}}
\sum_{\mathrm{trees}}
\frac{ \sum_{n\in \mathcal{N}_k} N(n)\,\Delta\mathrm{MSE}(n)}
{\sum_{n}N(n)\,\Delta\mathrm{MSE}(n)}\,,
\end{equation}

\noindent where $\mathcal{N}_k$ is the set of forks splitting on feature $k$, $N(n)$ is the number of samples at fork, $n$, and $\Delta\mathrm{MSE}(n)$ is the reduction in MSE achieved at that split.  Higher $I(k)$ indicates greater predictive power for feature $k$ \citep[see][for further discussion]{Bluck2022}. As with the classification analyses, all hyper-parameters are presented in Appendix~\ref{sec:App_C} to enable reproducibility.

\section{Results}
\label{sec:results}

\begin{figure*}
    \centering
    \includegraphics[width=\textwidth]{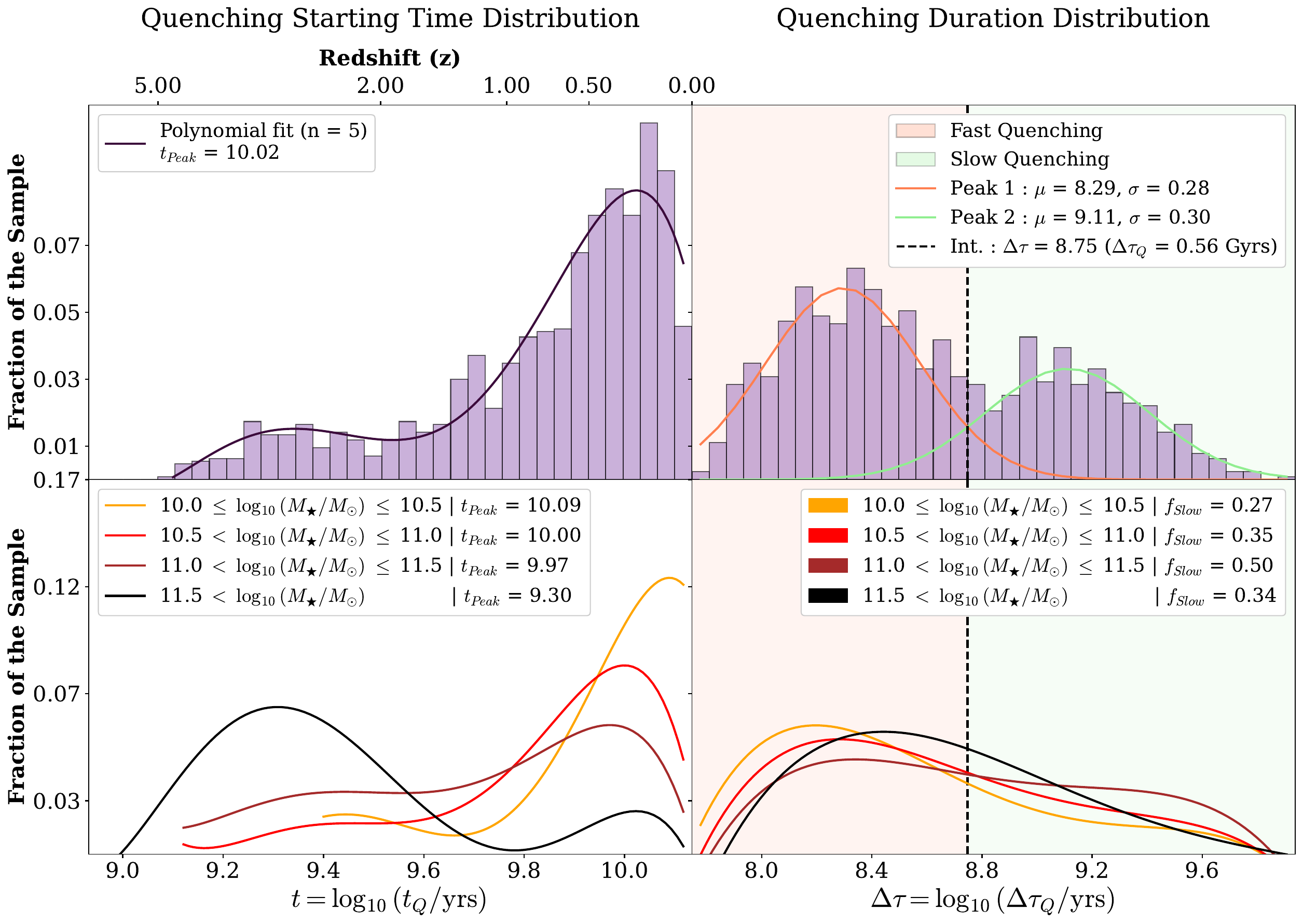}
    \caption{Distributions of quenching starting times and durations in IllustrisTNG. 
      \textit{Top-left-hand panel}: Distribution of galaxy quenching starting times as a function of the age of the Universe (bottom axis) and corresponding redshift (top axis). The histogram is overlaid with a 5th-order polynomial fit (purple curve), highlighting a broad peak around $\sim10$~Gyr ($z \sim 0.33$), indicating the typical time at which galaxies quench. 
     \textit{Bottom-left-hand panel}: Stellar mass-dependent distributions of quenching initiation times (i.e., quenching time distributions computed within stellar mass bins). Each curve is a 5th-order polynomial fit to the distribution of quenching start times within individual stellar mass bins. The systematic leftward shift of the distribution peaks with increasing stellar mass illustrates that more massive galaxies typically initiate quenching at earlier cosmic epochs.
    \textit{Top-right-hand panel}: Distribution of quenching durations, exhibiting a clear bimodal structure. Two Gaussian fits identify peaks centered at $\mu = 8.29$ and $\mu = 9.11$ in $\log_{10}(\Delta\tau_Q/\text{yrs})$, with corresponding standard deviations of $\sigma = 0.28$ and $\sigma = 0.30$, respectively. An intermediate threshold at $\log_{10}(\Delta\tau_Q/\text{yrs}) = 8.75$ (equivalent to 0.56 Gyr) is used to distinguish between fast- and slow-quenching populations. 
   \textit{Bottom-right-hand panel}: Stellar mass–dependent quenching duration distributions, each fitted with a 5th-order polynomial. The parameter $f_{\text{slow}}$ denotes the fraction of slow-quenching galaxies.}
    \label{QST_QDT}
\end{figure*}

\subsection{Quenching Times \& Durations}
\label{sec:QST_QDT}

To gain insight into the diverse evolutionary pathways leading to quiescence within the TNG simulation, we analyze the quenched galaxy population in detail by examining the distributions of quenching starting times ($t_Q$) and quenching duration times ($\Delta\tau_Q$). 

Figure~\ref{QST_QDT} provides a comprehensive overview of these distributions and reveals several key trends. The top-left panel presents the distribution of quenching starting times across cosmic time (bottom axis) and redshift (top axis). The histogram exhibits a pronounced peak at late cosmic times ($z \sim 0.3$), with only about $27\%$ of quenching originating at $z > 1$. This indicates that the bulk of galaxies in our sample undergo quenching relatively recently in the history of the Universe. This result is consistent with previous studies \citep[e.g.][]{Muzzin2013, Nelson2018}. 

We apply polynomial fits to the mass-binned $t_Q$ distributions (bottom-left panel) as a function of stellar mass, to investigate whether quenching start time is mass dependent. This analysis reveals a systematic dependence of quenching time on stellar mass, whereby more massive galaxies typically begin quenching at earlier epochs compared to their lower-mass counterparts. This result recovers the well-established phenomenon of `downsizing' in galaxy evolution, whereby massive systems exhaust their gas reservoirs and transition to quiescence earlier in cosmic history compared to lower mass systems \citep[e.g.,][]{Cowie1996, Heavens2004, Thomas2005}.

\begin{figure*}
    \centering
    \includegraphics[width=0.8\textwidth]{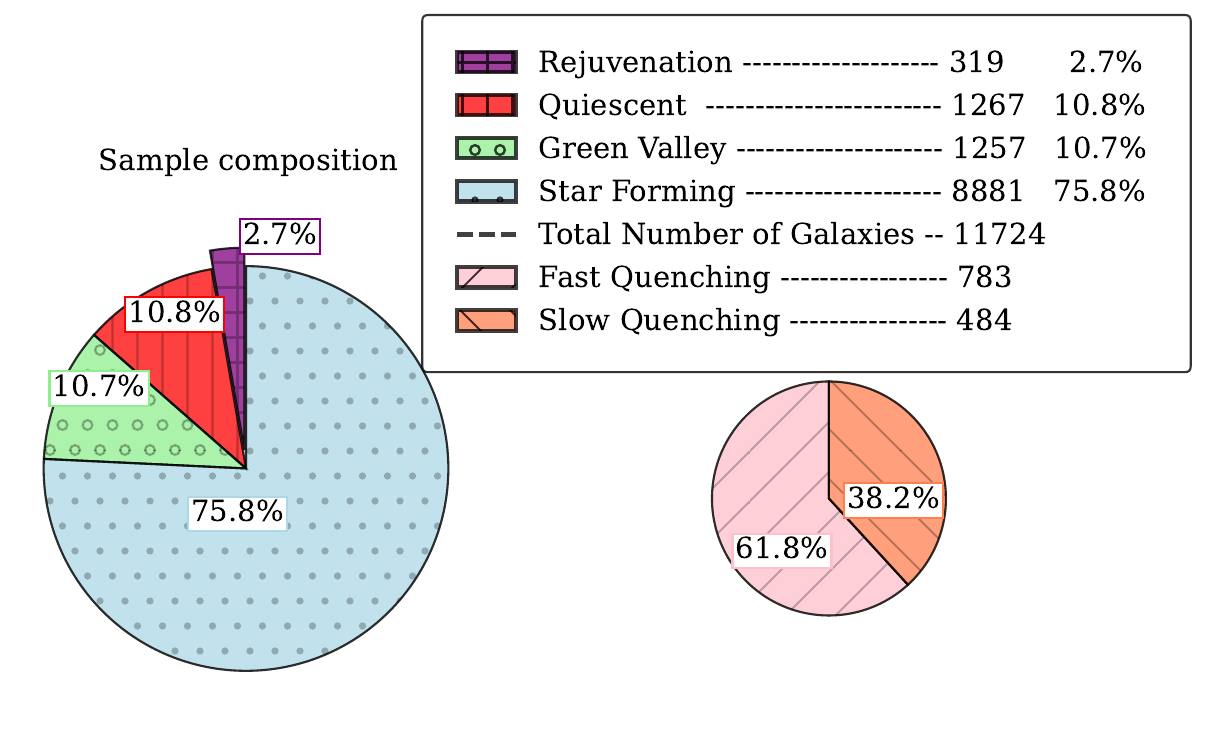}
    \caption{Fractional composition of our central galaxy sample (left-hand panel) and the breakdown of quenching timescales for quenched galaxies (right-hand panel). The left-hand panel shows that $\sim76\%$ remain star-forming, $\sim11\%$ are fully quenched, $\sim11\%$ lie in the green valley, and $\sim3\%$ experience rejuvenation (by $z = 0$).  The right-hand panel indicates that $\sim62\%$ of quenched galaxies follow a fast-quenching pathway, while $\sim38\%$ quench more slowly.}
    \label{Sample_Comp}
\end{figure*}

The top-right panel of Figure~\ref{QST_QDT} displays the distribution of quenching durations for the quenched population. The clear bimodal structure, captured by fitting two Gaussians to the distribution, indicates the existence of two distinct quenching channels: one corresponding to rapid (fast) quenching and another to more gradual (slow) quenching. To objectively separate these regimes, we adopt a threshold at $\log_{10}(\Delta \tau_{Q}/\mathrm{yrs}) = 8.75$ (equivalent to $\Delta\tau_Q = 0.56~\mathrm{Gyr}$), which is defined by the intersection point of the two Gaussian fits (i.e., the minimum of the bi-Gaussian model). Galaxies with quenching durations below this threshold are classified in this work as fast-quenching, while those above this threshold are classified as slow-quenching. We will explore the potential connection of mergers to both of these quenching modes in subsequent subsections.

The bottom-right panel explores the relationship between stellar mass and quenching duration. The fraction of slow-quenching galaxies shows an increase with stellar mass, suggesting a tendency for more massive systems to undergo somewhat longer quenching phases, although this trend is not strictly monotonic, particularly in the highest-mass bin. Rapid (fast) quenching still dominates at lower stellar masses. This weak mass dependence of quenching timescales is broadly consistent with trends reported in observational data \citep{Pacifici2016, Trussler2020}. 

However, recent observational studies suggest that the opposite trend can also occur, with evidence for rapid quenching in more massive systems and more gradual quenching at lower stellar masses. For instance, \citealt{Carnall2019} and \citealt{Bravo2023} find that massive galaxies in LEGA-C and GAMA samples tend to transition onto the red sequence on shorter timescales than less massive counterparts, while \citealt{Hahn2017} report similar behavior for SDSS central galaxies, with high-mass centrals quenching in less than $\sim$0.5~Gyr compared to $\gtrsim$1.5~Gyr for lower-mass systems. These findings indicate that both trends may coexist in nature, with quenching efficiency and duration depending sensitively on the interplay between mass, gas supply, and feedback mode.

The implications of these results are twofold. First, they highlight that quenching is not a monolithic process in TNG for central galaxies, but rather encompasses a spectrum of quenching times, with both fast and slow pathways contributing to the buildup of the quenched population. Second, the strong dependence of quenching time and duration on stellar mass suggests that different physical processes may govern star formation shutdown in low- and high-mass central galaxies.

Taken together, Figure~\ref{QST_QDT} quantifies the temporal landscape of galaxy quenching in our sample. Furthermore, it establishes a clear methodological foundation for linking quenching timescales to underlying physical processes, and for distinguishing between fundamentally different evolutionary pathways within the quenched population.

\subsection{Sample summary}
\label{sec:Sample_Summary}

To provide a concise overview of our sample and the classification scheme established this far, we present in Figure~\ref{Sample_Comp} a summary of the fractional composition of our central galaxy sample and the breakdown of quenching timescales among the quenched population. The left panel of the figure shows that the majority of galaxies ($\sim76\%$) remain star-forming at $z=0$, while $\sim11\%$ are classified as fully quenched, $\sim11\%$ occupy the green valley, and only a small fraction ($\sim3\%$) exhibit rejuvenation episodes. 

The right panel further divides the quenched category into fast- and slow-quenching sub-populations, illustrating that fast quenching accounts for approximately $62\%$ of quenched galaxies, while the remaining $38\%$ undergo a more gradual, slow quenching process. This summary highlights the prevalence of sustained star formation within our sample, the relative rarity of rejuvenation, and the substantial diversity in quenching timescales among those galaxies that do cease star formation. These well-characterized sub-populations provide a robust framework for further analysis of the physical processes governing galaxy evolution in the subsequent sections of this work.

\subsection{Are galaxy mergers sufficient to cause quenching?}
\label{sec:Given_Mergers}

\begin{figure*}
    \centering
    \includegraphics[width=0.7\textwidth]{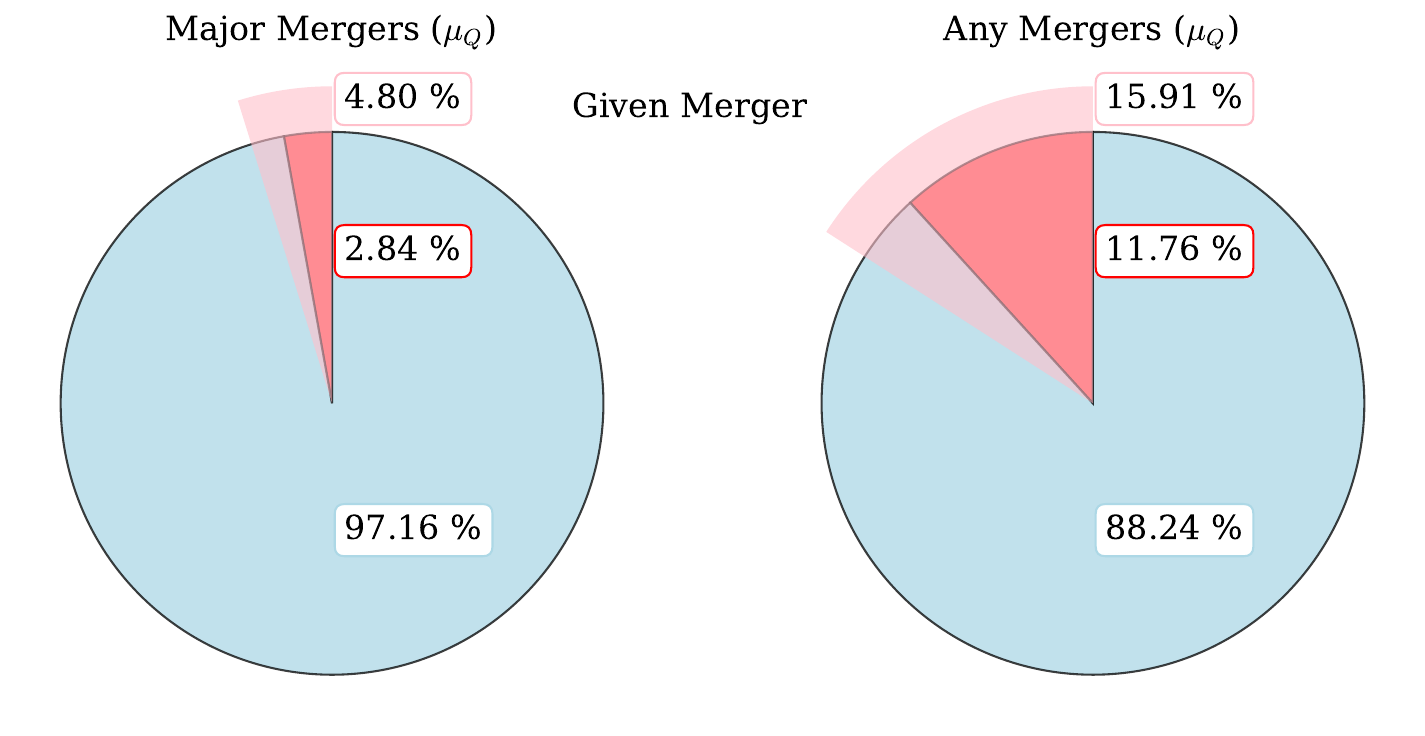}
    \caption{The fraction of mergers that are followed by quenching within various time windows. Left-hand panel: The fraction of all major mergers that take place within the quenching window, defined as spanning from 1\,Gyr before the onset of quenching to the completion of the transition. The extended light-shaded section of the pie chart represents the corresponding fraction when extending this window to 2\,Gyr before quenching begins. Right-hand panel: The same analysis as in the left panel but here considering any mergers. Since the fractions in both panels are low, this implies that mergers — especially major mergers — are not sufficient to cause quenching on their own. Nonetheless, they may still contribute to quenching in combination with other processes.}
    \label{Given_Merg}
\end{figure*}

A central question in galaxy evolution is whether mergers cause the quenching of star formation in central galaxies, either directly via engendering a starburst and gas depletion, or indirectly via triggering AGN feedback. Ideally, one would like to quantify the fraction of mergers that directly cause quenching, a quantity we denote as $\theta_M$. However, establishing true causality in this context is extremely challenging (even in simulations), as it requires disentangling complex, interconnected astrophysical processes that can contribute to quenching. While we cannot measure $\theta_M$ directly, we can place a robust upper limit on its value by considering the fraction of mergers that occur within the \wind[$\Delta t$]\text{ } for each galaxy (a window of time that spans $\Delta t$ before the onset of quenching and continues through quenching completion, see Section~\ref{sec:Merger_Quenchin_Window}).

To this end, we define $\mu_M$ as the fraction of mergers that occur in temporal proximity to a quenching event, specifically, those that take place within a window extending from 1Gyr (\wind[1]) or, in alternative tests, 2Gyr (\wind[2]), before the onset of quenching up to the completion of the quenching transition. By construction, $\theta_M \leq \mu_M$, since only mergers that occur within this time window could plausibly play a direct causal role in quenching (see section~\ref{sec:Merger_Quenchin_Window}.). Thus, $\mu_M$ provides an empirical ceiling on the true causal contribution of mergers: if $\mu_M$ itself is small, then the actual fraction of mergers that are truly responsible for quenching must be even smaller.

For this analysis, we consider both major mergers and all mergers, and report results separately for each scenario. This approach, which we refer to as the “given merger” framework, effectively asks: given that a merger occurs, how likely is quenching to follow? In other words, it tests whether mergers are sufficient to lead to quenching in the TNG simulation.

Figure~\ref{Given_Merg} summarizes our findings with two pie plots. The left-hand panel shows that, for major mergers, only about $3\%$ occur within \wind[1], rising up to $4.80\%$ when the window is extended to \wind[2]. The right-hand panel demonstrates that, when considering any merger, these fractions are somewhat higher ($11.76\%$ and $15.91\%$, respectively), but still constitute a clear minority. Hence, the overwhelming majority of mergers (regardless of type) do not lead to quenching within 1 (or even 2) Gyr. Since $\mu_M$ is already small in all cases, it follows that $\theta_M$, the true fraction of mergers that directly cause quenching, must be smaller still.

Crucially, this result demonstrates that mergers rarely lead to quenching in TNG. The vast majority of both major and minor mergers occur within star forming systems which remain star forming thereafter. Therefore, mergers (of any type) by themselves do not cause quenching. Nonetheless, it remains possible that, whilst not sufficient to cause quenching, they may be necessary to engender this transition.

\subsection{Are galaxy mergers necessary to cause quenching?}
\label{sec:Given_Quenching}

\begin{figure*}
    \centering
    \includegraphics[width=0.85\textwidth]{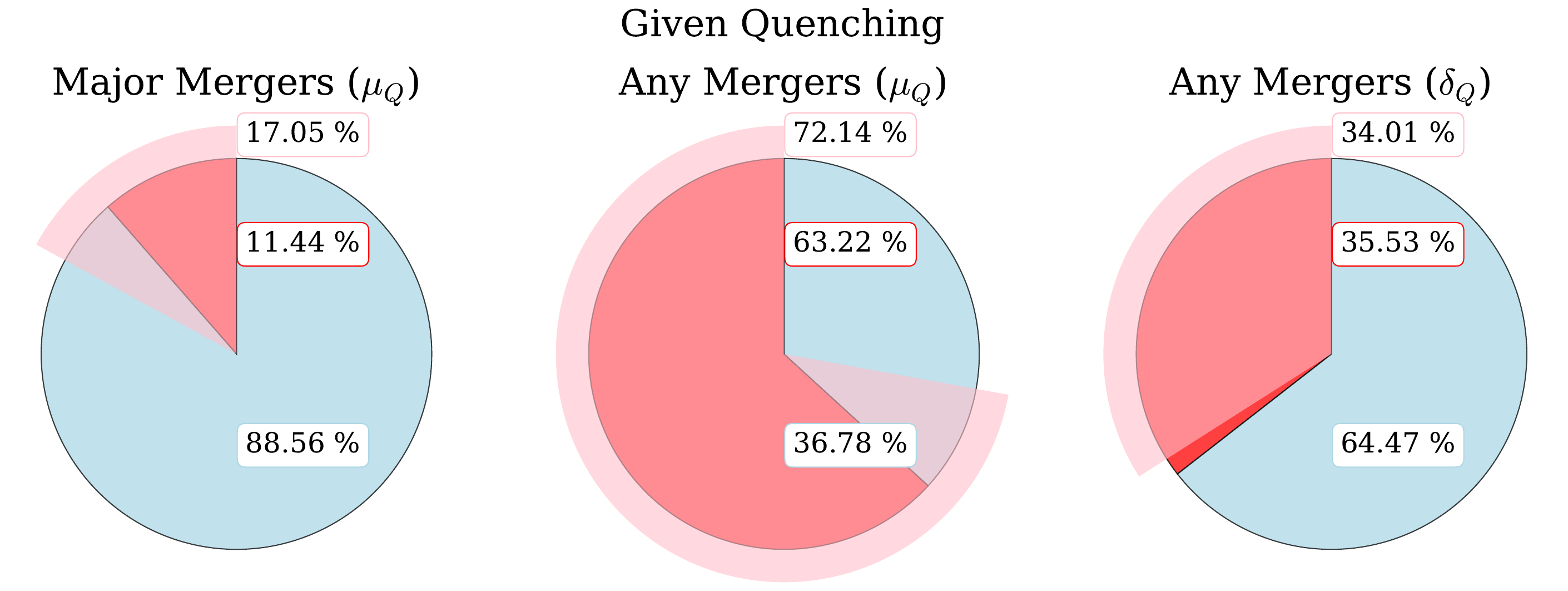}
    \caption{The fraction of quenched galaxies that experience at least one merger within a specific time window prior to their quenching event. \textit{Left-hand panel:} The fraction of quenched galaxies that undergo any major merger within a time window spanning from 1\,Gyr before the onset of quenching to the completion of the transition. The extended light-shaded section of the pie chart represents the corresponding fraction when extending this window to 2\,Gyr before quenching begins. \textit{Middle panel:} The same analysis as in the left panel but considering any mergers. \textit{Right-hand panel:} The same analysis as in the central panel but controlling for spurious effects from the probability of minor mergers per unit time in star forming systems. The low association fractions between quenching events and mergers, especially for major mergers, indicate that most quenching events occur independently of mergers.}
    \label{fig:Given_Q_No_SM_C}
\end{figure*}

While our previous analysis demonstrates that mergers are not sufficient to cause quenching in central galaxies within TNG (section~\ref{sec:Given_Mergers}), they could still play a necessary role in enabling quenching to occur. To test this possibility, we develop an analogous framework to that used for sufficiency, but now reverse the causal ordering. Explicitly, we ask: what fraction of quenching events present with at least one merger within the \wind?

Ideally, one would like to directly measure the fraction of quenched galaxies for which a merger is the true causal agent, which we denote as $\theta_Q$. However, as before, disentangling the complex, interconnected processes that contribute to quenching is extremely challenging, even in simulations. Therefore, we instead compute an empirical upper limit, $\mu_Q$, defined as the fraction of quenched galaxies that experience at least one merger within the association window specified in Section~\ref{sec:Merger_Quenchin_Window} (i.e., spanning from 1 or 2~Gyr before the onset of quenching to the completion of the transition). By construction, $\theta_Q \leq \mu_Q$, since only mergers occurring within this time window could plausibly be directly linked to quenching.

Applying this analysis, we find that for major mergers, $\mu_Q~=~11.44\%$ for a 1~Gyr window and $\mu_Q = 17.05\%$ for a 2~Gyr window. These values indicate that major mergers cannot be a necessary condition for quenching, as the vast majority of quenched galaxies do not experience a major merger within the relevant time windows. In contrast, considering any mergers (i.e., both major and minor), we obtain much larger upper limits of: $\mu_Q = 63.22\%$ (\wind[1]) and $\mu_Q = 72.14\%$ (\wind[2]). However, these elevated fractions may simply reflect the ubiquity of minor mergers in TNG, and do not necessarily provide meaningful insight into the true causal role of mergers, as $\theta_Q$ could in principle be anywhere in the range, $\theta_Q = \{0 - \mu_Q\}$.

To refine our estimate further, we control for spurious associations. Explicitly, we compare the merger histories of quenched galaxies to those of a matched control sample of star forming galaxies. For each quenched galaxy, we extract its individual \window and then artificially impose this window on the star forming control galaxies, recording how many experience at least one merger within this time interval (without quenching). We then define:

\begin{equation}
{\beta_Q = \frac{\sum_{i=1}^{N_Q} N_{{\rm SF}\text{-}m,i}}{N_Q N_{\rm SF}}}
\end{equation}

\noindent where, $N_Q$ is the number of quenched galaxies, $N_{\rm SF}$ is the number of control star-forming galaxies, and $N_{{\rm SF}\text{-}m,i}$ is the number of control galaxies experiencing at least one merger in the artificial window associated with quenched galaxy, $i$. This statistic, $\beta_Q$, quantifies the expected fraction of merger associations arising purely by chance (i.e., totally independent of quenching).

We then define a new upper limit ($\delta_Q$):

\begin{equation} 
\delta_Q = \mu_Q - \beta_Q
\end{equation}

\noindent which represents the fraction of quenching events associated with a merger in excess of what is expected from random temporal coincidence (as ascertained in purely star forming galaxies). \\\\

The following inequality: 

\begin{equation}
0 \leq \theta_Q \leq \delta_Q \leq \mu_Q
\end{equation}

\noindent follows from the way each statistic controls for confounding variables. The true causal fraction, $\theta_Q$, is the fraction of quenched galaxies for which a merger is actually the cause of quenching, requiring perfect control for all astrophysical processes and confounders. In contrast, $\mu_Q$ is a raw upper limit, set by the fraction of quenched galaxies that happen to experience a merger within the association window, controlling only for temporal coincidence and not for any other factors. Thus, $\mu_Q$ is always an overestimate. By refining $\mu_Q$ with an additional control for spurious associations, specifically, by subtracting $\beta_Q$, the expected chance overlap found in the control (star forming) sample, we obtain $\delta_Q$. This represents a more stringent upper limit, now controlling both for time and for chance coincidence, but still not for all possible confounding variables. Consequently, we must have $\theta_Q \leq \delta_Q \leq \mu_Q$, because $\delta_Q$ is closer to the true causal value, but not yet fully corrected from all other potential influences.

Using this approach, we find that for any mergers, $\delta_Q = 35.53\%$ and $\delta_Q = 34.01\%$ for \wind[1] and \wind[2], respectively. These values indicate that, after accounting for spurious associations, a majority of quenching events do not coincide with a relevant merger of any mass ratio.

Figure~\ref{fig:Given_Q_No_SM_C} summarizes these findings for both major mergers and all mergers, with and without control for chance associations. These values of $\mu_Q$ for major mergers and $\delta_Q$ for all mergers underscores that mergers are not essentially necessary prerequisite for quenching in the a majority of cases. Put simply, most of central galaxies quench in TNG without experiencing (major or minor) mergers in a causally relevant time window.

It is noteworthy that, while $\mu_Q$ increases as the association window is widened (as expected, since a broader window is more likely to encompass random mergers), $\delta_Q$ actually decreases slightly from 1~Gyr to 2~Gyr. This trend arises because, as we consider windows further back in time, the merger histories of quenched and control star-forming galaxies become increasingly similar. Ultimately, the probability of a random merger occurring in either population approaches equality as the window approaches infinity, reducing the excess found in the quenched sample. The fact that $\delta_Q$ does not increase (and even decreases) when the window is widened demonstrates that this methodology effectively controls for spurious merger associations, actually reinforcing the robustness of our conclusion.

\subsubsection{Controlling for the Merger Rate - Stellar Mass Relation}

\begin{figure}
    \centering
    \includegraphics[width=0.5\textwidth]{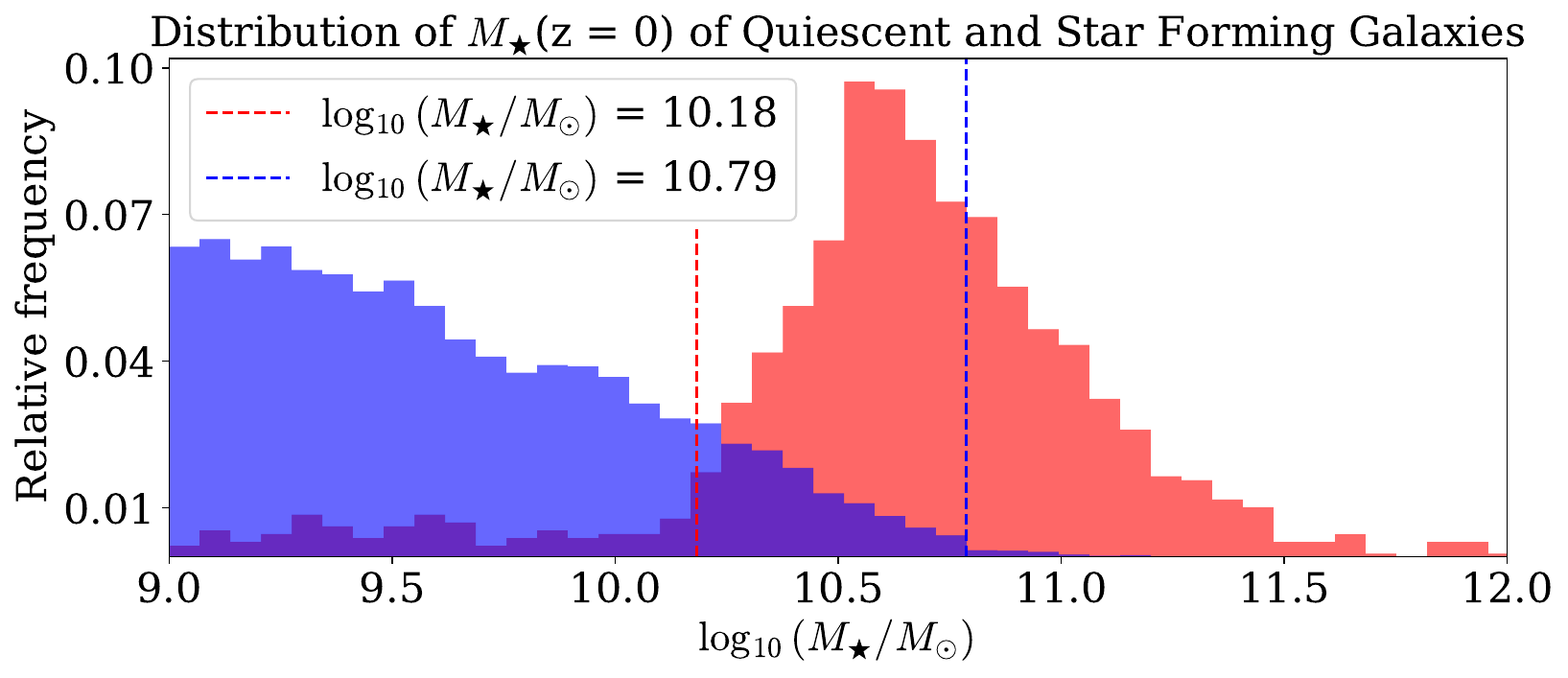}
    \caption{
    Distribution of stellar masses for star-forming and quenched galaxies in the full sample. The vertical dashed lines delimit the restricted range $10^{10.18} \leq M_\star / M_\odot \leq 10^{10.79}$, within which 668 quenched galaxies and 1233 star-forming galaxies are selected. This cut ensures a fairer comparison of merger statistics between the two populations.}
    \label{fig:SM_Dist}
\end{figure}

\begin{figure*}
    \centering
    \includegraphics[width=0.85\textwidth]{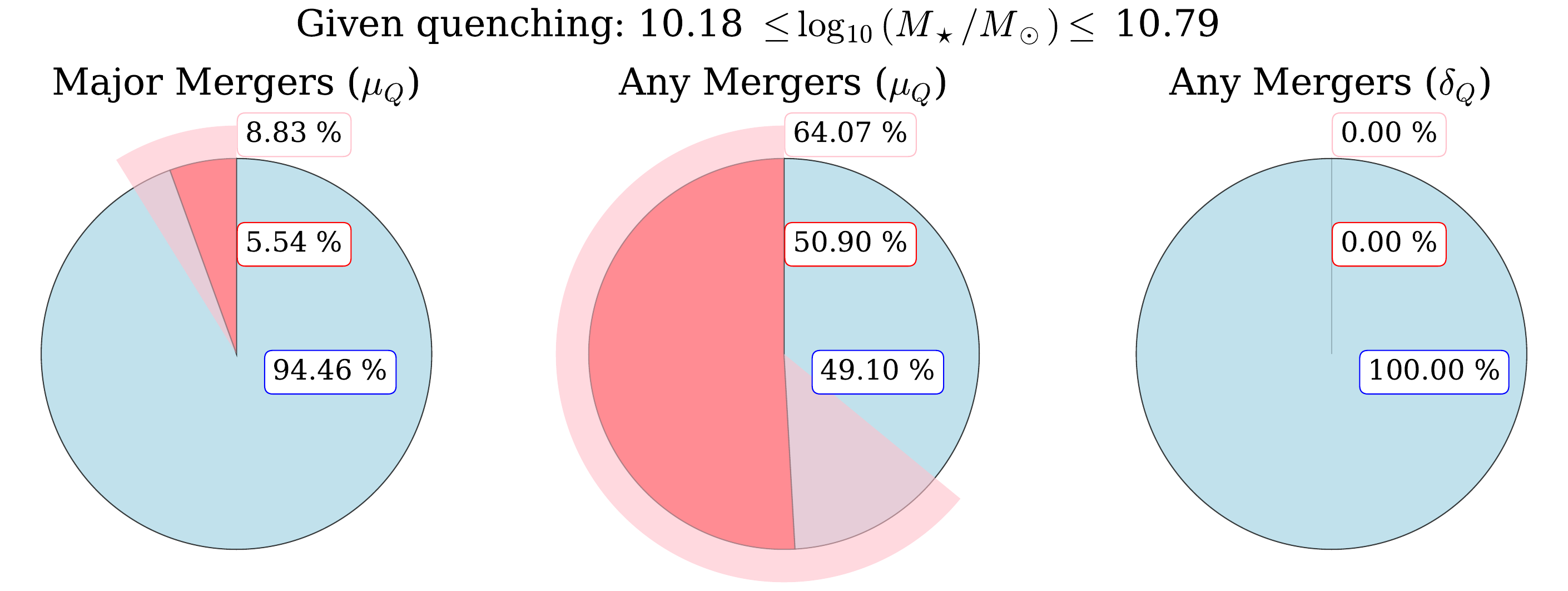}
    \caption{
        The fraction of quenched galaxies in the mass-controlled sample ($10^{10.18} \leq M_\star / M_\odot \leq 10^{10.79}$) that experience at least one merger within the association window, before and after correction for spurious associations. This figure is identical in structure to Fig. 6, but shown here for the reduced stellar mass range. The smaller values of $\mu_Q$ and $\delta_Q$ highlight the minor role of mergers in quenching when controlling by the merger rate - stellar mass relation.
    }
    \label{fig:Given_Q_SM_C}
\end{figure*}

The above analysis can be refined one step further by considering the role of stellar mass. We compute both the Pearson ($r$) and Spearman ($\rho$) correlation coefficients between the total number of mergers and stellar mass for our galaxy samples, finding $r = 0.59$ and $\rho = 0.73$ for star-forming galaxies, $r = 0.65$ and $\rho = 0.89$ for quiescent galaxies, and $r = 0.62$ and $\rho = 0.80$ for the combined sample. These values indicate moderate-strong positive correlations, especially within the quiescent population, suggesting that more massive galaxies tend to have experienced more mergers, regardless of whether or not they have quenched.

As shown in Figure~\ref{fig:SM_Dist}, star-forming galaxies in our sample tend to be less massive than quenched galaxies (as expected, e.g. \citealt{Knobel2015,Bluck2016,Donnari2021}). Because of the observed correlation between stellar mass and merger count, these systematic differences in mass distributions can influence the values of $\delta_Q$ reported previously.

To mitigate this mass bias, we recompute $\mu_Q$ and $\delta_Q$ for a restricted subset of both populations, with stellar masses in the range $10^{10.18} \leq M_\star / M_\odot \leq 10^{10.79}$, as illustrated in Figure~\ref{fig:SM_Dist}. The bin width in this histogram is determined using Freedman–Diaconis’s rule, $\Delta = 2\cdot\,{\rm IQR}(x)\,\cdot n^{-1/3} \approx 0.07 \,\text{dex}$ (where, IQR is the inter-quartile range across each sub-population). The vertical dashed lines mark the stellar mass limits selected such that all histogram bins within this range contain at least 20 galaxies in count space for their respective populations. This criterion ensures that both subsamples include statistically representative numbers of galaxies across the selected mass range. The blue line represents the upper mass limit chosen so that the star-forming sample satisfies this criterion, whereas the red line marks the lower mass limit chosen to satisfy it for the quenched sample. This cut ensures that the two populations are more directly comparable in their merger statistics, with star-forming controls computed from galaxies of similar mass to the quenched population.

Figure~\ref{fig:Given_Q_SM_C} summarizes the results of this mass-controlled analysis. For major mergers, we find $\mu_Q = 5.54\%$ for a 1 Gyr window and $8.83\%$ for a 2 Gyr window. For any mergers, $\mu_Q = 50.90\%$ (1 Gyr) and $64.07\%$ (2 Gyr), and after correcting for spurious associations, $\delta_Q = 0.00\%$ for both time windows. In fact, in this case, the raw corrected values of $\delta_Q$ are slightly negative, indicating a slight excess of spurious associations relative to the uncorrected signal. For consistency, however, we report $\delta_Q = \max(\delta_Q,\, 0)$, reflecting that no significant residual merger–quenching correlation remains after controlling for both mass and spurious matches.

These values are systematically lower than those for the full sample, demonstrating that the apparent merger–quenching association is, in part, driven by underlying differences in stellar mass. Within a fixed stellar mass bin, the merger histories of star-forming and quiescent galaxies become substantially more similar, and $\delta_Q$ is further reduced, essentially to zero. Therefore, there is no evidence at all for mergers (of any mass ratio) being necessary to enable quenching of central galaxies in TNG.

\subsubsection{Fast vs. Slow Quenching}

So far in this section we have established that mergers are neither sufficient nor necessary for central galaxy quenching. However, it remains possible that certain quenching events (e.g., fast or slow quenching) may require a merger trigger. To investigate this possibility, we extend this mass-controlled analysis by separately evaluating galaxies that undergo fast and slow quenching. 

\begin{figure*}
    \centering
    \includegraphics[width=0.85\textwidth]{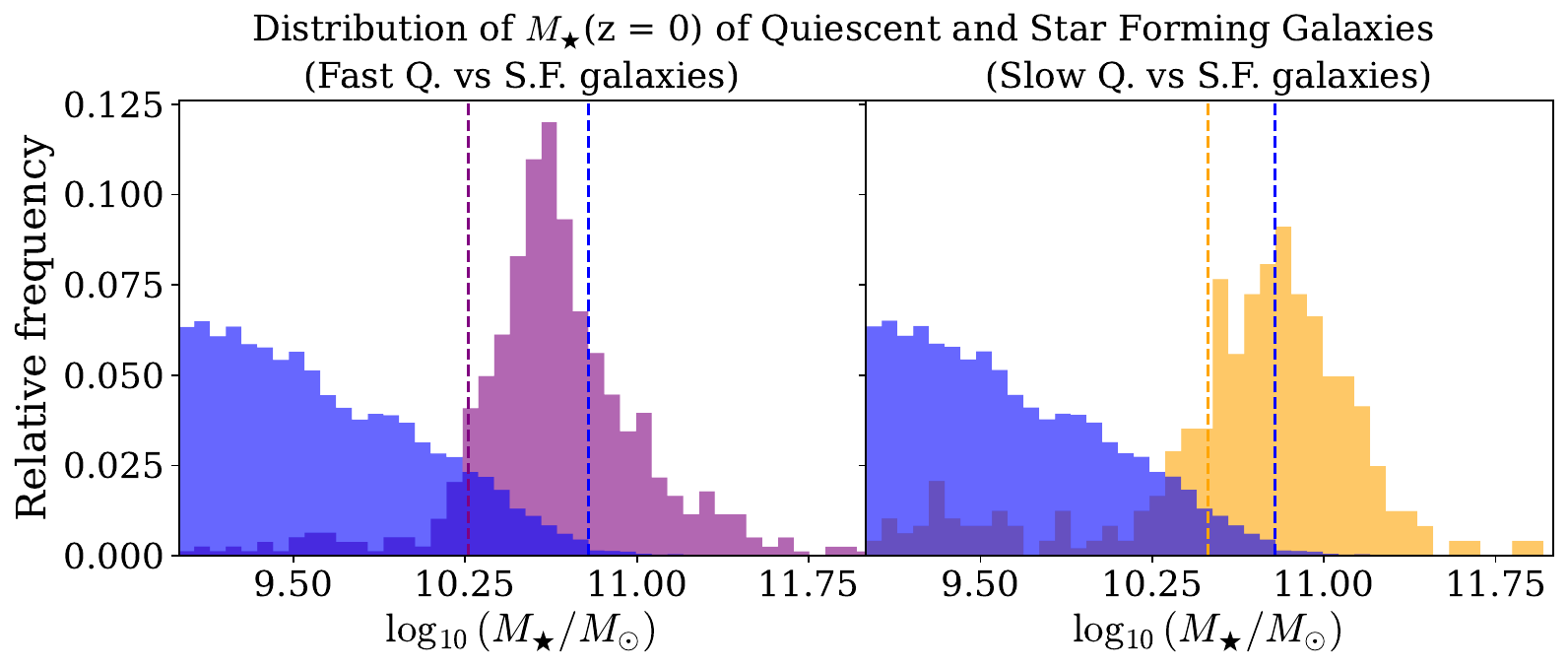}
    \caption{
     Distribution of stellar masses for fast- and slow-quenching galaxies compared to the star-forming population. Left: fast-quenching vs. star-forming galaxies. Right: slow-quenching vs. star-forming galaxies. The vertical dashed lines delimit the mass ranges used for each analysis. In both panels, the upper mass limit (blue line) is fixed at $\log_{10}(M_\star/M_\odot) = 10.79$, while the lower limits are set to $\log_{10}(M_\star/M_\odot) = 10.26$ for fast-quenching galaxies (purple line) and $\log_{10}(M_\star/M_\odot) = 10.50$ for slow-quenching galaxies (orange line). These cuts ensure that all bins contain at least 20 galaxies in count space. The resulting subsamples consist of 471 fast-quenching and 959 star-forming galaxies (left), and 131 slow-quenching and 331 star-forming galaxies (right). This approach enables a mass-controlled comparison of merger statistics across distinct quenching timescales.
    }
    \label{fig:Given_Fast_Slow_SM}
\end{figure*}

As in the previous case (of all quenching), we restrict the stellar mass range using a bin width of approximately 0.07 dex and impose a minimum count threshold of 20 galaxies per bin. The upper mass limit remains fixed at $10^{10.79}\,M_\odot$, while the lower limits are adjusted for each quenching mode to ensure adequate statistics. Specifically, the fast-quenching sample uses a lower cutoff at $10^{10.26}\,M_\odot$ (purple dashed line), yielding 471 fast-quenching and 959 star-forming galaxies (left panel of Figure~\ref{fig:Given_Fast_Slow_SM}). The slow-quenching sample uses a lower cutoff at $10^{10.50}\,M_\odot$ (orange dashed line), resulting in 131 slow-quenching and 331 star-forming galaxies (right panel of Figure~\ref{fig:Given_Fast_Slow_SM}). These cuts allow for a fair comparison of merger statistics across quenching modes, while controlling for stellar mass in a consistent manner.

\begin{figure*}
    \centering
    \includegraphics[width=0.85\textwidth]{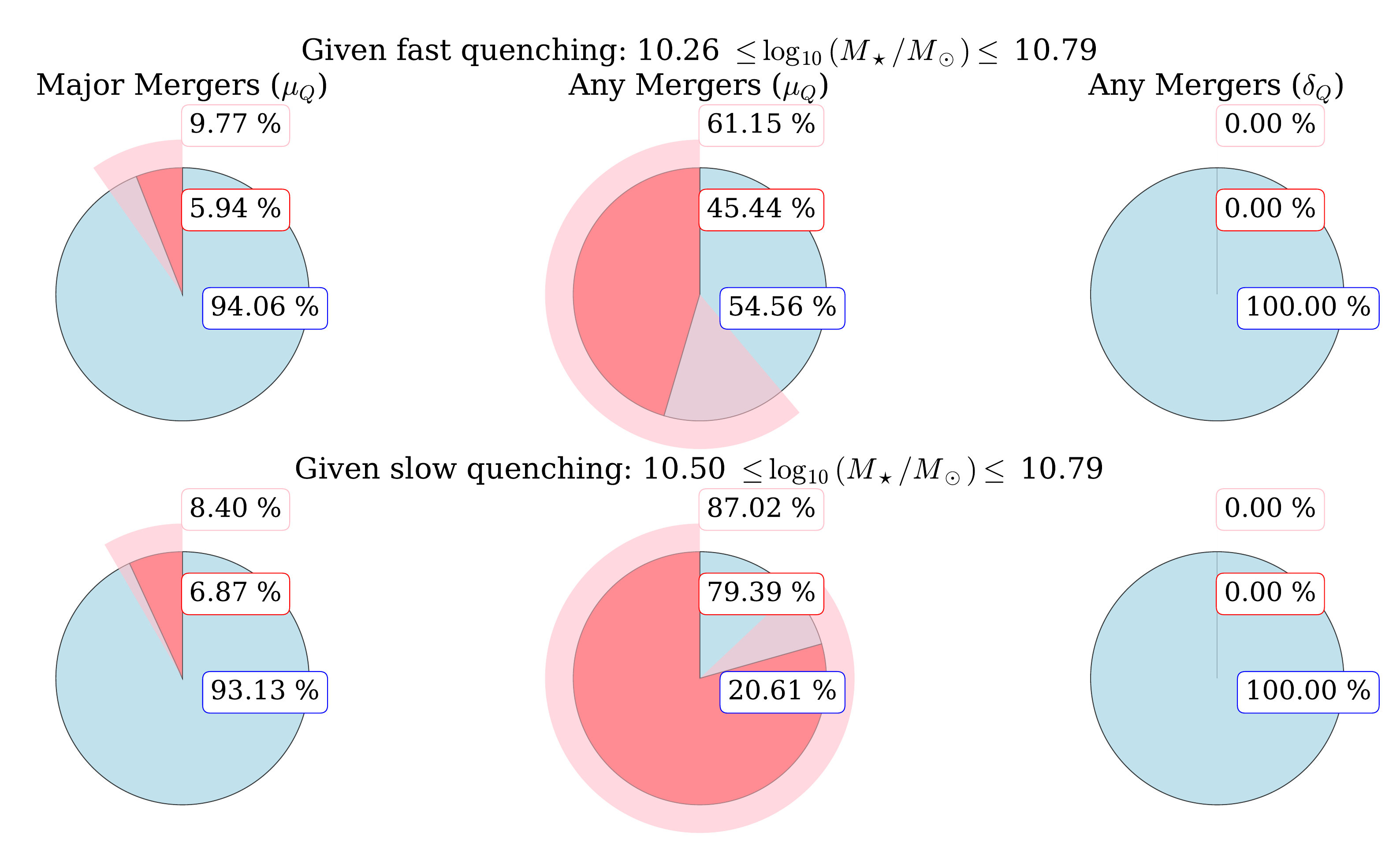}
    \caption{
    The fraction of fast- and slow-quenching galaxies in the mass-controlled samples that experience at least one merger within the association window, before and after correction for spurious associations. Top row: fast-quenching sample ($10^{10.26} \leq M_\star/M_\odot \leq 10^{10.79}$), based on 471 fast-quenching and 959 star-forming galaxies. Bottom row: slow-quenching sample ($10^{10.50} \leq M_\star/M_\odot \leq 10^{10.79}$), based on 131 slow-quenching and 331 star-forming galaxies. As in previous analyses, $\mu_Q$ denotes the raw merger fraction and $\delta_Q$ the corrected excess over star-forming controls. In both cases, $\delta_Q = 0.00\%$, as corrected values are slightly negative and thus clipped to zero. These results indicate that the link between mergers and quenching weakens to zero when controlling for both stellar mass and quenching timescales.
    }
    \label{fig:Given_Fast_Slow}
\end{figure*}

In Figure~\ref{fig:Given_Fast_Slow} we present the results of the mass-controlled merger analysis for fast vs. slow quenching galaxies, using both 1\,Gyr and 2\,Gyr association windows (as before). For fast quenching galaxies (top row), the fraction of quenching events associated with a major merger is just $\mu_Q = 5.94\%$ for a 1\,Gyr window, increasing to $9.77\%$ for a 2\,Gyr window. When considering any mergers, the corresponding values are $\mu_Q = 45.44\%$ (1\,Gyr) and $61.15\%$ (2\,Gyr). For slow-quenching galaxies (bottom row), the major merger fractions are $\mu_Q = 6.87\%$ (1\,Gyr) and $8.40\%$ (2\,Gyr), while for any mergers we find $\mu_Q = 79.39\%$ (1 Gyr) and $87.02\%$ (2 Gyr). In all cases, the corrected merger excess $\delta_Q$ is reported as $0.00\%$, since the actual values are slightly negative and are clipped to zero for consistency. 

Explicitly, this indicates that the number of minor mergers in the association window may be fully explained by chance, given the stellar mass dependent merger rate. These findings lead to the conclusion that, once stellar mass is controlled for, the role of mergers in both fast and slow quenching is minimal, with no significant excess relative to stellar mass matched star forming controls. Moreover, even without the need for careful control sampling, we find that major mergers are of little relevance for either the fast or slow quenching channels in TNG.

\begin{figure*}
    \centering
    \includegraphics[width=0.75\textwidth]{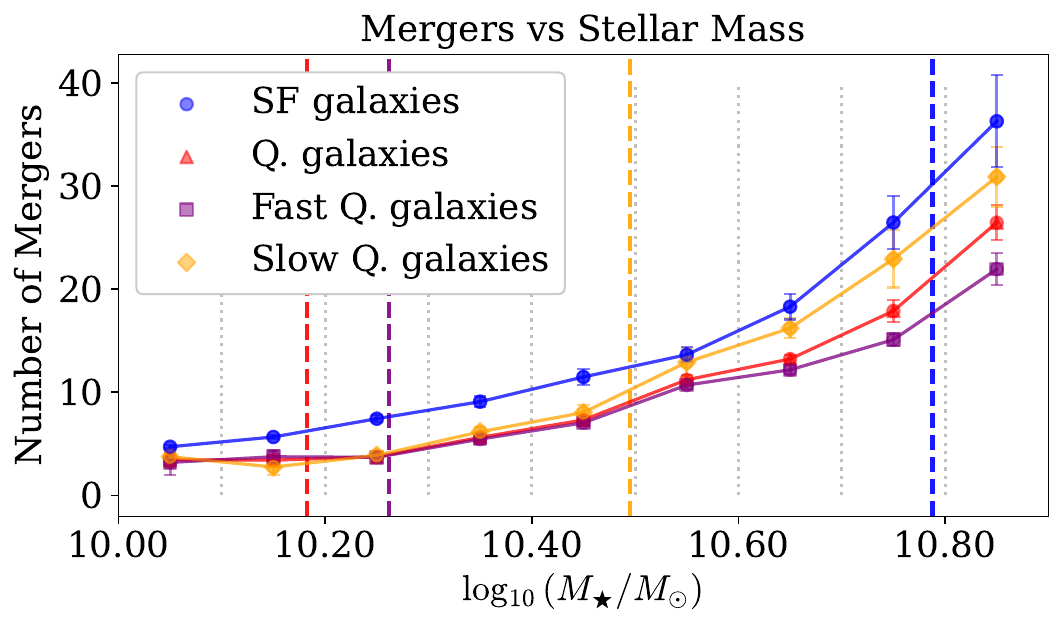}
    \caption{
    Average number of mergers per galaxy as a function of stellar mass for star-forming (blue), quenched (red), fast-quenching (purple), and slow-quenching (orange) populations. Each point represents the mean number of mergers in a stellar mass bin, with vertical error bars showing the standard error of the mean ($\sigma / \sqrt{n}$). Vertical gray dashed lines mark the bin edges, while the vertical colored lines reproduce the stellar mass limits applied in earlier analyses: the blue line is the universal upper limit ($\log_{10}(M_\star/M_\odot) = 10.79$), and the red, purple, and orange lines are the lower mass limits for quenched, fast-quenching, and slow-quenching samples, respectively. All populations exhibit a similar positive relationship between stellar mass and number of mergers, with star-forming galaxies showing slightly more mergers on average than their quiescent counterparts. This explains the slightly negative values of $\delta_Q$ seen previously. Importantly, this figure also demonstrates that mergers are not more frequent in quenched galaxies (of any kind) compared to star forming systems, once comparison is made at the same stellar mass.}
    \label{fig:Mergers_vs_SM}
\end{figure*}

Finally, in order to better understand the origin of the slightly negative corrected merger excesses ($\delta_Q$) observed in the previous analysis, we examine the average merger histories of each galaxy population as a function of stellar mass. Figure~\ref{fig:Mergers_vs_SM} shows the mean number of mergers per galaxy across stellar mass bins for star-forming, quenched, fast-quenching, and slow-quenching galaxies. The vertical gray dashed lines indicate the edges of the stellar mass bins used, while error bars represent the standard error on the mean within each bin. The vertical colored dashed lines reproduce the mass limits previously adopted: the universal upper limit for all samples (blue), and the lower mass limits specific to each quenching mode (red for all quenched, purple for fast-quenching, orange for slow-quenching).

This plot demonstrates that, once stellar mass is controlled for, all populations exhibit statistically similar merger rate–stellar mass relations, with star-forming galaxies showing higher average merger counts across all mass bins. However, star-forming galaxies predominantly occupy lower stellar masses, where the average merger counts are smaller. This mass distribution explains why a positive excess is still observed when stellar mass constraints are not imposed. This subtle reversal also accounts for the slightly negative $\delta_Q$ values obtained in the previous section, which were set to zero for consistency. Crucially, these results indicate that, once stellar mass is controlled for, quenched galaxies experience a comparable, or even slightly lower, number of mergers than star-forming systems. Taken together, this provides strong evidence that mergers, across all mass ratios, are not the primary drivers of quenching in TNG.

In summary of the entirety of this section so far, after controlling for both random temporal associations and stellar mass, our results robustly show that the overwhelming majority of quenching events do not involve a merger (of any mass ratio). Hence, mergers are not necessary for quenching in TNG. Moreover, as seen earlier, mergers are also not sufficient to cause quenching in TNG. Taken together, the results clearly challenge any scenario in which mergers are evoked as a direct quenching mechanism in this simulation.

\subsection{Random Forest Analyses: Is the merger history important for quenching and/or supermassive black hole growth?}

\begin{figure*}
    \centering
    \includegraphics[width=0.85\textwidth]{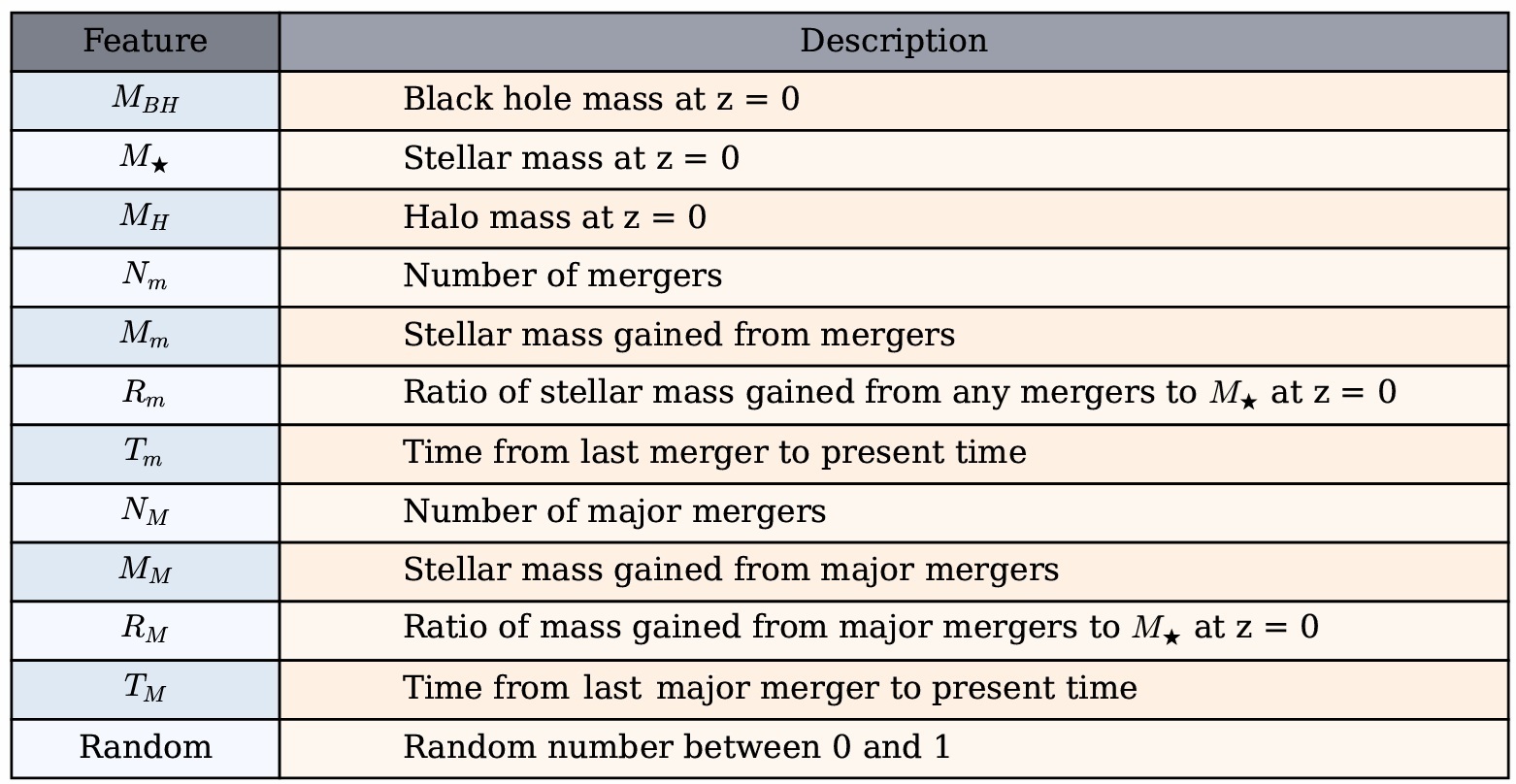}
    \caption{Summary of input features for the random forest classifier, comprising both merger-related and non-merger parameters to enable direct comparison of their relative importances. Non-merger features include present-day galaxy properties—black hole mass (\(M_{\mathrm{BH}}\)), stellar mass (\(M_\star\)), and halo mass (\(M_{\mathrm{H}}\))—while merger-related features cover merger counts (\(N_m\), \(N_M\)), stellar mass gained (\(M_m\), \(M_M\)), relative mass ratios (\(R_m\), \(R_M\)), and timing since the last merger (\(T_m\), \(T_M\)). A randomly generated feature (“Random”) is also included to establish a baseline for feature importance.}
    \label{fig:RF_Table}
\end{figure*}

In the previous section, we test the immediate necessity and sufficiency of mergers as direct triggers of galaxy quenching. We find that mergers alone are neither necessary nor sufficient for quenching central galaxies in IllustrisTNG. However, mergers could still influence galaxy quenching through a cumulative effect. For instance, mergers may lead to supermassive black hole growth, which can in turn lead to quenching via AGN feedback. Hence, mergers may yet be critical to the quenching of central galaxies in TNG, just in a more indirect manner than conceived of in previous theoretical work with idealized hydrodynamical simulations.

To investigate this important possibility comprehensively, we employ a Random Forest (RF) classifier to evaluate whether merger histories (summarized through various cumulative metrics) can effectively predict galaxy quenching outcomes. Moreover, we also investigate how predictive merger parameters are of supermassive black hole mass. Ultimately, we assess whether both quenching and supermassive black hole growth are controlled via mergers or secular processes. This enables deep insight into how the TNG model functions and offers a route to detailed observational tests moving forwards.

The random forest approach is ideal here for two reasons. First, this method provides a powerful route to control for an arbitrarily large set of nuisance parameters simultaneously, enabling robust insights beyond mere correlation (see \citealt{Bluck2022} for a detailed discussion). Second, due to the relative transparency of this method, the precise breakdown in importance of each parameter to solving the problem at hand is extractable in a straightforward manner, unlike in deep learning techniques. See Section \ref{RF} for details on the method.

\subsubsection{Random Forest Quenching Classifications}
\label{sec:RandomForestClass_Resutls}

\begin{figure*}
    \centering
    \includegraphics[width=0.95\textwidth]{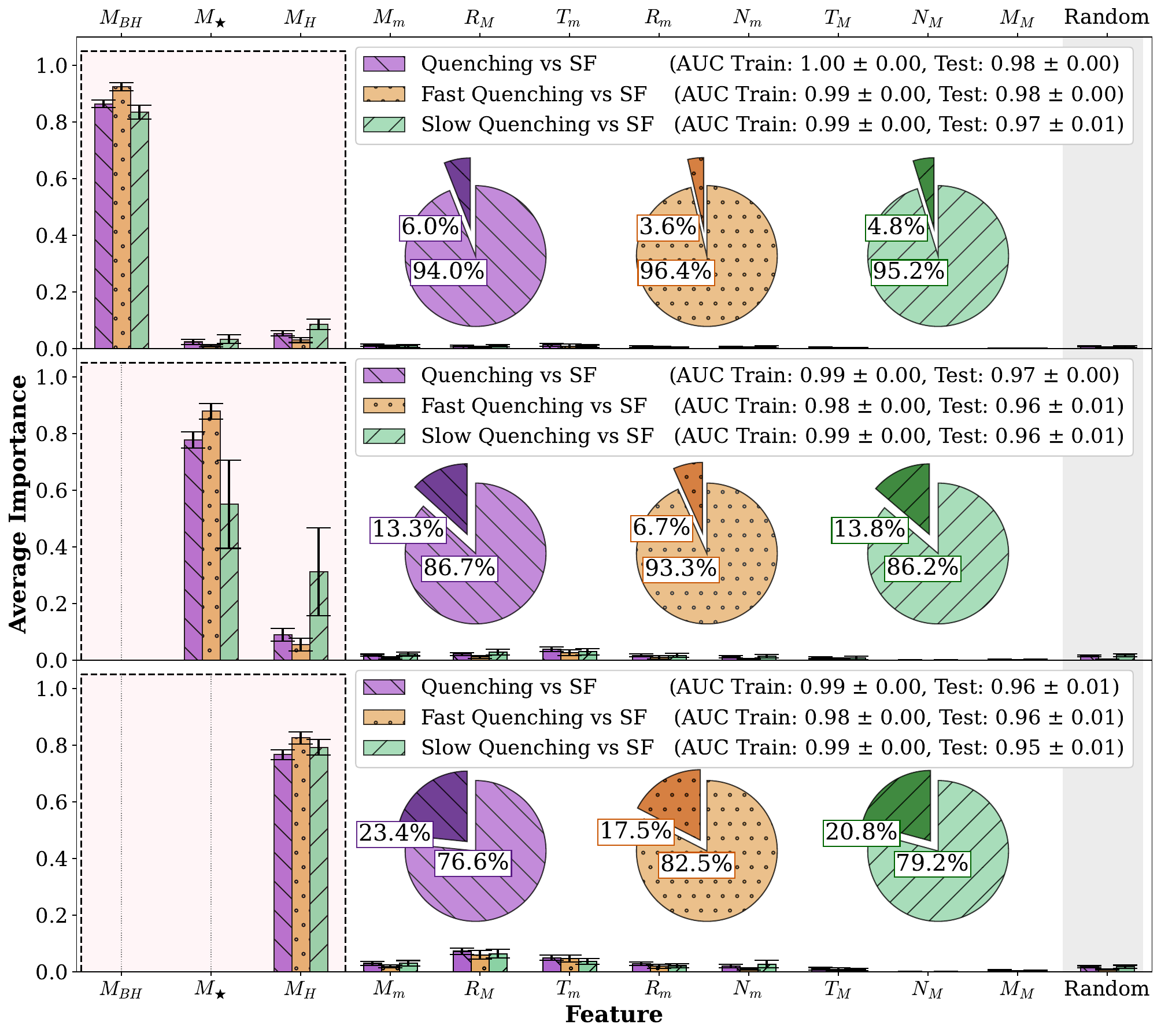}
    \caption{Random forest classification analyses to predict star forming vs. all quenched, star forming vs. fast quenched, and star forming vs. slow quenched categories (as labeled by the legends). The bar heights represent the mean importance of each feature across 100 independent runs, with error bars corresponding to the standard deviation of the importance values. The three panels correspond to different feature sets: (i) all parameters included, (ii) black hole mass ($M_{BH}$) excluded, and (iii) both black hole mass ($M_{BH}$) and stellar mass ($M_{\bigstar}$) excluded. The pie charts illustrate the cumulative feature importance, separated between merger-related parameters (i.e., $M_{M}$, $N_{M}$, $T_{M}$, $R_{M}$, $T_{m}$, $R_{m}$, $N_{m}$; left uncolored) and non-merger parameters (highlighted). In all cases, non-merger parameters overwhelmingly dominate the total feature importance, reinforcing the conclusion that mergers play a negligible role in predicting quenching, regardless of the inclusion or exclusion of $M_{BH}$ and $M_{\bigstar}$.}
    \label{fig:RF_All}
\end{figure*}

\begin{figure*}
    \centering
    \includegraphics[width=0.95\textwidth]{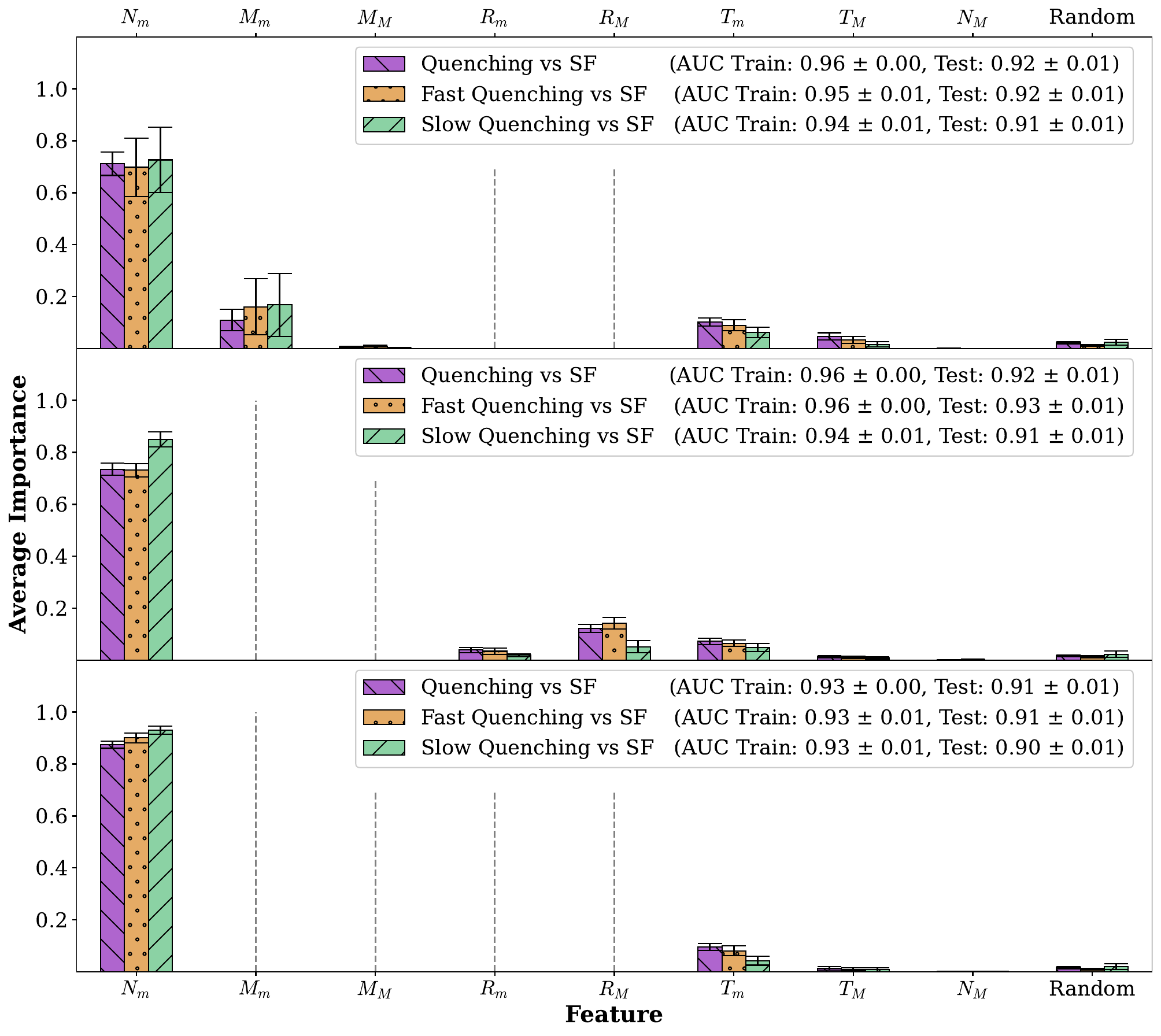}
    \caption{Random forest classification analyses restricted to merger-related parameters, with leakage control to avoid reconstructing intrinsic properties. As in Fig.~13, we classify star forming vs.\ all quenched, star forming vs.\ fast quenched, and star forming vs.\ slow quenched samples (see legends). Bar heights give the mean feature importance over 100 independent runs, and error bars show the standard deviation. The three panels correspond to feature sets designed to prevent recovery of present–day stellar mass from paired proxies, as follows. \textit{Top}: only the stellar–mass contributions from mergers, \(M_{m}\) and \(M_{M}\), are retained, with the fractional contributions \(R_{m}\) and \(R_{M}\) masked; \textit{Middle}: only \(R_{m}\) and \(R_{M}\) are retained, with \(M_{m}\) and \(M_{M}\) masked; \textit{Bottom}, all four \(\{M_{m}, M_{M}, R_{m}, R_{M}\}\) are masked. Masked predictors are indicated by vertical dashed bars. }
    \label{fig:RF_Only_Merg}
\end{figure*}

We train RF models to classify galaxies as: (i)~quenched versus star-forming, (ii)~fast-quenched versus star-forming, and (iii)~slow-quenched versus star-forming. This enables us to assess whether mergers may be more important for certain quenching modes. The input features consist of intrinsic galaxy properties, including central black hole mass at $z=0$ ($M_{\rm BH}$), stellar mass at $z=0$ ($M_{\star}$), and halo mass at $z=0$ ($M_{H}$). Additionally, we include a comprehensive list of merger history parameters: the total number of mergers ($N_{m}$), the total stellar mass gained from mergers ($M_{m}$), the fractional stellar mass growth from mergers (defined as the ratio of stellar mass gained from mergers to the stellar mass at $z=0$, $R_{m}$), and the time since the last merger event ($T_{m}$). Furthermore, we separate out major mergers specifically, using analogous parameters: the total number of major mergers ($N_{M}$), the stellar mass gained from major mergers ($M_{M}$), the fractional mass growth from major mergers (ratio of stellar mass gained from major mergers to the stellar mass at $z=0$, $R_{M}$), and time since the last major merger ($T_{M}$). Finally, a random number is also included as a control feature to set a baseline for negligible feature importance. See Figure~\ref{fig:RF_Table} for the full set of parameters.

The RF classification achieves high predictive accuracy across all scenarios, demonstrating strong discriminative power with testing-set AUC scores higher than 0.97 for quenched versus star-forming galaxies, and similarly high performance for fast and slow quenching subcategories. The consistently strong performance across all quenching definitions underscores the robustness of the classifier and its effective generalization.

In Figure~\ref{fig:RF_All} top-panel, we present results from the RF classification runs. The feature importance analysis reveals that supermassive black hole mass ($M_{\rm BH}$) dominates the predictive power of quenching in TNG (as seen previously in \citealt{Piotrowska2022, Bluck2023, Bluck2024b}). However, we extend these previous works by now including a comprehensive set of merger-related parameters. Merger-related parameters consistently exhibit very low importance, collectively contributing less than $\sim$6\% of the predictive signal in the RF classification. Individual merger parameters, such as the total stellar mass gained from mergers ($M_{m}$), the number of major mergers ($N_{M}$), or fractional mass growth from mergers ($R_{m}$), each show negligible importance compared to intrinsic properties. 

These results clearly demonstrate that the cumulative evolutionary effects from mergers do not substantially impact the likelihood of galaxy quenching, at fixed final intrinsic parameters (particularly supermassive black hole mass). Furthermore, these results are essentially identical in the fast and slow quenching sub-categories, indicating that mergers are no more important for the quenching of either of these variants in TNG.

Nevertheless, it remains possible that mergers are still important for quenching indirectly, as a route to growing supermassive black holes. Hence, we further investigate the potential indirect role of mergers by training RF models while systematically excluding the most important galaxy properties. Excluding black hole mass ($M_{\rm BH}$, Figure~\ref{fig:RF_All}, middle panel), stellar mass emerges as the dominant feature, with halo mass becoming secondary. Even in the absence of $M_{\rm BH}$, merger-related parameters continue to provide minimal predictive value, totaling at most 14\% cumulative importance. Notably, even direct indicators such as the stellar mass gained from mergers ($M_{m}$), or fractional mass growth from mergers ($R_{m}$), do not significantly increase in importance. 

Further excluding stellar mass ($M_{\star}$, Figure~\ref{fig:RF_All}, lower panel), halo mass alone effectively dominates the predictive power. While merger-related parameters increase slightly in cumulative importance to about 23\%, their individual contributions remain very low. Thus, even when forced into prominence by removing the primary intrinsic predictors, merger histories provide limited (or incremental) value for predicting quenching.

Continuing the analysis beyond Figure~\ref{fig:RF_All}, we re-run the random forest classifier using only parameters that encode the merger history: the number of mergers ($N_m$ and $N_M$), the stellar mass added by mergers ($M_m$ and $M_M$), the fractional mass gain ($R_m$ and $R_M$), and the time since the most recent merger ($T_m$ and $T_M$). Three separate models are built to avoid any possibility of leakage, i.e.\ the reconstruction of the present-day stellar mass from paired proxies. Including the total stellar mass gained from mergers ($M_m$ or $M_M$) and the fractional mass growth ($R_m$ or $R_M$) will allow the algorithm to infer $M_\star(z{=}0)=M_m/R_m$ (or $M_M/R_M$), thereby reintroducing a strong proxy for stellar and black hole mass. To prevent this, the first experiment keeps only $M_m$ and $M_M$ while masking $R_m$ and $R_M$, the second keeps only $R_m$ and $R_M$ while masking $M_m$ and $M_M$, and the third masks all four.

As summarized in Figure~\ref{fig:RF_Only_Merg}, even after removing the dominant intrinsic properties, the merger-only classifiers achieve surprisingly high test AUC values of approximately $0.90$--$0.93$ when at least one stellar mass growth indicator is retained. The fractional stellar mass gains, $R_m$ and $R_M$, yield test AUC values near $0.92$. Similarly, using only the total mass accreted through mergers produces comparable performance, with test AUCs around $0.92$. Even when all stellar growth proxies are masked, the classifier still achieves a test AUC of roughly $0.91$, indicating that the number of mergers and the time since the last merger still encode some information about quenching, although significantly less than that captured by secular parameters.

\begin{figure*}
    \centering
    \includegraphics[width=0.95\textwidth]{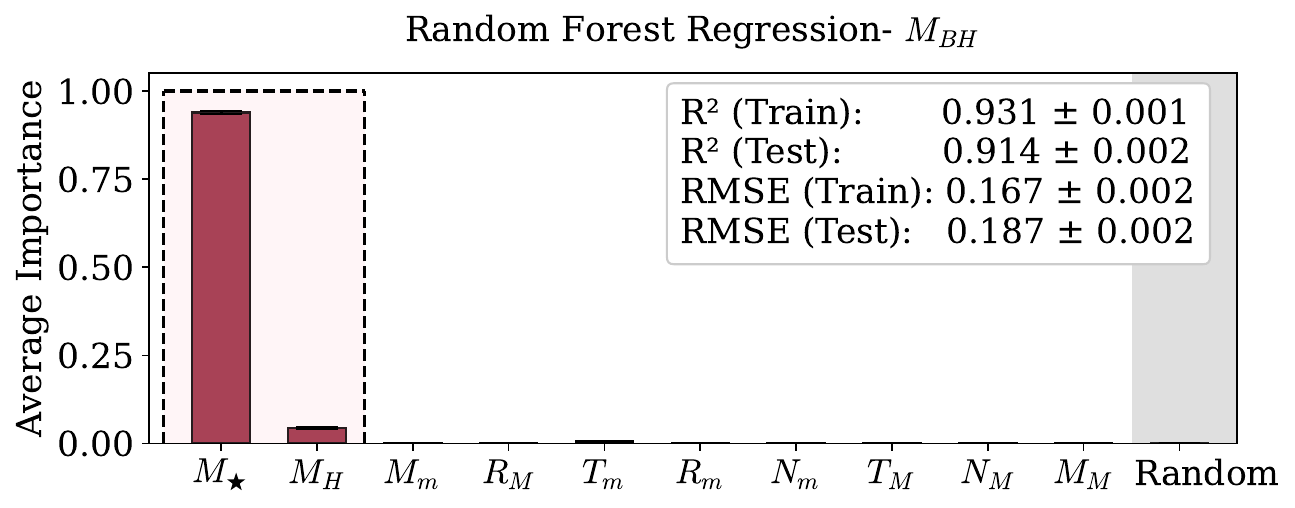}
    \caption{Random forest regression analysis to predict supermassive black hole masses, using the full predictor set (feature parameters as in Figure~\ref{fig:RF_All}). Bar heights show the mean feature importance over 100 independent runs; error bars indicate the standard deviation across the set. The synthetic \emph{Random} predictor provides a null baseline. The regressor attains \(R^{2}_{\mathrm{train}}=0.931\pm0.001\) and \(R^{2}_{\mathrm{test}}=0.914\pm0.002\), with \(\mathrm{RMSE}_{\mathrm{train}}=0.167\pm0.002\) and \(\mathrm{RMSE}_{\mathrm{test}}=0.187\pm0.002\). In IllustrisTNG, stellar mass is clearly the best predictor of supermassive black hole mass, with merger parameters having negligible importance.}
    \label{fig:RF_Regression}
\end{figure*}

Taken together, these results show that merger statistics have genuine predictive power, albeit at a lower level than the intrinsic galaxy properties considered in Figure~\ref{fig:RF_All}. However, this predictive power does not necessarily imply that mergers directly, or even indirectly, cause quenching. Instead, it reflects the fact that galaxies experiencing many mergers are generally more massive and host larger black holes. This is an instance of correlation without causation, which is established by the complete removal of merger-parameter importance once supermassive black hole mass is included in the input list to the classifier.  

Earlier in this study we find moderate correlations between stellar mass and the number of mergers (Pearson $r \approx 0.62$ and Spearman $\rho \approx 0.80$ for the combined sample), with even stronger correlations in the quiescent population. Since black hole mass correlates tightly with stellar mass, the merger parameters indirectly track the intrinsic properties that actually drive quenching.

The leakage controlled experiments therefore demonstrate that merger related quantities encode some predictive information because they correlate with galaxy stellar mass and black hole mass, not because mergers are a primary mechanism for quenching. When non-merger parameters are excluded, fractional mass gains and merger counts provide a moderate AUC because they act as proxies for mass growth. Consequently, even though there is a statistical link between merger history and quenching, the dominant driver of quenching in IllustrisTNG remains the mass of the central black hole, while mergers play at most an indirect role through their influence on the build up of stellar and black hole mass.

\subsubsection{Random Forest Regression}
\label{sec:RandomForestReg_Results}

\begin{figure*}
    \centering
    \includegraphics[width=0.99\textwidth]{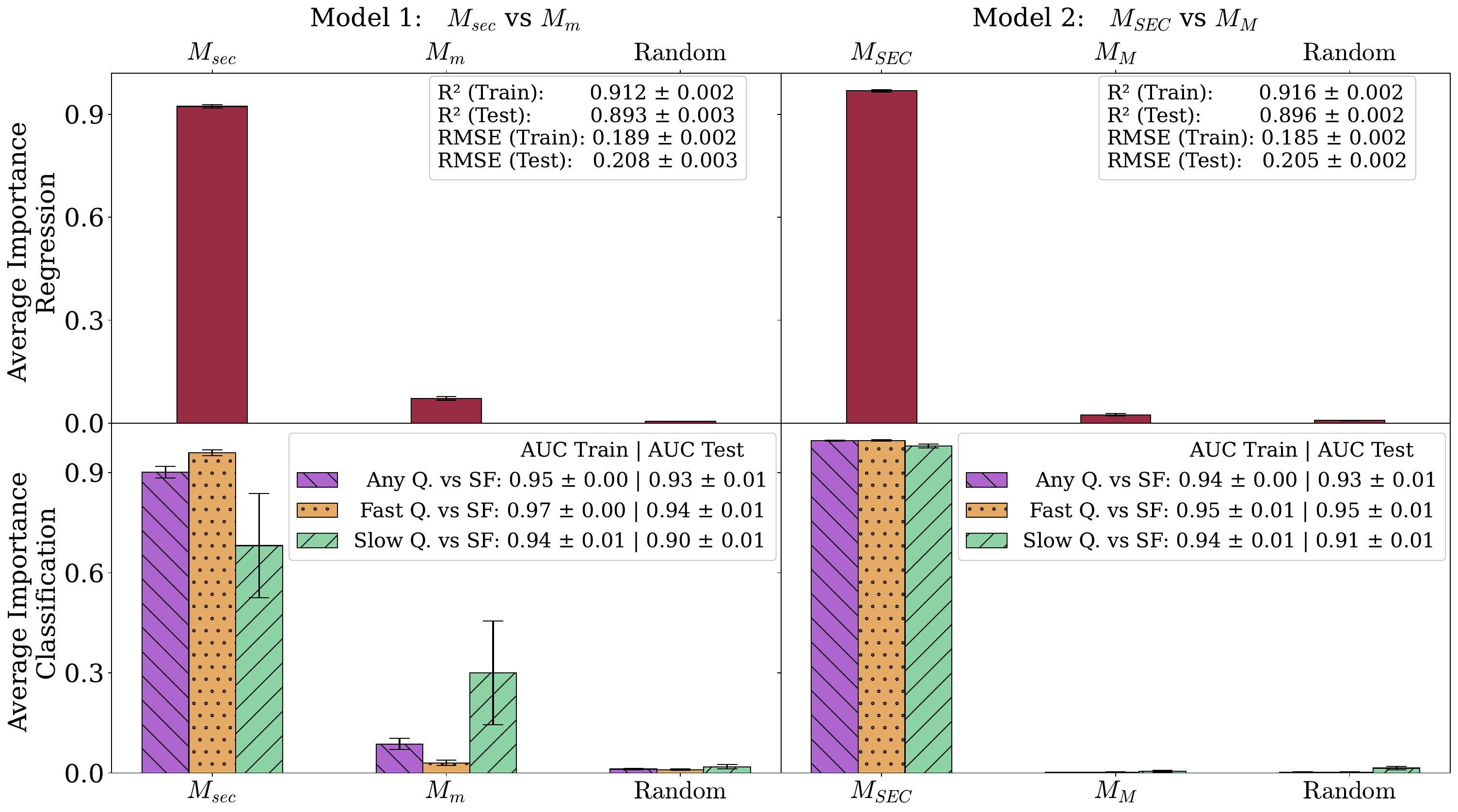}
    \caption{Random forest analyses with a `merger - secular' decomposition of present–day stellar mass, separating intrinsic growth from merger–built growth. We define \(M_{\mathrm{sec}}\equiv M_{\star}-M_{m}\) and \(M_{\mathrm{SEC}}\equiv M_{\star}-M_{M}\). \textit{Left} panels use \(\{M_{\mathrm{sec}},\,M_{m}\}\); \textit{right} panels use \(\{M_{\mathrm{SEC}},\,M_{M}\}\). \textit{Top} row: regression of $M_{\mathrm{BH}}$; \textit{bottom} row: classifications as in Figs.~10 and 11 (SF vs.\ all Q, SF vs.\ fast Q, SF vs.\ slow Q). Bar heights indicate the mean feature importance over 100 runs, and error bars show the standard deviation; a synthetic \textit{Random} predictor provides a null baseline. Secular stellar mass is clearly the best predictor of both supermassive black hole mass and galaxy quenching. Hence, both of these connected processes are clearly not merger dependent.}
    \label{fig:RF_M_sec}
\end{figure*}

Our previous analyses have shown that galaxy quenching in IllustrisTNG is most strongly tied to the mass of the central black hole (in agreement with \citealt{Piotrowska2022, Bluck2023, Bluck2024b}). The final place in the causal chain that mergers could be relevant for quenching in this simulation is via helping to grow the supermassive black holes, needed to quench central galaxies. 

To investigate this possibility, we train a random forest regressor to predict $M_{\rm BH}$ from the full set of predictor variables. We present the results from this analysis in Figure~\ref{fig:RF_Regression}. The regression achieves $R^2_{\rm train} = 0.931 \pm 0.001$ and $R^2_{\rm test} = 0.914 \pm 0.002$, with ${\rm RMSE}_{\rm test} = 0.167 \pm 0.002$ and ${\rm RMSE}_{\rm train} = 0.187 \pm 0.002$.  Black hole mass is therefore predicted with high fidelity from galaxy properties.  The feature importance distribution is dominated by stellar mass $M_\star$, with halo mass $M_H$ contributing at the few per cent level. All merger related parameters contribute negligibly to predicting final supermassive black hole masses. The synthetic random predictor yields zero importance, which confirms that the regression is not overfitting noise.  These findings indicate that the mass of the central black hole is most connected to the total stellar mass, which arises from both in situ star formation and via mergers. 

Since stellar mass is the single most predictive feature for inferring black hole mass in TNG, and black hole mass is in turn the strongest predictor of quenching, it becomes interesting to attempt to understand how the different channels of stellar mass growth influence these outcomes.  

In IllustrisTNG, a galaxy’s present day stellar mass increases through two distinct processes: in situ star formation and accretion of stars through mergers.  We therefore decompose $M_\star$ into a secular component, $M_{\rm sec}$, which reflects the mass formed within the galaxy, and an externally built component, given by the cumulative stellar mass gained via mergers (of all mass ratios).  For the full merger history we define $M_{\rm sec} = M_\star - M_m$, where $M_m$ is the total stellar mass acquired through all mergers. Additionally, we perform a similar factorization for major mergers, defining $M_{\rm SEC} = M_\star - M_M$, where $M_M$ is the mass gained in major mergers.  

In Figure~\ref{fig:RF_M_sec} we present two regression models that use these decompositions to predict $M_{\rm BH}$.  Model 1 employs $\{M_{\rm sec},\,M_m\}$ and Model 2 employs $\{M_{\rm SEC},\,M_M\}$.  In both cases the regressor achieves high fidelity: $R^2_{\rm test} = 0.893 \pm 0.003$ for Model 1 and $0.896 \pm 0.002$ for Model 2.  The feature importance distributions show that the secular component ($M_{\rm sec}$ or $M_{\rm SEC}$) carries nearly all of the predictive power, while the merger built component contributes only a few per cent. Therefore, the growth of supermassive black holes in TNG is highly likely to be governed primarily via secular accretion, not via mergers (which is consistent with results found in \citealt{Weinberger2018, Pillepich2018}).

To assess how these components affect quenching directly, we apply the same feature sets to the classification tasks used in Figs. \ref{fig:RF_All} and \ref{fig:RF_Only_Merg}.  The mean test AUCs for star–forming versus quenched, fast–quenched, and slow–quenched samples are 0.93, 0.94, and 0.90 for Model 1 and 0.93, 0.95, and 0.91 for Model 2, indicating a very good classification performance.  Once again, the secular component is responsible for almost all of the predictive signal, and the merger built component adds little further information.

Taken together, these regression and classification results show that the growth of supermassive black holes and the quenching of galaxies are linked primarily to secular evolution in TNG.  The portion of stellar mass gained through mergers has only a very minor influence on final $M_{\rm BH}$ and on the quenching likelihood. This is highly consistent with the broader conclusion of this work that mergers play at most a very minor role in setting a galaxy’s quenching state. Ultimately, both the evolution of supermassive black holes and the quenching of central galaxies in TNG are governed almost exclusively by secular processes. Therefore, mergers are neither directly nor indirectly responsible for central galaxy quenching in this contemporary simulation.

\section{Discussion}
\label{sec:discussions}

The canonical picture of merger-driven quenching posits that gas-rich major mergers funnel cold gas toward the galactic center, igniting a short-lived central starburst, and subsequently triggering radiatively efficient quasar-mode AGN feedback. In turn, this may rapidly expel or heat the remaining gas in the ISM, thereby quenching star formation \citep[e.g.,][]{Springel2005, DiMatteo2005, Hopkins2006, Hopkins2008}. Idealized hydrodynamical simulations of isolated mergers have long supported this paradigm, demonstrating that, in the absence of fresh gas inflows, a single quasar-driven outburst can suffice to permanently shut down star formation. 

Observationally, several lines of evidence have reinforced this view: enhanced star-formation rates (SFRs) in galaxy pairs and ongoing mergers \citep[e.g.,][]{Ellison2008, Scudder2012, Patton2013}, increased AGN incidence in interacting systems \citep[e.g.,][]{Ellison2011, Silverman2011, Ellison2019}, and a higher prevalence of post-starburst (PSB) features in recently merged galaxies \citep[e.g.,][]{Wild2009, Pawlik2018}. Together, these findings have led to the widespread belief that mergers serve as a common, and possibly even dominant, channel for rapid quenching of central galaxies.

However, recent analyses of large-volume cosmological hydrodynamical simulations challenge this narrative. For instance, \citealt{Quai2023} examine merger histories in Illustris, EAGLE, and TNG, finding that rapid quenching following mergers is rare, and that star-forming post-mergers typically do not quench within short timescales. These studies tested the sufficiency of mergers—whether they tend to trigger quenching—and concluded that such events are neither typical nor efficient at halting star formation in contemporary simulations. Nonetheless, these works left open the questions of whether mergers might be necessary to drive quenching (in combination with other effects), or whether they might indirectly influence quenching via hierarchical assembly pathways (especially the growth of supermassive black holes).

Our analysis goes beyond prior work by assessing not only the sufficiency but also the potential necessity of mergers in the quenching of central galaxies in TNG. Rather than focusing solely on star forming postmergers, we consider the entire quenched galaxy population and employ both wide and narrow merger–quenching association windows to pick up on any causal association. Moreover, we adopt a machine learning approach to compare the predictive power of merger-related features (such as the mass gained from mergers) against key secular parameters (such as stellar mass, halo mass, and black hole mass). This allows us to quantify the relative contribution of mergers to the quenching process and test whether they add independent predictive value over the secular parameters. In this manner, our methodology offers a more holistic assessment of the merger – quenching connection in a fully cosmological context.

Our findings from TNG clearly indicate that the vast majority of mergers (either major or minor) do not lead to quenching within 1 - 2\,Gyr. Furthermore, the vast majority of quenching events involve no major mergers within 1 - 2\,Gyr prior to entering (or during) the green valley phase. On the pother hand minor mergers are typical during quenching but, crucially, are no more prevalent than in star forming systems. These trends persist across the full redshift and stellar mass ranges of our TNG central galaxy sample, and are robust to varying classifier definitions. 

Importantly, the lack of strong merger – quenching correlations is not merely a null result as it contradicts the predictions of the canonical merger–quasar–quench paradigm \citep[e.g.,][]{Hopkins2006, Hopkins2008}. In TNG, mergers are neither sufficient nor necessary to cause quenching. Furthermore, the merger history as a whole is not predictive of quenching status (as ascertained via the random forest analyses). This implies that quenching in the TNG simulation is governed predominantly by secular processes, particularly those linked to black hole growth. This underscores a fundamental departure from the classic picture: even though mergers occur in great numbers in TNG, they are not the cause of quenching.

This stark departure from the classical hypothesis arises in part because idealized simulations and cosmological simulations implement AGN feedback differently. In idealized merger simulations \citep[e.g.,][]{Springel2005, Hopkins2006}, feedback is implemented in closed-box systems where mergers rapidly deplete the cold gas reservoir. In contrast, cosmological simulations, like TNG, model galaxies as open systems embedded in the cosmic web, where quenching requires both gas depletion and the suppression of future inflows. 

TNG achieves this through low-Eddington ratio, kinetic radio-mode feedback \citep{Weinberger2017, Weinberger2018, Zinger2020}, which does not require high central gas concentrations. Hence, this operates completely independently of merger events. While both paradigms invoke AGN feedback, they differ fundamentally in implementation: the canonical model relies on merger-triggered quasar-mode winds, whereas TNG relies on long-term preventative radio-mode feedback. The latter is not expected to correlate with merger activity directly. Instead, the radio-mode feedback of TNG is governed by black hole mass primarily. Interestingly, we find in this work that black holes themselves are largely grown via secular evolution in TNG, rather than in galaxy mergers.

In addition to radio-mode kinetic feedback, TNG also operates a quasar mode, although this is known to be of little impact on quenching (see, e.g., \citealt{Weinberger2018, Zinger2020, Piotrowska2022}). The quasar-mode AGN feedback in TNG is coupled thermally, injecting energy isotropically into the surrounding gas \citep[see,][]{Weinberger2017}. This configuration is relatively inefficient at driving large-scale outflows, as it distributes energy uniformly rather than concentrating it to induce strong turbulence. 

In contrast, the quasar-mode feedback in EAGLE (\citealt{Schaye2015, Crain2015}) injects energy stochastically, delivering discrete thermal energy bursts to randomly selected gas particles, each heated by a fixed temperature increment. This method has been shown to generate more turbulent and vigorous outflows, enhancing the potential for quenching. This difference partly explains why EAGLE shows evidence of quasar-induced quenching, even in the absence of a bespoke radio-mode feedback framework. However, EAGLE galaxies also exhibit more frequent rejuvenation episodes, suggesting that its quasar-mode is less effective at maintaining long-term quiescence \citep[see,][]{Goubert2025}. In contrast, TNG’s radio-mode, while less violent, suppresses gas accretion over longer timescales and exhibits without substantial rejuvenation of galaxies. 

This paper clearly demonstrates that a reasonably successful cosmological model (i.e., TNG), which reproduces the multi-epoch stellar mass functions, the cosmological star formation rate density evolution, and the approximate quenched fraction dependence on mass (see, e.g., \citealt{Nelson2018, Pillepich2018, Weinberger2018, Zinger2020, Terrazas2020, Bluck2016, Piotrowska2022}) can be formed without any merger - quenching relationship. However, it remains to be seen whether this is in line with nature, or not. If it is, then the merger - quasar - quench paradigm is over. If it is not, there is a potential way to improve the TNG model by making its quasar-mode AGN feedback more effective, borrowing from the relative success in EAGLE via adding enhanced stochasticity in its implementation. Nonetheless, the need for retaining the preventative kinetic mode is also well motivated, in order to prevent excessive late-time rejuvenation, as found with quasar-mode AGN feedback operating alone in EAGLE.

In observations, elevated merger fractions in PSB galaxies are widely reported \citep[e.g.,][]{Pawlik2018, Wilkinson2022, Ellison2024}. Additionally, time-resolved postmerger analyses show very large relative PSB excesses over mass- and redshift-matched controls, peaking a few hundred 100 Myr after coalescence (typically $\sim$0.2 - 0.5\,Gyr), at factors of $\sim$30 - 100, depending on PSB definition \citep[e.g.,][]{Ellison2024}. At the same time, the \textit{absolute} PSB fractions in post‑mergers are modest when averaged over the full timeline. For example, summing the post‑merger counts in \citealt{Ellison2024} yields overall averages of $\sim$4.5\% for a broader PCA selection and $\sim$1.2\% for a strict E$+$A selection, even though the same study reports peaks approaching $\sim$10–20\% (PCA) and $\sim$7\% (E$+$A) shortly after coalescence. 

Moreover, a non-negligible minority of PSBs lack visible merger signatures, with \citealt{Wilkinson2022} finding only $\sim 19\%$ to $42\%$ of PSBs showing tidal features under their methods, consistent with missed faint or faded disturbances and with multiple rapid-quenching channels \citep[see also][]{Wild2010, Maltby2018}. Morphology-based merger identification can miss subtle or late-stage interactions, and even deep-learning classifiers have completeness – purity trade-offs, so image-based merger fractions are likely lower limits. Observations of enhanced AGN incidence in pairs and post-mergers are robust \citep[e.g.,][]{Ellison2008, Ellison2011, Ellison2019}, although the causal role of this AGN feedback in permanently shutting down star formation remains hotly debated. Offsets between the lifetimes of tidal features, merger-induced starbursts, and rapidly varying AGN emission hinder one-to-one temporal associations, and at higher redshift elevated gas fractions and viewing-angle dependence further bias the morphological observability of mergers \citep[e.g.,][]{Lotz2008, Mantha2019}.

Overall, current data support a nuanced view. Mergers substantially boost the probability of rapid quenching traced by PSB spectra within a short post-coalescence window, yet mergers are neither strictly necessary nor sufficient. Many mergers do not produce a PSB on average over the full timeline, and a significant subset of PSBs show no visible merger evidence \citep[e.g.,][]{Wilkinson2022, Ellison2024}. Moreover, a large (most likely dominant) fraction of quenched galaxies reach quiescence through slower modes and, hence, are not traced via these observational techniques. Furthermore, most merger studies focus on rare, dramatic events, potentially overlooking the importance of slow, secular processes.

Crucially, we know of no published work which compares the merger incidence in quenching galaxies, or the quenching incidence of merging galaxies, controlling for the key parameter of supermassive black hole mass (or its proxies).  This is a natural place to extend the analysis of mergers in observational data and begin to test whether mergers are essential in nature for quenching, or rather merely a bi-product of galaxy evolution. In parallel with this approach, one can also begin to test whether high-luminosity AGN are common in quenching galaxies, or not. Some recent work suggests that they are not, in line with the TNG prediction (see, \citealt{Almaini2025}). In any case, the TNG simulation predicts a total decoupling of quenching from both merging and quasars, which may in principle be tested observationally across cosmic time. Indeed, we feel this is the most pressing observational test of this theoretical paradigm needed to advance our understanding of quenching at this point.

Ultimately, a definitive observational test of the merger - quenching paradigm will likely require forward modeling from simulations like TNG and EAGLE into observable space, accounting for key limitations such as PSF broadening, signal-to-noise effects, and timescale mismatches. Additionally, this approach will require careful control samples based around the predicted key parameters, most notably supermassive black hole mass. At low-$z$, both the merger rate and the quasar rate are very low (e.g., \citealt{Patton2008, Aird2015}), suggesting that a definitive test of this paradigm must be conducted at higher redshifts. Nevertheless, it is essential to have spectroscopic redshifts for large samples of galaxies drawn from large contiguous areas on-sky to properly constrain pair and merger fractions, which is currently lacking. Fortunately, the VLT-MOONRISE survey (\citealt{Maiolino2020, Cirasuolo2020}) will address this observational limitation at cosmic noon. This will enable fundamentally novel tests between the secular and merger driven quenching hypotheses at the relevant epochs for the first time.

\section{Summary}
\label{sec:summary}

In this paper we analyze the quenching of central galaxies in IllustrisTNG100-1 (TNG) by combining star formation histories, redshift- and mass-dependent sSFR classification, explicit measurements of quenching onset and duration times, and a reconstruction of (major \& minor) merger histories for more than $11,000$ systems with $M_\star \ge 10^9\,{\rm M_\odot}$ by $z=0$. We test whether mergers are sufficient or necessary for central galaxy quenching. Additionally, we evaluate the predictive power of the cumulative merger history over quenching in TNG by assessing the performance of merger related parameters, relative to intrinsic galaxy properties, in various random forest models.

Our results show that the onset of quenching spans a wide range of cosmic times, with a peak at $t \simeq 10\,{\rm Gyr}$ after the Big Bang ($z \approx 0.33$). More massive galaxies begin to quench earlier than their less massive counterparts, in line with downsizing in galaxy formation. Quenching durations are strongly bimodal in the simulation, presenting with a fast channel (with $\Delta\tau_Q < 0.56\,{\rm Gyr}$) and a slow channel (with $\Delta\tau_Q \ge 0.56\,{\rm Gyr}$). The slow quenching channel is more common at higher $M_\star$, with the fast quenching channel being more common at lower $M_\star$. 

Only a small fraction of major mergers temporally align with quenching. For major mergers, about $3$ per cent lead to quenching within a $1\,{\rm Gyr}$ window, and about $5$ per cent within a $2\,{\rm Gyr}$ window. For any mergers, the corresponding fractions are about $12$ and $16$ per cent, respectively. This clearly implies that the vast majority of mergers (of any mass ratio) do not give rise to quenching. Therefore, merging is not sufficient for quenching in TNG. 

Additionally, we assess whether, despite not being sufficient, mergers are nonetheless necessary for quenching to occur in TNG. Given quenching, the raw upper limits remain low for major mergers ($\mu_Q \simeq 11$ - $17$ per cent across the two windows). This implies that major mergers are rarely present prior to (or during) quenching events in TNG. Hence, major mergers are not necessary for quenching to occur, at least in the vast majority of cases. 

On the other hand, the any-merger fractions are substantially higher, but largely reflect random coincidence. After subtracting the expected chance overlap with star–forming controls, the excess association is about $35.53$ per cent for a $1\,{\rm Gyr}$ window and about $34.01$ per cent for a $2\,{\rm Gyr}$ window. This drops precipitously when we further restrict to stellar mass matched star forming control samples, yielding a null merger excess ($\simeq 0$), within both association windows. Similar results are obtained when analyzing quenching modes separately (i.e., given fast- and slow-quenching). This further implies that mergers are not necessary for either type of quenching event.

These tests clearly show that mergers are not sufficient for quenching (since the vast majority of mergers occur without quenching following). Moreover, mergers are not necessary for quenching either (since the vast majority of quenching events occur without contemporaneous merger events).

The analyses described above rule out mergers as a direct trigger for central quenching in TNG (at least in the vast majority of cases). However, the cumulative impact of mergers on galaxies (and especially their central supermassive black holes) may still be important for quenching in an indirect sense. 

Utilizing a random Forest classifier, we establish that supermassive black hole mass is clearly the most predictive parameter of central galaxy quenching in TNG (in agreement with \citealt{Piotrowska2022, Bluck2023, Bluck2024b}). We extend this prior conclusion by considering a comprehensive set of cumulative merger metrics. Taken together, merger parameters contribute negligibly to predicting central galaxy quiescence, even when black hole mass (and, additionally, stellar mass) is withheld from the classifier. This strongly implies that the cumulative merger history is not closely linked to quenching in TNG.

Finally, we investigate whether supermassive black holes grow primarily via merger-driven or secular processes during galaxy evolution. Via a Random Forest regression analysis we find that stellar mass is the most effective predictor of central black hole mass. Moreover, we find that it is particularly the {\it secular} stellar mass (i.e., the mass that formed in situ) which dominates the growth of supermassive black holes, and best predicts quenching. Hence, even indirectly, mergers are not a critical aspect of quenching in the TNG simulation. This conclusion is in stark contrast to earlier ideas of quenching, summarized in the `merger-quasar-quench' paradigm (e.g., \citealt{Springel2005, DiMatteo2005, Hopkins2006, Hopkins2008}).

A physical interpretation may be formulated as follows. In a cosmological setting, where gas supply from cooling of the CGM and direct accretion from the ISM is expected, outflow events (no matter how powerful) cannot permanently quench a galaxy by themselves. To achieve long-term quiescence, the hot gas halo must be stabilized against cooling and collapse. In TNG this is achieved via low-Eddington ratio kinetic feedback, which heats the hot gas halo, preventing cooling, and ultimately leads to quenching via starvation of gas supply (see, e.g., \citealt{Weinberger2018, Zinger2020, Piotrowska2022}). 

Hence, the key to quenching central galaxies in TNG is hot halo stabilization via kinetic AGN feedback, not outflows driven by quasar-mode AGN feedback. The former is regulated primarily via secular processes, particularly the cooling rate of the hot gas halo. Thus, mergers are not relevant for quenching in this scenario. Moreover, we find that in TNG the growth of supermassive black holes themselves are also regulated primarily in a secular manner, likely via disc instabilities. Therefore, even in an indirect sense, mergers and quenching are fundamentally decoupled in the TNG model.

Future observational work is needed to test whether the secular evolution of supermassive black holes and quenching predicted by the TNG simulation is an accurate description of nature, or not. Nevertheless, TNG manifestly demonstrates that one may quench central galaxies in approximately the correct abundance without any significant dependence on mergers, either directly (i.e., as a trigger) or indirectly (i.e., as a route to growing supermassive black holes). It is particularly important to appreciate that this is a radically different view of galaxy evolution to that presented in the commonly held `merger-quasar-quench' paradigm, which has been hugely influential over the past two decades.

\section*{Acknowledgments}

{\bf We thank the anonymous referee for a deeply insightful and helpful review, which significantly benefited this work.} We are also grateful for helpful discussion with Roberto Maiolino. AFLB gratefully acknowledges support from an NSF research grant: NSF-AST 2408009. PG also acknowledges support from NSF-AST 2408009.

\section*{Data Availability}

All of the data used in this paper are publicly available from: https://www.tng-project.org/ . Further information on the data products are provided in \citealt{Pillepich2018, Nelson2018, Springel2018, Naiman2018, Marinacci2018}.


\bibliographystyle{mnras}
\bibliography{A_references}

@ARTICLE{Aird2015,
       author = {{Aird}, J. and {Coil}, A.~L. and {Georgakakis}, A. and {Nandra}, K. and {Barro}, G. and {P{\'e}rez-Gonz{\'a}lez}, P.~G.},
        title = "{The evolution of the X-ray luminosity functions of unabsorbed and absorbed AGNs out to z{\ensuremath{\sim}} 5}",
      journal = {\mnras},
         year = 2015,
        month = aug,
       volume = {451},
       number = {2},
        pages = {1892-1927},
          doi = {10.1093/mnras/stv1062},
archivePrefix = {arXiv},
       eprint = {1503.01120},
 primaryClass = {astro-ph.HE},
       adsurl = {https://ui.adsabs.harvard.edu/abs/2015MNRAS.451.1892A}
}

@ARTICLE{Almaini2025,
       author = {{Almaini}, Omar and {Wild}, Vivienne and {Maltby}, David and {Taylor}, Elizabeth and {Rowlands}, Kate and {de Lisle}, Thomas and {Alatalo}, Katherine and {Harrold}, Jimi and {Hewitt}, Guillaume and {Patil}, Pallavi and {Skarbinski}, Maya},
        title = "{No evidence for excess AGN activity in recently quenched massive galaxies at cosmic noon}",
      journal = {\mnras},
         year = 2025,
        month = jun,
       volume = {539},
       number = {4},
        pages = {3568-3581},
          doi = {10.1093/mnras/staf659},
archivePrefix = {arXiv},
       eprint = {2504.15342},
 primaryClass = {astro-ph.GA},
       adsurl = {https://ui.adsabs.harvard.edu/abs/2025MNRAS.539.3568A}
}

@ARTICLE{Barnes1992,
       author = {{Barnes}, Joshua E.},
        title = "{Transformations of Galaxies. I. Mergers of Equal-Mass Stellar Disks}",
      journal = {\apj},
         year = 1992,
        month = jul,
       volume = {393},
        pages = {484},
          doi = {10.1086/171522},
       adsurl = {https://ui.adsabs.harvard.edu/abs/1992ApJ...393..484B}}

@ARTICLE{Bluck2014,
       author = {{Bluck}, Asa F.~L. and {Mendel}, J. Trevor and {Ellison}, Sara L. and {Moreno}, Jorge and {Simard}, Luc and {Patton}, David R. and {Starkenburg}, Else},
        title = "{Bulge mass is king: the dominant role of the bulge in determining the fraction of passive galaxies in the Sloan Digital Sky Survey}",
      journal = {\mnras},
         year = 2014,
        month = jun,
       volume = {441},
       number = {1},
        pages = {599-629},
          doi = {10.1093/mnras/stu594},
archivePrefix = {arXiv},
       eprint = {1403.5269},
 primaryClass = {astro-ph.GA},
       adsurl = {https://ui.adsabs.harvard.edu/abs/2014MNRAS.441..599B}}

@ARTICLE{Bluck2016,
       author = {{Bluck}, Asa F.~L. and {Mendel}, J. Trevor and {Ellison}, Sara L. and {Patton}, David R. and {Simard}, Luc and {Henriques}, Bruno M.~B. and {Torrey}, Paul and {Teimoorinia}, Hossen and {Moreno}, Jorge and {Starkenburg}, Else},
        title = "{The impact of galactic properties and environment on the quenching of central and satellite galaxies: a comparison between SDSS, Illustris and L-Galaxies}",
      journal = {\mnras},
         year = 2016,
        month = nov,
       volume = {462},
       number = {3},
        pages = {2559-2586},
          doi = {10.1093/mnras/stw1665},
archivePrefix = {arXiv},
       eprint = {1607.03318},
 primaryClass = {astro-ph.GA},
       adsurl = {https://ui.adsabs.harvard.edu/abs/2016MNRAS.462.2559B}}

@ARTICLE{Bluck2020a,
       author = {{Bluck}, Asa F.~L. and {Maiolino}, Roberto and {S{\'a}nchez}, Sebastian F. and {Ellison}, Sara L. and {Thorp}, Mallory D. and {Piotrowska}, Joanna M. and {Teimoorinia}, Hossen and {Bundy}, Kevin A.},
        title = "{Are galactic star formation and quenching governed by local, global, or environmental phenomena?}",
      journal = {\mnras},
         year = 2020,
        month = feb,
       volume = {492},
       number = {1},
        pages = {96-139},
          doi = {10.1093/mnras/stz3264},
archivePrefix = {arXiv},
       eprint = {1911.08857},
 primaryClass = {astro-ph.GA},
       adsurl = {https://ui.adsabs.harvard.edu/abs/2020MNRAS.492...96B}}

@ARTICLE{Bluck2020b,
       author = {{Bluck}, Asa F.~L. and {Maiolino}, Roberto and {Piotrowska}, Joanna M. and {Trussler}, James and {Ellison}, Sara L. and {S{\'a}nchez}, Sebastian F. and {Thorp}, Mallory D. and {Teimoorinia}, Hossen and {Moreno}, Jorge and {Conselice}, Christopher J.},
        title = "{How do central and satellite galaxies quench? - Insights from spatially resolved spectroscopy in the MaNGA survey}",
      journal = {\mnras},
         year = 2020,
        month = nov,
       volume = {499},
       number = {1},
        pages = {230-268},
          doi = {10.1093/mnras/staa2806},
archivePrefix = {arXiv},
       eprint = {2009.05341},
 primaryClass = {astro-ph.GA},
       adsurl = {https://ui.adsabs.harvard.edu/abs/2020MNRAS.499..230B}}

@ARTICLE{Bluck2022,
       author = {{Bluck}, Asa F.~L. and {Maiolino}, Roberto and {Brownson}, Simcha and {Conselice}, Christopher J. and {Ellison}, Sara L. and {Piotrowska}, Joanna M. and {Thorp}, Mallory D.},
        title = "{The quenching of galaxies, bulges, and disks since cosmic noon. A machine learning approach for identifying causality in astronomical data}",
      journal = {\aap},
         year = 2022,
        month = mar,
       volume = {659},
          eid = {A160},
        pages = {A160},
          doi = {10.1051/0004-6361/202142643},
archivePrefix = {arXiv},
       eprint = {2201.07814},
 primaryClass = {astro-ph.GA},
       adsurl = {https://ui.adsabs.harvard.edu/abs/2022A&A...659A.160B}}

@ARTICLE{Bluck2023,
       author = {{Bluck}, Asa F.~L. and {Piotrowska}, Joanna M. and {Maiolino}, Roberto},
        title = "{The Fundamental Signature of Star Formation Quenching from AGN Feedback: A Critical Dependence of Quiescence on Supermassive Black Hole Mass, Not Accretion Rate}",
      journal = {\apj},
         year = 2023,
        month = feb,
       volume = {944},
       number = {1},
          eid = {108},
        pages = {108},
          doi = {10.3847/1538-4357/acac7c},
archivePrefix = {arXiv},
       eprint = {2301.03677},
 primaryClass = {astro-ph.GA},
       adsurl = {https://ui.adsabs.harvard.edu/abs/2023ApJ...944..108B}}

@ARTICLE{Bondi1944,
       author = {{Bondi}, H. and {Hoyle}, F.},
        title = "{On the mechanism of accretion by stars}",
      journal = {\mnras},
         year = 1944,
        month = jan,
       volume = {104},
        pages = {273},
          doi = {10.1093/mnras/104.5.273},
       adsurl = {https://ui.adsabs.harvard.edu/abs/1944MNRAS.104..273B}
}

@ARTICLE{Bravo2023,
       author = {{Bravo}, Mat{\'\i}as and {Robotham}, Aaron S.~G. and {Lagos}, Claudia del P. and {Davies}, Luke J.~M. and {Bellstedt}, Sabine and {Thorne}, Jessica E.},
        title = "{Galaxy quenching time-scales from a forensic reconstruction of their colour evolution}",
      journal = {\mnras},
         year = 2023,
        month = jul,
       volume = {522},
       number = {3},
        pages = {4481-4498},
          doi = {10.1093/mnras/stad1234},
archivePrefix = {arXiv},
       eprint = {2301.03702},
 primaryClass = {astro-ph.GA},
       adsurl = {https://ui.adsabs.harvard.edu/abs/2023MNRAS.522.4481B}
}

@ARTICLE{Breiman2001,
       author = {{Breiman}, Leo},
        title = "{Random Forests.}",
      journal = {Machine Learning},
         year = 2001,
        month = jan,
       volume = {45},
        pages = {5-32},
          doi = {10.1023/A:1010933404324},
       adsurl = {https://ui.adsabs.harvard.edu/abs/2001MachL..45....5B}
}

@ARTICLE{Brinchmann2004,
       author = {{Brinchmann}, J. and {Charlot}, S. and {White}, S.~D.~M. and {Tremonti}, C. and {Kauffmann}, G. and {Heckman}, T. and {Brinkmann}, J.},
        title = "{The physical properties of star-forming galaxies in the low-redshift Universe}",
      journal = {\mnras},
         year = 2004,
        month = jul,
       volume = {351},
       number = {4},
        pages = {1151-1179},
          doi = {10.1111/j.1365-2966.2004.07881.x},
archivePrefix = {arXiv},
       eprint = {astro-ph/0311060},
 primaryClass = {astro-ph},
       adsurl = {https://ui.adsabs.harvard.edu/abs/2004MNRAS.351.1151B}
}

@ARTICLE{Brownson2022,
       author = {{Brownson}, Simcha and {Bluck}, Asa F.~L. and {Maiolino}, Roberto and {Jones}, Gareth C.},
        title = "{What drives galaxy quenching? A deep connection between galaxy kinematics and quenching in the local Universe}",
      journal = {\mnras},
         year = 2022,
        month = apr,
       volume = {511},
       number = {2},
        pages = {1913-1941},
          doi = {10.1093/mnras/stab3749},
archivePrefix = {arXiv},
       eprint = {2201.02484},
 primaryClass = {astro-ph.GA},
       adsurl = {https://ui.adsabs.harvard.edu/abs/2022MNRAS.511.1913B}
}

@ARTICLE{Carnall2019,
       author = {{Carnall}, A.~C. and {McLure}, R.~J. and {Dunlop}, J.~S. and {Cullen}, F. and {McLeod}, D.~J. and {Wild}, V. and {Johnson}, B.~D. and {Appleby}, S. and {Dav{\'e}}, R. and {Amorin}, R. and {Bolzonella}, M. and {Castellano}, M. and {Cimatti}, A. and {Cucciati}, O. and {Gargiulo}, A. and {Garilli}, B. and {Marchi}, F. and {Pentericci}, L. and {Pozzetti}, L. and {Schreiber}, C. and {Talia}, M. and {Zamorani}, G.},
        title = "{The VANDELS survey: the star-formation histories of massive quiescent galaxies at 1.0 < z < 1.3}",
      journal = {\mnras},
         year = 2019,
        month = nov,
       volume = {490},
       number = {1},
        pages = {417-439},
          doi = {10.1093/mnras/stz2544},
archivePrefix = {arXiv},
       eprint = {1903.11082},
 primaryClass = {astro-ph.GA},
       adsurl = {https://ui.adsabs.harvard.edu/abs/2019MNRAS.490..417C}
}

@ARTICLE{Cirasuolo2020,
       author = {{Cirasuolo}, M. and {Fairley}, A. and {Rees}, P. and {Gonzalez}, O.~A. and {Taylor}, W. and {Maiolino}, R. and {Afonso}, J. and {Evans}, C. and {Flores}, H. and {Lilly}, S. and {Oliva}, E. and {Paltani}, S. and {Vanzi}, L. and {Abreu}, M. and {Accardo}, M. and {Adams}, N. and {{\'A}lvarez M{\'e}ndez}, D. and {Amans}, J. -P. and {Amarantidis}, S. and {Atek}, H. and {Atkinson}, D. and {Banerji}, M. and {Barrett}, J. and {Barrientos}, F. and {Bauer}, F. and {Beard}, S. and {B{\'e}chet}, C. and {Belfiore}, A. and {Bellazzini}, M. and {Benoist}, C. and {Best}, P. and {Biazzo}, K. and {Black}, M. and {Boettger}, D. and {Bonifacio}, P. and {Bowler}, R. and {Bragaglia}, A. and {Brierley}, S. and {Brinchmann}, J. and {Brinkmann}, M. and {Buat}, V. and {Buitrago}, F. and {Burgarella}, D. and {Burningham}, B. and {Buscher}, D. and {Cabral}, A. and {Caffau}, E. and {Cardoso}, L. and {Carnall}, A. and {Carollo}, M. and {Castillo}, R. and {Castignani}, G. and {Catelan}, M. and {Cicone}, C. and {Cimatti}, A. and {Cioni}, M. -R.~L. and {Clementini}, G. and {Cochrane}, W. and {Coelho}, J. and {Colling}, M. and {Contini}, T. and {Contreras}, R. and {Conzelmann}, R. and {Cresci}, G. and {Cropper}, M. and {Cucciati}, O. and {Cullen}, F. and {Cumani}, C. and {Curti}, M. and {Da Silva}, A. and {Daddi}, E. and {Dalessandro}, E. and {Dalessio}, F. and {Dauvin}, L. and {Davidson}, G. and {de Laverny}, P. and {Delplancke-Str{\"o}bele}, F. and {De Lucia}, G. and {Del Vecchio}, C. and {Dessauges-Zavadsky}, M. and {Di Matteo}, P. and {Dole}, H. and {Drass}, H. and {Dunlop}, J. and {D{\"u}nner}, R. and {Eales}, S. and {Ellis}, R. and {Enriques}, B. and {Fasola}, G. and {Ferguson}, A. and {Ferruzzi}, D. and {Fisher}, M. and {Flores}, M. and {Fontana}, A. and {Forchi}, V. and {Francois}, P. and {Franzetti}, P. and {Gargiulo}, A. and {Garilli}, B. and {Gaudemard}, J. and {Gieles}, M. and {Gilmore}, G. and {Ginolfi}, M. and {Gomes}, J.~M. and {Guinouard}, I. and {Gutierrez}, P. and {Haigron}, R. and {Hammer}, F. and {Hammersley}, P. and {Haniff}, C. and {Harrison}, C. and {Haywood}, M. and {Hill}, V. and {Hubin}, N. and {Humphrey}, A. and {Ibata}, R. and {Infante}, L. and {Ives}, D. and {Ivison}, R. and {Iwert}, O. and {Jablonka}, P. and {Jakob}, G. and {Jarvis}, M. and {King}, D. and {Kneib}, J. -P. and {Laporte}, P. and {Lawrence}, A. and {Lee}, D. and {Li Causi}, G. and {Lorenzoni}, S. and {Lucatello}, S. and {Luco}, Y. and {Macleod}, A. and {Magliocchetti}, M. and {Magrini}, L. and {Mainieri}, V. and {Maire}, C. and {Mannucci}, F. and {Martin}, N. and {Matute}, I. and {Maurogordato}, S. and {McGee}, S. and {Mcleod}, D. and {McLure}, R. and {McMahon}, R. and {Melse}, B. -T. and {Messias}, H. and {Mucciarelli}, A. and {Nisini}, B. and {Nix}, J. and {Norberg}, P. and {Oesch}, P. and {Oliveira}, A. and {Origlia}, L. and {Padilla}, N. and {Palsa}, R. and {Pancino}, E. and {Papaderos}, P. and {Pappalardo}, C. and {Parry}, I. and {Pasquini}, L. and {Peacock}, J. and {Pedichini}, F. and {Pello}, R. and {Peng}, Y. and {Pentericci}, L. and {Pfuhl}, O. and {Piazzesi}, R. and {Popovic}, D. and {Pozzetti}, L. and {Puech}, M. and {Puzia}, T. and {Raichoor}, A. and {Randich}, S. and {Recio-Blanco}, A. and {Reis}, S. and {Reix}, F. and {Renzini}, A. and {Rodrigues}, M. and {Rojas}, F. and {Rojas-Arriagada}, {\'A}. and {Rota}, S. and {Royer}, F. and {Sacco}, G. and {Sanchez-Janssen}, R. and {Sanna}, N. and {Santos}, P. and {Sarzi}, M. and {Schaerer}, D. and {Schiavon}, R. and {Schnell}, R. and {Schultheis}, M. and {Scodeggio}, M. and {Serjeant}, S. and {Shen}, T. -C. and {Simmonds}, C. and {Smoker}, J. and {Sobral}, D. and {Sordet}, M. and {Sp{\'e}rone}, D.},
        title = "{MOONS: The New Multi-Object Spectrograph for the VLT}",
      journal = {The Messenger},
         year = 2020,
        month = jun,
       volume = {180},
        pages = {10-17},
          doi = {10.18727/0722-6691/5195},
archivePrefix = {arXiv},
       eprint = {2009.00628},
 primaryClass = {astro-ph.IM},
       adsurl = {https://ui.adsabs.harvard.edu/abs/2020Msngr.180...10C}
}

@ARTICLE{Cowie1996,
       author = {{Cowie}, Lennox L. and {Songaila}, Antoinette and {Hu}, Esther M. and {Cohen}, J.~G.},
        title = "{New Insight on Galaxy Formation and Evolution From Keck Spectroscopy of the Hawaii Deep Fields}",
      journal = {\aj},
         year = 1996,
        month = sep,
       volume = {112},
        pages = {839},
          doi = {10.1086/118058},
archivePrefix = {arXiv},
       eprint = {astro-ph/9606079},
 primaryClass = {astro-ph},
       adsurl = {https://ui.adsabs.harvard.edu/abs/1996AJ....112..839C}
}

@ARTICLE{Crain2015,
       author = {{Crain}, Robert A. and {Schaye}, Joop and {Bower}, Richard G. and {Furlong}, Michelle and {Schaller}, Matthieu and {Theuns}, Tom and {Dalla Vecchia}, Claudio and {Frenk}, Carlos S. and {McCarthy}, Ian G. and {Helly}, John C. and {Jenkins}, Adrian and {Rosas-Guevara}, Yetli M. and {White}, Simon D.~M. and {Trayford}, James W.},
        title = "{The EAGLE simulations of galaxy formation: calibration of subgrid physics and model variations}",
      journal = {\mnras},
         year = 2015,
        month = jun,
       volume = {450},
       number = {2},
        pages = {1937-1961},
          doi = {10.1093/mnras/stv725},
archivePrefix = {arXiv},
       eprint = {1501.01311},
 primaryClass = {astro-ph.GA},
       adsurl = {https://ui.adsabs.harvard.edu/abs/2015MNRAS.450.1937C}
}

@ARTICLE{Croton2006,
       author = {{Croton}, Darren J. and {Springel}, Volker and {White}, Simon D.~M. and {De Lucia}, G. and {Frenk}, C.~S. and {Gao}, L. and {Jenkins}, A. and {Kauffmann}, G. and {Navarro}, J.~F. and {Yoshida}, N.},
        title = "{The many lives of active galactic nuclei: cooling flows, black holes and the luminosities and colours of galaxies}",
      journal = {\mnras},
         year = 2006,
        month = jan,
       volume = {365},
       number = {1},
        pages = {11-28},
          doi = {10.1111/j.1365-2966.2005.09675.x},
archivePrefix = {arXiv},
       eprint = {astro-ph/0508046},
 primaryClass = {astro-ph},
       adsurl = {https://ui.adsabs.harvard.edu/abs/2006MNRAS.365...11C}
}

@ARTICLE{DiMatteo2005,
       author = {{Di Matteo}, Tiziana and {Springel}, Volker and {Hernquist}, Lars},
        title = "{Energy input from quasars regulates the growth and activity of black holes and their host galaxies}",
      journal = {\nat},
         year = 2005,
        month = feb,
       volume = {433},
       number = {7026},
        pages = {604-607},
          doi = {10.1038/nature03335},
archivePrefix = {arXiv},
       eprint = {astro-ph/0502199},
 primaryClass = {astro-ph},
       adsurl = {https://ui.adsabs.harvard.edu/abs/2005Natur.433..604D}
}

@ARTICLE{DiMatteo2008,
       author = {{Di Matteo}, Tiziana and {Colberg}, J{\"o}rg and {Springel}, Volker and {Hernquist}, Lars and {Sijacki}, Debora},
        title = "{Direct Cosmological Simulations of the Growth of Black Holes and Galaxies}",
      journal = {\apj},
         year = 2008,
        month = mar,
       volume = {676},
       number = {1},
        pages = {33-53},
          doi = {10.1086/524921},
archivePrefix = {arXiv},
       eprint = {0705.2269},
 primaryClass = {astro-ph},
       adsurl = {https://ui.adsabs.harvard.edu/abs/2008ApJ...676...33D}
}

@ARTICLE{Donnari2021,
       author = {{Donnari}, Martina and {Pillepich}, Annalisa and {Joshi}, Gandhali D. and {Nelson}, Dylan and {Genel}, Shy and {Marinacci}, Federico and {Rodriguez-Gomez}, Vicente and {Pakmor}, R{\"u}diger and {Torrey}, Paul and {Vogelsberger}, Mark and {Hernquist}, Lars},
        title = "{Quenched fractions in the IllustrisTNG simulations: the roles of AGN feedback, environment, and pre-processing}",
      journal = {\mnras},
         year = 2021,
        month = jan,
       volume = {500},
       number = {3},
        pages = {4004-4024},
          doi = {10.1093/mnras/staa3006},
archivePrefix = {arXiv},
       eprint = {2008.00005},
 primaryClass = {astro-ph.GA},
       adsurl = {https://ui.adsabs.harvard.edu/abs/2021MNRAS.500.4004D}
}

@ARTICLE{Driver2006,
       author = {{Driver}, S.~P. and {Allen}, P.~D. and {Graham}, Alister. W. and {Cameron}, E. and {Liske}, J. and {Ellis}, S.~C. and {Cross}, N.~J.~G. and {De Propris}, R. and {Phillipps}, S. and {Couch}, W.~J.},
        title = "{The Millennium Galaxy Catalogue: morphological classification and bimodality in the colour-concentration plane}",
      journal = {\mnras},
         year = 2006,
        month = may,
       volume = {368},
       number = {1},
        pages = {414-434},
          doi = {10.1111/j.1365-2966.2006.10126.x},
archivePrefix = {arXiv},
       eprint = {astro-ph/0602240},
 primaryClass = {astro-ph},
       adsurl = {https://ui.adsabs.harvard.edu/abs/2006MNRAS.368..414D}
}

@ARTICLE{Ellison2008,
       author = {{Ellison}, Sara L. and {Patton}, David R. and {Simard}, Luc and {McConnachie}, Alan W.},
        title = "{Galaxy Pairs in the Sloan Digital Sky Survey. I. Star Formation, Active Galactic Nucleus Fraction, and the Mass-Metallicity Relation}",
      journal = {\aj},
         year = 2008,
        month = may,
       volume = {135},
       number = {5},
        pages = {1877-1899},
          doi = {10.1088/0004-6256/135/5/1877},
archivePrefix = {arXiv},
       eprint = {0803.0161},
 primaryClass = {astro-ph},
       adsurl = {https://ui.adsabs.harvard.edu/abs/2008AJ....135.1877E}
}

@ARTICLE{Ellison2011,
       author = {{Ellison}, Sara L. and {Patton}, David R. and {Mendel}, J. Trevor and {Scudder}, Jillian M.},
        title = "{Galaxy pairs in the Sloan Digital Sky Survey - IV. Interactions trigger active galactic nuclei}",
      journal = {\mnras},
         year = 2011,
        month = dec,
       volume = {418},
       number = {3},
        pages = {2043-2053},
          doi = {10.1111/j.1365-2966.2011.19624.x},
archivePrefix = {arXiv},
       eprint = {1108.2711},
 primaryClass = {astro-ph.CO},
       adsurl = {https://ui.adsabs.harvard.edu/abs/2011MNRAS.418.2043E}
}

@ARTICLE{Ellison2013,
       author = {{Ellison}, Sara L. and {Mendel}, J. Trevor and {Patton}, David R. and {Scudder}, Jillian M.},
        title = "{Galaxy pairs in the Sloan Digital Sky Survey - VIII. The observational properties of post-merger galaxies}",
      journal = {\mnras},
         year = 2013,
        month = nov,
       volume = {435},
       number = {4},
        pages = {3627-3638},
          doi = {10.1093/mnras/stt1562},
archivePrefix = {arXiv},
       eprint = {1308.3707},
 primaryClass = {astro-ph.CO},
       adsurl = {https://ui.adsabs.harvard.edu/abs/2013MNRAS.435.3627E}
}

@ARTICLE{Ellison2019,
       author = {{Ellison}, Sara L. and {Viswanathan}, Akshara and {Patton}, David R. and {Bottrell}, Connor and {McConnachie}, Alan W. and {Gwyn}, Stephen and {Cuillandre}, Jean-Charles},
        title = "{A definitive merger-AGN connection at z {\ensuremath{\sim}} 0 with CFIS: mergers have an excess of AGN and AGN hosts are more frequently disturbed}",
      journal = {\mnras},
         year = 2019,
        month = aug,
       volume = {487},
       number = {2},
        pages = {2491-2504},
          doi = {10.1093/mnras/stz1431},
archivePrefix = {arXiv},
       eprint = {1905.08830},
 primaryClass = {astro-ph.GA},
       adsurl = {https://ui.adsabs.harvard.edu/abs/2019MNRAS.487.2491E}
}

@ARTICLE{Ellison2024,
       author = {{Ellison}, Sara and {Ferreira}, Leonardo and {Wild}, Vivienne and {Wilkinson}, Scott and {Rowlands}, Kate and {Patton}, David R.},
        title = "{Galaxy evolution in the post-merger regime. II {\textendash} Post-merger quenching peaks within 500 Myr of coalescence}",
      journal = {The Open Journal of Astrophysics},
         year = 2024,
        month = dec,
       volume = {7},
          eid = {121},
        pages = {121},
          doi = {10.33232/001c.127779},
archivePrefix = {arXiv},
       eprint = {2410.06357},
 primaryClass = {astro-ph.GA},
       adsurl = {https://ui.adsabs.harvard.edu/abs/2024OJAp....7E.121E}
}

@ARTICLE{Fabian2012,
       author = {{Fabian}, A.~C.},
        title = "{Observational Evidence of Active Galactic Nuclei Feedback}",
      journal = {\araa},
         year = 2012,
        month = sep,
       volume = {50},
        pages = {455-489},
          doi = {10.1146/annurev-astro-081811-125521},
archivePrefix = {arXiv},
       eprint = {1204.4114},
 primaryClass = {astro-ph.CO},
       adsurl = {https://ui.adsabs.harvard.edu/abs/2012ARA&A..50..455F}
}

@ARTICLE{Gensior2020,
       author = {{Gensior}, Jindra and {Kruijssen}, J.~M. Diederik and {Keller}, Benjamin W.},
        title = "{Heart of darkness: the influence of galactic dynamics on quenching star formation in galaxy spheroids}",
      journal = {\mnras},
         year = 2020,
        month = jun,
       volume = {495},
       number = {1},
        pages = {199-223},
          doi = {10.1093/mnras/staa1184},
archivePrefix = {arXiv},
       eprint = {2002.01484},
 primaryClass = {astro-ph.GA},
       adsurl = {https://ui.adsabs.harvard.edu/abs/2020MNRAS.495..199G}
}

@ARTICLE{Goubert2024,
       author = {{Goubert}, Paul H. and {Bluck}, Asa F.~L. and {Piotrowska}, Joanna M. and {Maiolino}, Roberto},
        title = "{The role of environment and AGN feedback in quenching local galaxies: comparing cosmological hydrodynamical simulations to the SDSS}",
      journal = {\mnras},
         year = 2024,
        month = mar,
       volume = {528},
       number = {3},
        pages = {4891-4921},
          doi = {10.1093/mnras/stae269},
archivePrefix = {arXiv},
       eprint = {2401.12953},
 primaryClass = {astro-ph.GA},
       adsurl = {https://ui.adsabs.harvard.edu/abs/2024MNRAS.528.4891G}
}

@ARTICLE{Goubert2025,
       author = {{Goubert}, Paul H. and {Bluck}, Asa F.~L. and {Piotrowska}, Joanna M. and {Torrey}, Paul and {Maiolino}, Roberto and {Franco}, Thomas Pinto and {Casimiro}, Camilo and {Cea}, Nicolas},
        title = "{Environmental versus intrinsic quenching at cosmic noon: predictions from cosmological hydrodynamical simulations for VLT-MOONRISE}",
      journal = {\mnras},
         year = 2025,
        month = nov,
       volume = {543},
       number = {3},
        pages = {2006-2034},
          doi = {10.1093/mnras/staf1554},
archivePrefix = {arXiv},
       eprint = {2509.09626},
 primaryClass = {astro-ph.GA},
       adsurl = {https://ui.adsabs.harvard.edu/abs/2025MNRAS.543.2006G}
}

@ARTICLE{Gunn1972,
       author = {{Gunn}, James E. and {Gott}, III, J. Richard},
        title = "{On the Infall of Matter Into Clusters of Galaxies and Some Effects on Their Evolution}",
      journal = {\apj},
         year = 1972,
        month = aug,
       volume = {176},
        pages = {1},
          doi = {10.1086/151605},
       adsurl = {https://ui.adsabs.harvard.edu/abs/1972ApJ...176....1G}
}

@ARTICLE{Hahn2017,
       author = {{Hahn}, ChangHoon and {Tinker}, Jeremy L. and {Wetzel}, Andrew},
        title = "{Star Formation Quenching Timescale of Central Galaxies in a Hierarchical Universe}",
      journal = {\apj},
         year = 2017,
        month = may,
       volume = {841},
       number = {1},
          eid = {6},
        pages = {6},
          doi = {10.3847/1538-4357/aa6d6b},
archivePrefix = {arXiv},
       eprint = {1609.04398},
 primaryClass = {astro-ph.GA},
       adsurl = {https://ui.adsabs.harvard.edu/abs/2017ApJ...841....6H}
}

@ARTICLE{Heavens2004,
       author = {{Heavens}, Alan and {Panter}, Benjamin and {Jimenez}, Raul and {Dunlop}, James},
        title = "{The star-formation history of the Universe from the stellar populations of nearby galaxies}",
      journal = {\nat},
         year = 2004,
        month = apr,
       volume = {428},
       number = {6983},
        pages = {625-627},
          doi = {10.1038/nature02474},
archivePrefix = {arXiv},
       eprint = {astro-ph/0403293},
 primaryClass = {astro-ph},
       adsurl = {https://ui.adsabs.harvard.edu/abs/2004Natur.428..625H}
}

@ARTICLE{Hopkins2006,
       author = {{Hopkins}, Philip F. and {Hernquist}, Lars and {Cox}, Thomas J. and {Di Matteo}, Tiziana and {Robertson}, Brant and {Springel}, Volker},
        title = "{A Unified, Merger-driven Model of the Origin of Starbursts, Quasars, the Cosmic X-Ray Background, Supermassive Black Holes, and Galaxy Spheroids}",
      journal = {\apjs},
         year = 2006,
        month = mar,
       volume = {163},
       number = {1},
        pages = {1-49},
          doi = {10.1086/499298},
archivePrefix = {arXiv},
       eprint = {astro-ph/0506398},
 primaryClass = {astro-ph},
       adsurl = {https://ui.adsabs.harvard.edu/abs/2006ApJS..163....1H}
}

@ARTICLE{Hopkins2008,
       author = {{Hopkins}, Philip F. and {Hernquist}, Lars and {Cox}, Thomas J. and {Kere{\v{s}}}, Du{\v{s}}an},
        title = "{A Cosmological Framework for the Co-Evolution of Quasars, Supermassive Black Holes, and Elliptical Galaxies. I. Galaxy Mergers and Quasar Activity}",
      journal = {\apjs},
         year = 2008,
        month = apr,
       volume = {175},
       number = {2},
        pages = {356-389},
          doi = {10.1086/524362},
archivePrefix = {arXiv},
       eprint = {0706.1243},
 primaryClass = {astro-ph},
       adsurl = {https://ui.adsabs.harvard.edu/abs/2008ApJS..175..356H}
}

@ARTICLE{Hoyle1939,
       author = {{Hoyle}, F. and {Lyttleton}, R.~A.},
        title = "{The effect of interstellar matter on climatic variation}",
      journal = {Proceedings of the Cambridge Philosophical Society},
         year = 1939,
        month = jan,
       volume = {35},
       number = {3},
        pages = {405},
          doi = {10.1017/S0305004100021150},
       adsurl = {https://ui.adsabs.harvard.edu/abs/1939PCPS...35..405H}
}

@ARTICLE{Knobel2015,
       author = {{Knobel}, Christian and {Lilly}, Simon J. and {Woo}, Joanna and {Kova{\v{c}}}, Katarina},
        title = "{Quenching of Star Formation in Sloan Digital Sky Survey Groups: Centrals, Satellites, and Galactic Conformity}",
      journal = {\apj},
         year = 2015,
        month = feb,
       volume = {800},
       number = {1},
          eid = {24},
        pages = {24},
          doi = {10.1088/0004-637X/800/1/24},
archivePrefix = {arXiv},
       eprint = {1408.2553},
 primaryClass = {astro-ph.GA},
       adsurl = {https://ui.adsabs.harvard.edu/abs/2015ApJ...800...24K}
}

@ARTICLE{Larson1980,
       author = {{Larson}, R.~B. and {Tinsley}, B.~M. and {Caldwell}, C.~N.},
        title = "{The evolution of disk galaxies and the origin of S0 galaxies}",
      journal = {\apj},
         year = 1980,
        month = may,
       volume = {237},
        pages = {692-707},
          doi = {10.1086/157917},
       adsurl = {https://ui.adsabs.harvard.edu/abs/1980ApJ...237..692L}
}

@ARTICLE{Lilly1996,
       author = {{Lilly}, S.~J. and {Le Fevre}, O. and {Hammer}, F. and {Crampton}, David},
        title = "{The Canada-France Redshift Survey: The Luminosity Density and Star Formation History of the Universe to Z approximately 1}",
      journal = {\apjl},
         year = 1996,
        month = mar,
       volume = {460},
        pages = {L1},
          doi = {10.1086/309975},
archivePrefix = {arXiv},
       eprint = {astro-ph/9601050},
 primaryClass = {astro-ph},
       adsurl = {https://ui.adsabs.harvard.edu/abs/1996ApJ...460L...1L}
}

@ARTICLE{Lotz2008,
       author = {{Lotz}, Jennifer M. and {Jonsson}, Patrik and {Cox}, T.~J. and {Primack}, Joel R.},
        title = "{Galaxy merger morphologies and time-scales from simulations of equal-mass gas-rich disc mergers}",
      journal = {\mnras},
         year = 2008,
        month = dec,
       volume = {391},
       number = {3},
        pages = {1137-1162},
          doi = {10.1111/j.1365-2966.2008.14004.x},
archivePrefix = {arXiv},
       eprint = {0805.1246},
 primaryClass = {astro-ph},
       adsurl = {https://ui.adsabs.harvard.edu/abs/2008MNRAS.391.1137L}
}

@ARTICLE{Lotz2010,
       author = {{Lotz}, Jennifer M. and {Jonsson}, Patrik and {Cox}, T.~J. and {Primack}, Joel R.},
        title = "{The effect of mass ratio on the morphology and time-scales of disc galaxy mergers}",
      journal = {\mnras},
         year = 2010,
        month = may,
       volume = {404},
       number = {2},
        pages = {575-589},
          doi = {10.1111/j.1365-2966.2010.16268.x},
archivePrefix = {arXiv},
       eprint = {0912.1590},
 primaryClass = {astro-ph.CO},
       adsurl = {https://ui.adsabs.harvard.edu/abs/2010MNRAS.404..575L}
}

@ARTICLE{Madau2014,
       author = {{Madau}, Piero and {Dickinson}, Mark},
        title = "{Cosmic Star-Formation History}",
      journal = {\araa},
         year = 2014,
        month = aug,
       volume = {52},
        pages = {415-486},
          doi = {10.1146/annurev-astro-081811-125615},
archivePrefix = {arXiv},
       eprint = {1403.0007},
 primaryClass = {astro-ph.CO},
       adsurl = {https://ui.adsabs.harvard.edu/abs/2014ARA&A..52..415M}
}

@ARTICLE{Maiolino2012,
       author = {{Maiolino}, R. and {Gallerani}, S. and {Neri}, R. and {Cicone}, C. and {Ferrara}, A. and {Genzel}, R. and {Lutz}, D. and {Sturm}, E. and {Tacconi}, L.~J. and {Walter}, F. and {Feruglio}, C. and {Fiore}, F. and {Piconcelli}, E.},
        title = "{Evidence of strong quasar feedback in the early Universe}",
      journal = {\mnras},
         year = 2012,
        month = sep,
       volume = {425},
       number = {1},
        pages = {L66-L70},
          doi = {10.1111/j.1745-3933.2012.01303.x},
archivePrefix = {arXiv},
       eprint = {1204.2904},
 primaryClass = {astro-ph.CO},
       adsurl = {https://ui.adsabs.harvard.edu/abs/2012MNRAS.425L..66M}
}

@ARTICLE{Maiolino2020,
       author = {{Maiolino}, R. and {Cirasuolo}, M. and {Afonso}, J. and {Bauer}, F.~E. and {Bowler}, R. and {Cucciati}, O. and {Daddi}, E. and {De Lucia}, G. and {Evans}, C. and {Flores}, H. and {Gargiulo}, A. and {Garilli}, B. and {Jablonka}, P. and {Jarvis}, M. and {Kneib}, J. -P. and {Lilly}, S. and {Looser}, T. and {Magliocchetti}, M. and {Man}, Z. and {Mannucci}, F. and {Maurogordato}, S. and {McLure}, R.~J. and {Norberg}, P. and {Oesch}, P. and {Oliva}, E. and {Paltani}, S. and {Pappalardo}, C. and {Peng}, Y. and {Pentericci}, L. and {Pozzetti}, L. and {Renzini}, A. and {Rodrigues}, M. and {Royer}, F. and {Serjeant}, S. and {Vanzi}, L. and {Wild}, V. and {Zamorani}, G.},
        title = "{MOONRISE: The Main MOONS GTO Extragalactic Survey}",
      journal = {The Messenger},
         year = 2020,
        month = jun,
       volume = {180},
        pages = {24-29},
          doi = {10.18727/0722-6691/5197},
archivePrefix = {arXiv},
       eprint = {2009.00644},
 primaryClass = {astro-ph.GA},
       adsurl = {https://ui.adsabs.harvard.edu/abs/2020Msngr.180...24M}
}

@ARTICLE{Maltby2018,
       author = {{Maltby}, David T. and {Almaini}, Omar and {Wild}, Vivienne and {Hatch}, Nina A. and {Hartley}, William G. and {Simpson}, Chris and {Rowlands}, Kate and {Socolovsky}, Miguel},
        title = "{The structure of post-starburst galaxies at 0.5 < z < 2: evidence for two distinct quenching routes at different epochs}",
      journal = {\mnras},
         year = 2018,
        month = oct,
       volume = {480},
       number = {1},
        pages = {381-401},
          doi = {10.1093/mnras/sty1794},
archivePrefix = {arXiv},
       eprint = {1807.01325},
 primaryClass = {astro-ph.GA},
       adsurl = {https://ui.adsabs.harvard.edu/abs/2018MNRAS.480..381M}
}

@ARTICLE{Mantha2019,
       author = {{Mantha}, Kameswara Bharadwaj and {McIntosh}, Daniel H. and {Ciaschi}, Cody P. and {Evan}, Rubyet and {Ferguson}, Henry C. and {Fries}, Logan B. and {Guo}, Yicheng and {Koekemoer}, Anton M. and {Landry}, Luther D. and {McGrath}, Elizabeth J. and {Simons}, Raymond C. and {Snyder}, Gregory F. and {Thompson}, Scott E. and {Bell}, Eric F. and {Ceverino}, Daniel and {Hathi}, Nimish P. and {Pacifici}, Camilla and {Primack}, Joel R. and {Rafelski}, Marc and {Rodriguez-Gomez}, Vicente},
        title = "{Studying the physical properties of tidal features - I. Extracting morphological substructure in CANDELS observations and VELA simulations}",
      journal = {\mnras},
         year = 2019,
        month = jun,
       volume = {486},
       number = {2},
        pages = {2643-2659},
          doi = {10.1093/mnras/stz872},
archivePrefix = {arXiv},
       eprint = {1903.11099},
 primaryClass = {astro-ph.GA},
       adsurl = {https://ui.adsabs.harvard.edu/abs/2019MNRAS.486.2643M}
}

@ARTICLE{Marinacci2018,
       author = {{Marinacci}, Federico and {Vogelsberger}, Mark and {Pakmor}, R{\"u}diger and {Torrey}, Paul and {Springel}, Volker and {Hernquist}, Lars and {Nelson}, Dylan and {Weinberger}, Rainer and {Pillepich}, Annalisa and {Naiman}, Jill and {Genel}, Shy},
        title = "{First results from the IllustrisTNG simulations: radio haloes and magnetic fields}",
      journal = {\mnras},
         year = 2018,
        month = nov,
       volume = {480},
       number = {4},
        pages = {5113-5139},
          doi = {10.1093/mnras/sty2206},
archivePrefix = {arXiv},
       eprint = {1707.03396},
 primaryClass = {astro-ph.CO},
       adsurl = {https://ui.adsabs.harvard.edu/abs/2018MNRAS.480.5113M}
}

@ARTICLE{Martig2009,
       author = {{Martig}, Marie and {Bournaud}, Fr{\'e}d{\'e}ric and {Teyssier}, Romain and {Dekel}, Avishai},
        title = "{Morphological Quenching of Star Formation: Making Early-Type Galaxies Red}",
      journal = {\apj},
         year = 2009,
        month = dec,
       volume = {707},
       number = {1},
        pages = {250-267},
          doi = {10.1088/0004-637X/707/1/250},
archivePrefix = {arXiv},
       eprint = {0905.4669},
 primaryClass = {astro-ph.CO},
       adsurl = {https://ui.adsabs.harvard.edu/abs/2009ApJ...707..250M}
}

@ARTICLE{Mihos1996,
       author = {{Mihos}, J. Christopher and {Hernquist}, Lars},
        title = "{Gasdynamics and Starbursts in Major Mergers}",
      journal = {\apj},
         year = 1996,
        month = jun,
       volume = {464},
        pages = {641},
          doi = {10.1086/177353},
archivePrefix = {arXiv},
       eprint = {astro-ph/9512099},
 primaryClass = {astro-ph},
       adsurl = {https://ui.adsabs.harvard.edu/abs/1996ApJ...464..641M}
}

@ARTICLE{Muratov2015,
       author = {{Muratov}, Alexander L. and {Kere{\v{s}}}, Du{\v{s}}an and {Faucher-Gigu{\`e}re}, Claude-Andr{\'e} and {Hopkins}, Philip F. and {Quataert}, Eliot and {Murray}, Norman},
        title = "{Gusty, gaseous flows of FIRE: galactic winds in cosmological simulations with explicit stellar feedback}",
      journal = {\mnras},
         year = 2015,
        month = dec,
       volume = {454},
       number = {3},
        pages = {2691-2713},
          doi = {10.1093/mnras/stv2126},
archivePrefix = {arXiv},
       eprint = {1501.03155},
 primaryClass = {astro-ph.GA},
       adsurl = {https://ui.adsabs.harvard.edu/abs/2015MNRAS.454.2691M}
}

@ARTICLE{Muzzin2013,
       author = {{Muzzin}, Adam and {Marchesini}, Danilo and {Stefanon}, Mauro and {Franx}, Marijn and {McCracken}, Henry J. and {Milvang-Jensen}, Bo and {Dunlop}, James S. and {Fynbo}, J.~P.~U. and {Brammer}, Gabriel and {Labb{\'e}}, Ivo and {van Dokkum}, Pieter G.},
        title = "{The Evolution of the Stellar Mass Functions of Star-forming and Quiescent Galaxies to z = 4 from the COSMOS/UltraVISTA Survey}",
      journal = {\apj},
         year = 2013,
        month = nov,
       volume = {777},
       number = {1},
          eid = {18},
        pages = {18},
          doi = {10.1088/0004-637X/777/1/18},
archivePrefix = {arXiv},
       eprint = {1303.4409},
 primaryClass = {astro-ph.CO},
       adsurl = {https://ui.adsabs.harvard.edu/abs/2013ApJ...777...18M}
}

@ARTICLE{Naab2003,
       author = {{Naab}, Thorsten and {Burkert}, Andreas},
        title = "{Statistical Properties of Collisionless Equal- and Unequal-Mass Merger Remnants of Disk Galaxies}",
      journal = {\apj},
         year = 2003,
        month = nov,
       volume = {597},
       number = {2},
        pages = {893-906},
          doi = {10.1086/378581},
archivePrefix = {arXiv},
       eprint = {astro-ph/0110179},
 primaryClass = {astro-ph},
       adsurl = {https://ui.adsabs.harvard.edu/abs/2003ApJ...597..893N}
}

@ARTICLE{Naiman2018,
       author = {{Naiman}, Jill P. and {Pillepich}, Annalisa and {Springel}, Volker and {Ramirez-Ruiz}, Enrico and {Torrey}, Paul and {Vogelsberger}, Mark and {Pakmor}, R{\"u}diger and {Nelson}, Dylan and {Marinacci}, Federico and {Hernquist}, Lars and {Weinberger}, Rainer and {Genel}, Shy},
        title = "{First results from the IllustrisTNG simulations: a tale of two elements - chemical evolution of magnesium and europium}",
      journal = {\mnras},
         year = 2018,
        month = jun,
       volume = {477},
       number = {1},
        pages = {1206-1224},
          doi = {10.1093/mnras/sty618},
archivePrefix = {arXiv},
       eprint = {1707.03401},
 primaryClass = {astro-ph.GA},
       adsurl = {https://ui.adsabs.harvard.edu/abs/2018MNRAS.477.1206N}
}

@ARTICLE{Nelson2018,
       author = {{Nelson}, Dylan and {Pillepich}, Annalisa and {Springel}, Volker and {Weinberger}, Rainer and {Hernquist}, Lars and {Pakmor}, R{\"u}diger and {Genel}, Shy and {Torrey}, Paul and {Vogelsberger}, Mark and {Kauffmann}, Guinevere and {Marinacci}, Federico and {Naiman}, Jill},
        title = "{First results from the IllustrisTNG simulations: the galaxy colour bimodality}",
      journal = {\mnras},
         year = 2018,
        month = mar,
       volume = {475},
       number = {1},
        pages = {624-647},
          doi = {10.1093/mnras/stx3040},
archivePrefix = {arXiv},
       eprint = {1707.03395},
 primaryClass = {astro-ph.GA},
       adsurl = {https://ui.adsabs.harvard.edu/abs/2018MNRAS.475..624N}
}

@ARTICLE{Nelson2015,
       author = {{Nelson}, D. and {Pillepich}, A. and {Genel}, S. and {Vogelsberger}, M. and {Springel}, V. and {Torrey}, P. and {Rodriguez-Gomez}, V. and {Sijacki}, D. and {Snyder}, G.~F. and {Griffen}, B. and {Marinacci}, F. and {Blecha}, L. and {Sales}, L. and {Xu}, D. and {Hernquist}, L.},
        title = "{The illustris simulation: Public data release}",
      journal = {Astronomy and Computing},
         year = 2015,
        month = nov,
       volume = {13},
        pages = {12-37},
          doi = {10.1016/j.ascom.2015.09.003},
archivePrefix = {arXiv},
       eprint = {1504.00362},
 primaryClass = {astro-ph.CO},
       adsurl = {https://ui.adsabs.harvard.edu/abs/2015A&C....13...12N}
}

@ARTICLE{Pacifici2016,
       author = {{Pacifici}, Camilla and {Kassin}, Susan A. and {Weiner}, Benjamin J. and {Holden}, Bradford and {Gardner}, Jonathan P. and {Faber}, Sandra M. and {Ferguson}, Henry C. and {Koo}, David C. and {Primack}, Joel R. and {Bell}, Eric F. and {Dekel}, Avishai and {Gawiser}, Eric and {Giavalisco}, Mauro and {Rafelski}, Marc and {Simons}, Raymond C. and {Barro}, Guillermo and {Croton}, Darren J. and {Dav{\'e}}, Romeel and {Fontana}, Adriano and {Grogin}, Norman A. and {Koekemoer}, Anton M. and {Lee}, Seong-Kook and {Salmon}, Brett and {Somerville}, Rachel and {Behroozi}, Peter},
        title = "{The Evolution of Star Formation Histories of Quiescent Galaxies}",
      journal = {\apj},
         year = 2016,
        month = nov,
       volume = {832},
       number = {1},
          eid = {79},
        pages = {79},
          doi = {10.3847/0004-637X/832/1/79},
archivePrefix = {arXiv},
       eprint = {1609.03572},
 primaryClass = {astro-ph.GA},
       adsurl = {https://ui.adsabs.harvard.edu/abs/2016ApJ...832...79P}
}

@ARTICLE{Patton2008,
       author = {{Patton}, D.~R. and {Atfield}, J.~E.},
        title = "{The Luminosity Dependence of the Galaxy Merger Rate}",
      journal = {\apj},
         year = 2008,
        month = sep,
       volume = {685},
       number = {1},
        pages = {235-246},
          doi = {10.1086/590542},
archivePrefix = {arXiv},
       eprint = {0806.0018},
 primaryClass = {astro-ph},
       adsurl = {https://ui.adsabs.harvard.edu/abs/2008ApJ...685..235P}
}

@ARTICLE{Patton2013,
       author = {{Patton}, David R. and {Torrey}, Paul and {Ellison}, Sara L. and {Mendel}, J. Trevor and {Scudder}, Jillian M.},
        title = "{Galaxy pairs in the Sloan Digital Sky Survey - VI. The orbital extent of enhanced star formation in interacting galaxies}",
      journal = {\mnras},
         year = 2013,
        month = jun,
       volume = {433},
       number = {1},
        pages = {L59-L63},
          doi = {10.1093/mnrasl/slt058},
archivePrefix = {arXiv},
       eprint = {1305.1595},
 primaryClass = {astro-ph.CO},
       adsurl = {https://ui.adsabs.harvard.edu/abs/2013MNRAS.433L..59P}
}

@ARTICLE{Patton2020,
       author = {{Patton}, David R. and {Wilson}, Kieran D. and {Metrow}, Colin J. and {Ellison}, Sara L. and {Torrey}, Paul and {Brown}, Westley and {Hani}, Maan H. and {McAlpine}, Stuart and {Moreno}, Jorge and {Woo}, Joanna},
        title = "{Interacting galaxies in the IllustrisTNG simulations - I: Triggered star formation in a cosmological context}",
      journal = {\mnras},
         year = 2020,
        month = jun,
       volume = {494},
       number = {4},
        pages = {4969-4985},
          doi = {10.1093/mnras/staa913},
archivePrefix = {arXiv},
       eprint = {2003.00289},
 primaryClass = {astro-ph.GA},
       adsurl = {https://ui.adsabs.harvard.edu/abs/2020MNRAS.494.4969P}
}

@ARTICLE{Pawlik2018,
       author = {{Pawlik}, M.~M. and {Taj Aldeen}, L. and {Wild}, V. and {Mendez-Abreu}, J. and {Lah{\'e}n}, N. and {Johansson}, P.~H. and {Jimenez}, N. and {Lucas}, W. and {Zheng}, Y. and {Walcher}, C.~J. and {Rowlands}, K.},
        title = "{The origins of post-starburst galaxies at z < 0.05}",
      journal = {\mnras},
         year = 2018,
        month = jun,
       volume = {477},
       number = {2},
        pages = {1708-1743},
          doi = {10.1093/mnras/sty589},
archivePrefix = {arXiv},
       eprint = {1803.02201},
 primaryClass = {astro-ph.GA},
       adsurl = {https://ui.adsabs.harvard.edu/abs/2018MNRAS.477.1708P}
}

@ARTICLE{Pedregosa2011,
       author = {{Pedregosa}, Fabian and {Varoquaux}, Ga{\"e}l and {Gramfort}, Alexandre and {Michel}, Vincent and {Thirion}, Bertrand and {Grisel}, Olivier and {Blondel}, Mathieu and {M{\"u}ller}, Andreas and {Nothman}, Joel and {Louppe}, Gilles and {Prettenhofer}, Peter and {Weiss}, Ron and {Dubourg}, Vincent and {Vanderplas}, Jake and {Passos}, Alexandre and {Cournapeau}, David and {Brucher}, Matthieu and {Perrot}, Matthieu and {Duchesnay}, {\'E}douard},
        title = "{Scikit-learn: Machine Learning in Python}",
      journal = {Journal of Machine Learning Research},
         year = 2011,
        month = oct,
       volume = {12},
        pages = {2825-2830},
          doi = {10.48550/arXiv.1201.0490},
archivePrefix = {arXiv},
       eprint = {1201.0490},
 primaryClass = {cs.LG},
       adsurl = {https://ui.adsabs.harvard.edu/abs/2011JMLR...12.2825P}
}

@ARTICLE{Pillepich2018,
       author = {{Pillepich}, Annalisa and {Springel}, Volker and {Nelson}, Dylan and {Genel}, Shy and {Naiman}, Jill and {Pakmor}, R{\"u}diger and {Hernquist}, Lars and {Torrey}, Paul and {Vogelsberger}, Mark and {Weinberger}, Rainer and {Marinacci}, Federico},
        title = "{Simulating galaxy formation with the IllustrisTNG model}",
      journal = {\mnras},
         year = 2018,
        month = jan,
       volume = {473},
       number = {3},
        pages = {4077-4106},
          doi = {10.1093/mnras/stx2656},
archivePrefix = {arXiv},
       eprint = {1703.02970},
 primaryClass = {astro-ph.GA},
       adsurl = {https://ui.adsabs.harvard.edu/abs/2018MNRAS.473.4077P}
}

@ARTICLE{Piotrowska2022,
       author = {{Piotrowska}, Joanna M. and {Bluck}, Asa F.~L. and {Maiolino}, Roberto and {Peng}, Yingjie},
        title = "{On the quenching of star formation in observed and simulated central galaxies: evidence for the role of integrated AGN feedback}",
      journal = {\mnras},
         year = 2022,
        month = may,
       volume = {512},
       number = {1},
        pages = {1052-1090},
          doi = {10.1093/mnras/stab3673},
archivePrefix = {arXiv},
       eprint = {2112.07672},
 primaryClass = {astro-ph.GA},
       adsurl = {https://ui.adsabs.harvard.edu/abs/2022MNRAS.512.1052P}
}

@ARTICLE{Planck2016,
       author = {{Planck Collaboration} and {Ade}, P.~A.~R. and {Aghanim}, N. and {Arnaud}, M. and {Ashdown}, M. and {Aumont}, J. and {Baccigalupi}, C. and {Banday}, A.~J. and {Barreiro}, R.~B. and {Bartlett}, J.~G. and {Bartolo}, N. and {Battaner}, E. and {Battye}, R. and {Benabed}, K. and {Beno{\^\i}t}, A. and {Benoit-L{\'e}vy}, A. and {Bernard}, J.-P. and {Bersanelli}, M. and {Bielewicz}, P. and {Bock}, J.~J. and {Bonaldi}, A. and {Bonavera}, L. and {Bond}, J.~R. and {Borrill}, J. and {Bouchet}, F.~R. and {Boulanger}, F. and {Bucher}, M. and {Burigana}, C. and {Butler}, R.~C. and {Calabrese}, E. and {Cardoso}, J.-F. and {Catalano}, A. and {Challinor}, A. and {Chamballu}, A. and {Chary}, R.-R. and {Chiang}, H.~C. and {Chluba}, J. and {Christensen}, P.~R. and {Church}, S. and {Clements}, D.~L. and {Colombi}, S. and {Colombo}, L.~P.~L. and {Combet}, C. and {Coulais}, A. and {Crill}, B.~P. and {Curto}, A. and {Cuttaia}, F. and {Danese}, L. and {Davies}, R.~D. and {Davis}, R.~J. and {de Bernardis}, P. and {de Rosa}, A. and {de Zotti}, G. and {Delabrouille}, J. and {D{\'e}sert}, F.-X. and {Di Valentino}, E. and {Dickinson}, C. and {Diego}, J.~M. and {Dolag}, K. and {Dole}, H. and {Donzelli}, S. and {Dor{\'e}}, O. and {Douspis}, M. and {Ducout}, A. and {Dunkley}, J. and {Dupac}, X. and {Efstathiou}, G. and {Elsner}, F. and {En{\ss}lin}, T.~A. and {Eriksen}, H.~K. and {Farhang}, M. and {Fergusson}, J. and {Finelli}, F. and {Forni}, O. and {Frailis}, M. and {Fraisse}, A.~A. and {Franceschi}, E. and {Frejsel}, A. and {Galeotta}, S. and {Galli}, S. and {Ganga}, K. and {Gauthier}, C. and {Gerbino}, M. and {Ghosh}, T. and {Giard}, M. and {Giraud-H{\'e}raud}, Y. and {Giusarma}, E. and {Gjerl{\o}w}, E. and {Gonz{\'a}lez-Nuevo}, J. and {G{\'o}rski}, K.~M. and {Gratton}, S. and {Gregorio}, A. and {Gruppuso}, A. and {Gudmundsson}, J.~E. and {Hamann}, J. and {Hansen}, F.~K. and {Hanson}, D. and {Harrison}, D.~L. and {Helou}, G. and {Henrot-Versill{\'e}}, S. and {Hern{\'a}ndez-Monteagudo}, C. and {Herranz}, D. and {Hildebrandt}, S.~R. and {Hivon}, E. and {Hobson}, M. and {Holmes}, W.~A. and {Hornstrup}, A. and {Hovest}, W. and {Huang}, Z. and {Huffenberger}, K.~M. and {Hurier}, G. and {Jaffe}, A.~H. and {Jaffe}, T.~R. and {Jones}, W.~C. and {Juvela}, M. and {Keih{\"a}nen}, E. and {Keskitalo}, R. and {Kisner}, T.~S. and {Kneissl}, R. and {Knoche}, J. and {Knox}, L. and {Kunz}, M. and {Kurki-Suonio}, H. and {Lagache}, G. and {L{\"a}hteenm{\"a}ki}, A. and {Lamarre}, J.-M. and {Lasenby}, A. and {Lattanzi}, M. and {Lawrence}, C.~R. and {Leahy}, J.~P. and {Leonardi}, R. and {Lesgourgues}, J. and {Levrier}, F. and {Lewis}, A. and {Liguori}, M. and {Lilje}, P.~B. and {Linden-V{\o}rnle}, M. and {L{\'o}pez-Caniego}, M. and {Lubin}, P.~M. and {Mac{\'\i}as-P{\'e}rez}, J.~F. and {Maggio}, G. and {Maino}, D. and {Mandolesi}, N. and {Mangilli}, A. and {Marchini}, A. and {Maris}, M. and {Martin}, P.~G. and {Martinelli}, M. and {Mart{\'\i}nez-Gonz{\'a}lez}, E. and {Masi}, S. and {Matarrese}, S. and {McGehee}, P. and {Meinhold}, P.~R. and {Melchiorri}, A. and {Melin}, J.-B. and {Mendes}, L. and {Mennella}, A. and {Migliaccio}, M. and {Millea}, M. and {Mitra}, S. and {Miville-Desch{\^e}nes}, M.-A. and {Moneti}, A. and {Montier}, L. and {Morgante}, G. and {Mortlock}, D. and {Moss}, A. and {Munshi}, D. and {Murphy}, J.~A. and {Naselsky}, P. and {Nati}, F. and {Natoli}, P. and {Netterfield}, C.~B. and {N{\o}rgaard-Nielsen}, H.~U. and {Noviello}, F. and {Novikov}, D. and {Novikov}, I. and {Oxborrow}, C.~A. and {Paci}, F. and {Pagano}, L. and {Pajot}, F. and {Paladini}, R. and {Paoletti}, D. and {Partridge}, B. and {Pasian}, F. and {Patanchon}, G. and {Pearson}, T.~J. and {Perdereau}, O. and {Perotto}, L. and {Perrotta}, F. and {Pettorino}, V. and {Piacentini}, F. and {Piat}, M. and {Pierpaoli}, E. and {Pietrobon}, D. and {Plaszczynski}, S. and {Pointecouteau}, E. and {Polenta}, G. and {Popa}, L. and {Pratt}, G.~W. and {Pr{\'e}zeau}, G.},
        title = "{Planck 2015 results. XIII. Cosmological parameters}",
      journal = {\aap},
         year = 2016,
        month = sep,
       volume = {594},
          eid = {A13},
        pages = {A13},
          doi = {10.1051/0004-6361/201525830},
archivePrefix = {arXiv},
       eprint = {1502.01589},
 primaryClass = {astro-ph.CO},
       adsurl = {https://ui.adsabs.harvard.edu/abs/2016A&A...594A..13P}
}

@ARTICLE{Quai2021,
       author = {{Quai}, Salvatore and {Hani}, Maan H. and {Ellison}, Sara L. and {Patton}, David R. and {Woo}, Joanna},
        title = "{Interacting galaxies in the IllustrisTNG simulations - III. (The rarity of) quenching in post-merger galaxies}",
      journal = {\mnras},
         year = 2021,
        month = jun,
       volume = {504},
       number = {2},
        pages = {1888-1901},
          doi = {10.1093/mnras/stab988},
archivePrefix = {arXiv},
       eprint = {2104.03327},
 primaryClass = {astro-ph.GA},
       adsurl = {https://ui.adsabs.harvard.edu/abs/2021MNRAS.504.1888Q}
}

@ARTICLE{Quai2023,
       author = {{Quai}, Salvatore and {Byrne-Mamahit}, Shoshannah and {Ellison}, Sara L. and {Patton}, David R. and {Hani}, Maan H.},
        title = "{The interconnection between galaxy mergers, AGN activity, and rapid quenching of star formation in simulated post-merger galaxies}",
      journal = {\mnras},
         year = 2023,
        month = feb,
       volume = {519},
       number = {2},
        pages = {2119-2137},
          doi = {10.1093/mnras/stac3713},
archivePrefix = {arXiv},
       eprint = {2212.10598},
 primaryClass = {astro-ph.GA},
       adsurl = {https://ui.adsabs.harvard.edu/abs/2023MNRAS.519.2119Q}
}

@ARTICLE{RodriguezGomez2015,
       author = {{Rodriguez-Gomez}, Vicente and {Genel}, Shy and {Vogelsberger}, Mark and {Sijacki}, Debora and {Pillepich}, Annalisa and {Sales}, Laura V. and {Torrey}, Paul and {Snyder}, Greg and {Nelson}, Dylan and {Springel}, Volker and {Ma}, Chung-Pei and {Hernquist}, Lars},
        title = "{The merger rate of galaxies in the Illustris simulation: a comparison with observations and semi-empirical models}",
      journal = {\mnras},
         year = 2015,
        month = may,
       volume = {449},
       number = {1},
        pages = {49-64},
          doi = {10.1093/mnras/stv264},
archivePrefix = {arXiv},
       eprint = {1502.01339},
 primaryClass = {astro-ph.GA},
       adsurl = {https://ui.adsabs.harvard.edu/abs/2015MNRAS.449...49R}
}

@ARTICLE{Santini2009,
       author = {{Santini}, P. and {Fontana}, A. and {Grazian}, A. and {Salimbeni}, S. and {Fiore}, F. and {Fontanot}, F. and {Boutsia}, K. and {Castellano}, M. and {Cristiani}, S. and {de Santis}, C. and {Gallozzi}, S. and {Giallongo}, E. and {Menci}, N. and {Nonino}, M. and {Paris}, D. and {Pentericci}, L. and {Vanzella}, E.},
        title = "{Star formation and mass assembly in high redshift galaxies}",
      journal = {\aap},
         year = 2009,
        month = sep,
       volume = {504},
       number = {3},
        pages = {751-767},
          doi = {10.1051/0004-6361/200811434},
archivePrefix = {arXiv},
       eprint = {0905.0683},
 primaryClass = {astro-ph.CO},
       adsurl = {https://ui.adsabs.harvard.edu/abs/2009A&A...504..751S}
}

@ARTICLE{Savitzky1964,
       author = {{Savitzky}, A. and {Golay}, M.~J.~E.},
        title = "{Smoothing and differentiation of data by simplified least squares procedures}",
      journal = {Analytical Chemistry},
         year = 1964,
        month = jan,
       volume = {36},
        pages = {1627-1639},
          doi = {10.1021/ac60214a047},
       adsurl = {https://ui.adsabs.harvard.edu/abs/1964AnaCh..36.1627S}
}

@ARTICLE{Schaye2015,
       author = {{Schaye}, Joop and {Crain}, Robert A. and {Bower}, Richard G. and {Furlong}, Michelle and {Schaller}, Matthieu and {Theuns}, Tom and {Dalla Vecchia}, Claudio and {Frenk}, Carlos S. and {McCarthy}, I.~G. and {Helly}, John C. and {Jenkins}, Adrian and {Rosas-Guevara}, Y.~M. and {White}, Simon D.~M. and {Baes}, Maarten and {Booth}, C.~M. and {Camps}, Peter and {Navarro}, Julio F. and {Qu}, Yan and {Rahmati}, Alireza and {Sawala}, Till and {Thomas}, Peter A. and {Trayford}, James},
        title = "{The EAGLE project: simulating the evolution and assembly of galaxies and their environments}",
      journal = {\mnras},
         year = 2015,
        month = jan,
       volume = {446},
       number = {1},
        pages = {521-554},
          doi = {10.1093/mnras/stu2058},
archivePrefix = {arXiv},
       eprint = {1407.7040},
 primaryClass = {astro-ph.GA},
       adsurl = {https://ui.adsabs.harvard.edu/abs/2015MNRAS.446..521S}
}

@ARTICLE{Scudder2012,
       author = {{Scudder}, Jillian M. and {Ellison}, Sara L. and {Torrey}, Paul and {Patton}, David R. and {Mendel}, J. Trevor},
        title = "{Galaxy pairs in the Sloan Digital Sky Survey - V. Tracing changes in star formation rate and metallicity out to separations of 80 kpc}",
      journal = {\mnras},
         year = 2012,
        month = oct,
       volume = {426},
       number = {1},
        pages = {549-565},
          doi = {10.1111/j.1365-2966.2012.21749.x},
archivePrefix = {arXiv},
       eprint = {1207.4791},
 primaryClass = {astro-ph.CO},
       adsurl = {https://ui.adsabs.harvard.edu/abs/2012MNRAS.426..549S}
}

@ARTICLE{Silk2012,
       author = {{Silk}, Joseph and {Mamon}, Gary A.},
        title = "{The current status of galaxy formation}",
      journal = {Research in Astronomy and Astrophysics},
         year = 2012,
        month = aug,
       volume = {12},
       number = {8},
        pages = {917-946},
          doi = {10.1088/1674-4527/12/8/004},
archivePrefix = {arXiv},
       eprint = {1207.3080},
 primaryClass = {astro-ph.CO},
       adsurl = {https://ui.adsabs.harvard.edu/abs/2012RAA....12..917S}
}

@ARTICLE{Sijacki2007,
       author = {{Sijacki}, Debora and {Springel}, Volker and {Di Matteo}, Tiziana and {Hernquist}, Lars},
        title = "{A unified model for AGN feedback in cosmological simulations of structure formation}",
      journal = {\mnras},
         year = 2007,
        month = sep,
       volume = {380},
       number = {3},
        pages = {877-900},
          doi = {10.1111/j.1365-2966.2007.12153.x},
archivePrefix = {arXiv},
       eprint = {0705.2238},
 primaryClass = {astro-ph},
       adsurl = {https://ui.adsabs.harvard.edu/abs/2007MNRAS.380..877S}
}

@ARTICLE{Silverman2011,
       author = {{Silverman}, J.~D. and {Kampczyk}, P. and {Jahnke}, K. and {Andrae}, R. and {Lilly}, S.~J. and {Elvis}, M. and {Civano}, F. and {Mainieri}, V. and {Vignali}, C. and {Zamorani}, G. and {Nair}, P. and {Le F{\`e}vre}, O. and {de Ravel}, L. and {Bardelli}, S. and {Bongiorno}, A. and {Bolzonella}, M. and {Cappi}, A. and {Caputi}, K. and {Carollo}, C.~M. and {Contini}, T. and {Coppa}, G. and {Cucciati}, O. and {de la Torre}, S. and {Franzetti}, P. and {Garilli}, B. and {Halliday}, C. and {Hasinger}, G. and {Iovino}, A. and {Knobel}, C. and {Koekemoer}, A.~M. and {Kova{\v{c}}}, K. and {Lamareille}, F. and {Le Borgne}, J.-F. and {Le Brun}, V. and {Maier}, C. and {Mignoli}, M. and {Pello}, R. and {P{\'e}rez-Montero}, E. and {Ricciardelli}, E. and {Peng}, Y. and {Scodeggio}, M. and {Tanaka}, M. and {Tasca}, L. and {Tresse}, L. and {Vergani}, D. and {Zucca}, E. and {Brusa}, M. and {Cappelluti}, N. and {Comastri}, A. and {Finoguenov}, A. and {Fu}, H. and {Gilli}, R. and {Hao}, H. and {Ho}, L.~C. and {Salvato}, M.},
        title = "{The Impact of Galaxy Interactions on Active Galactic Nucleus Activity in zCOSMOS}",
      journal = {\apj},
         year = 2011,
        month = dec,
       volume = {743},
       number = {1},
          eid = {2},
        pages = {2},
          doi = {10.1088/0004-637X/743/1/2},
archivePrefix = {arXiv},
       eprint = {1109.1292},
 primaryClass = {astro-ph.CO},
       adsurl = {https://ui.adsabs.harvard.edu/abs/2011ApJ...743....2S}
}

@ARTICLE{Springel2005,
       author = {{Springel}, Volker and {Di Matteo}, Tiziana and {Hernquist}, Lars},
        title = "{Modelling feedback from stars and black holes in galaxy mergers}",
      journal = {\mnras},
         year = 2005,
        month = aug,
       volume = {361},
       number = {3},
        pages = {776-794},
          doi = {10.1111/j.1365-2966.2005.09238.x},
archivePrefix = {arXiv},
       eprint = {astro-ph/0411108},
 primaryClass = {astro-ph},
       adsurl = {https://ui.adsabs.harvard.edu/abs/2005MNRAS.361..776S}
}

@ARTICLE{Springel2010,
       author = {{Springel}, Volker},
        title = "{E pur si muove: Galilean-invariant cosmological hydrodynamical simulations on a moving mesh}",
      journal = {\mnras},
         year = 2010,
        month = jan,
       volume = {401},
       number = {2},
        pages = {791-851},
          doi = {10.1111/j.1365-2966.2009.15715.x},
archivePrefix = {arXiv},
       eprint = {0901.4107},
 primaryClass = {astro-ph.CO},
       adsurl = {https://ui.adsabs.harvard.edu/abs/2010MNRAS.401..791S}
}

@ARTICLE{Springel2018,
       author = {{Springel}, Volker and {Pakmor}, R{\"u}diger and {Pillepich}, Annalisa and {Weinberger}, Rainer and {Nelson}, Dylan and {Hernquist}, Lars and {Vogelsberger}, Mark and {Genel}, Shy and {Torrey}, Paul and {Marinacci}, Federico and {Naiman}, Jill},
        title = "{First results from the IllustrisTNG simulations: matter and galaxy clustering}",
      journal = {\mnras},
         year = 2018,
        month = mar,
       volume = {475},
       number = {1},
        pages = {676-698},
          doi = {10.1093/mnras/stx3304},
archivePrefix = {arXiv},
       eprint = {1707.03397},
 primaryClass = {astro-ph.GA},
       adsurl = {https://ui.adsabs.harvard.edu/abs/2018MNRAS.475..676S}
}

@ARTICLE{Steinier1972,
       author = {{Steinier}, Jean. and {Termonia}, Yves. and {Deltour}, Jules.},
        title = "{Smoothing and differentiation of data by simplified least square procedure}",
      journal = {Analytical Chemistry},
         year = 1972,
        month = sep,
       volume = {44},
       number = {11},
        pages = {1906-1909},
          doi = {10.1021/ac60319a045},
       adsurl = {https://ui.adsabs.harvard.edu/abs/1972AnaCh..44.1906S}
}

@ARTICLE{Strateva2001,
       author = {{Strateva}, Iskra and {Ivezi{\'c}}, {\v{Z}}eljko and {Knapp}, Gillian R. and {Narayanan}, Vijay K. and {Strauss}, Michael A. and {Gunn}, James E. and {Lupton}, Robert H. and {Schlegel}, David and {Bahcall}, Neta A. and {Brinkmann}, Jon and {Brunner}, Robert J. and {Budav{\'a}ri}, Tam{\'a}s and {Csabai}, Istv{\'a}n and {Castander}, Francisco Javier and {Doi}, Mamoru and {Fukugita}, Masataka and {Gy{\H{o}}ry}, Zsuzsanna and {Hamabe}, Masaru and {Hennessy}, Greg and {Ichikawa}, Takashi and {Kunszt}, Peter Z. and {Lamb}, Don Q. and {McKay}, Timothy A. and {Okamura}, Sadanori and {Racusin}, Judith and {Sekiguchi}, Maki and {Schneider}, Donald P. and {Shimasaku}, Kazuhiro and {York}, Donald},
        title = "{Color Separation of Galaxy Types in the Sloan Digital Sky Survey Imaging Data}",
      journal = {\aj},
         year = 2001,
        month = oct,
       volume = {122},
       number = {4},
        pages = {1861-1874},
          doi = {10.1086/323301},
archivePrefix = {arXiv},
       eprint = {astro-ph/0107201},
 primaryClass = {astro-ph},
       adsurl = {https://ui.adsabs.harvard.edu/abs/2001AJ....122.1861S}
}

@ARTICLE{Teimoorinia2016,
       author = {{Teimoorinia}, Hossen and {Bluck}, Asa F.~L. and {Ellison}, Sara L.},
        title = "{An artificial neural network approach for ranking quenching parameters in central galaxies}",
      journal = {\mnras},
         year = 2016,
        month = apr,
       volume = {457},
       number = {2},
        pages = {2086-2106},
          doi = {10.1093/mnras/stw036},
archivePrefix = {arXiv},
       eprint = {1601.01258},
 primaryClass = {astro-ph.GA},
       adsurl = {https://ui.adsabs.harvard.edu/abs/2016MNRAS.457.2086T}
}

@ARTICLE{Terrazas2020,
       author = {{Terrazas}, Bryan A. and {Bell}, Eric F. and {Pillepich}, Annalisa and {Nelson}, Dylan and {Somerville}, Rachel S. and {Genel}, Shy and {Weinberger}, Rainer and {Habouzit}, M{\'e}lanie and {Li}, Yuan and {Hernquist}, Lars and {Vogelsberger}, Mark},
        title = "{The relationship between black hole mass and galaxy properties: examining the black hole feedback model in IllustrisTNG}",
      journal = {\mnras},
         year = 2020,
        month = apr,
       volume = {493},
       number = {2},
        pages = {1888-1906},
          doi = {10.1093/mnras/staa374},
archivePrefix = {arXiv},
       eprint = {1906.02747},
 primaryClass = {astro-ph.GA},
       adsurl = {https://ui.adsabs.harvard.edu/abs/2020MNRAS.493.1888T}
}

@ARTICLE{Thomas2005,
       author = {{Thomas}, Daniel and {Maraston}, Claudia and {Bender}, Ralf and {Mendes de Oliveira}, Claudia},
        title = "{The Epochs of Early-Type Galaxy Formation as a Function of Environment}",
      journal = {\apj},
         year = 2005,
        month = mar,
       volume = {621},
       number = {2},
        pages = {673-694},
          doi = {10.1086/426932},
archivePrefix = {arXiv},
       eprint = {astro-ph/0410209},
 primaryClass = {astro-ph},
       adsurl = {https://ui.adsabs.harvard.edu/abs/2005ApJ...621..673T}
}

@ARTICLE{Toomre1972,
       author = {{Toomre}, Alar and {Toomre}, Juri},
        title = "{Galactic Bridges and Tails}",
      journal = {\apj},
         year = 1972,
        month = dec,
       volume = {178},
        pages = {623-666},
          doi = {10.1086/151823},
       adsurl = {https://ui.adsabs.harvard.edu/abs/1972ApJ...178..623T}
}

@ARTICLE{Trussler2020,
       author = {{Trussler}, James and {Maiolino}, Roberto and {Maraston}, Claudia and {Peng}, Yingjie and {Thomas}, Daniel and {Goddard}, Daniel and {Lian}, Jianhui},
        title = "{Both starvation and outflows drive galaxy quenching}",
      journal = {\mnras},
         year = 2020,
        month = feb,
       volume = {491},
       number = {4},
        pages = {5406-5434},
          doi = {10.1093/mnras/stz3286},
archivePrefix = {arXiv},
       eprint = {1811.09283},
 primaryClass = {astro-ph.GA},
       adsurl = {https://ui.adsabs.harvard.edu/abs/2020MNRAS.491.5406T}
}

@ARTICLE{vanDenBosch2008,
       author = {{van den Bosch}, Frank C. and {Aquino}, Daniel and {Yang}, Xiaohu and {Mo}, H.~J. and {Pasquali}, Anna and {McIntosh}, Daniel H. and {Weinmann}, Simone M. and {Kang}, Xi},
        title = "{The importance of satellite quenching for the build-up of the red sequence of present-day galaxies}",
      journal = {\mnras},
         year = 2008,
        month = jun,
       volume = {387},
       number = {1},
        pages = {79-91},
          doi = {10.1111/j.1365-2966.2008.13230.x},
archivePrefix = {arXiv},
       eprint = {0710.3164},
 primaryClass = {astro-ph},
       adsurl = {https://ui.adsabs.harvard.edu/abs/2008MNRAS.387...79V}
}

@ARTICLE{Vogelsberger2014a,
       author = {{Vogelsberger}, Mark and {Genel}, Shy and {Springel}, Volker and {Torrey}, Paul and {Sijacki}, Debora and {Xu}, Dandan and {Snyder}, Greg and {Nelson}, Dylan and {Hernquist}, Lars},
        title = "{Introducing the Illustris Project: simulating the coevolution of dark and visible matter in the Universe}",
      journal = {\mnras},
         year = 2014,
        month = oct,
       volume = {444},
       number = {2},
        pages = {1518-1547},
          doi = {10.1093/mnras/stu1536},
archivePrefix = {arXiv},
       eprint = {1405.2921},
 primaryClass = {astro-ph.CO},
       adsurl = {https://ui.adsabs.harvard.edu/abs/2014MNRAS.444.1518V}
}

@ARTICLE{Vogelsberger2014b,
       author = {{Vogelsberger}, M. and {Genel}, S. and {Springel}, V. and {Torrey}, P. and {Sijacki}, D. and {Xu}, D. and {Snyder}, G. and {Bird}, S. and {Nelson}, D. and {Hernquist}, L.},
        title = "{Properties of galaxies reproduced by a hydrodynamic simulation}",
      journal = {\nat},
         year = 2014,
        month = may,
       volume = {509},
       number = {7499},
        pages = {177-182},
          doi = {10.1038/nature13316},
archivePrefix = {arXiv},
       eprint = {1405.1418},
 primaryClass = {astro-ph.CO},
       adsurl = {https://ui.adsabs.harvard.edu/abs/2014Natur.509..177V}
}

@ARTICLE{Weinberger2017,
       author = {{Weinberger}, Rainer and {Springel}, Volker and {Hernquist}, Lars and {Pillepich}, Annalisa and {Marinacci}, Federico and {Pakmor}, R{\"u}diger and {Nelson}, Dylan and {Genel}, Shy and {Vogelsberger}, Mark and {Naiman}, Jill and {Torrey}, Paul},
        title = "{Simulating galaxy formation with black hole driven thermal and kinetic feedback}",
      journal = {\mnras},
         year = 2017,
        month = mar,
       volume = {465},
       number = {3},
        pages = {3291-3308},
          doi = {10.1093/mnras/stw2944},
archivePrefix = {arXiv},
       eprint = {1607.03486},
 primaryClass = {astro-ph.GA},
       adsurl = {https://ui.adsabs.harvard.edu/abs/2017MNRAS.465.3291W}
}

@ARTICLE{Weinberger2018,
       author = {{Weinberger}, Rainer and {Springel}, Volker and {Pakmor}, R{\"u}diger and {Nelson}, Dylan and {Genel}, Shy and {Pillepich}, Annalisa and {Vogelsberger}, Mark and {Marinacci}, Federico and {Naiman}, Jill and {Torrey}, Paul and {Hernquist}, Lars},
        title = "{Supermassive black holes and their feedback effects in the IllustrisTNG simulation}",
      journal = {\mnras},
         year = 2018,
        month = sep,
       volume = {479},
       number = {3},
        pages = {4056-4072},
          doi = {10.1093/mnras/sty1733},
archivePrefix = {arXiv},
       eprint = {1710.04659},
 primaryClass = {astro-ph.GA},
       adsurl = {https://ui.adsabs.harvard.edu/abs/2018MNRAS.479.4056W}
}

@ARTICLE{Whitaker2011,
       author = {{Whitaker}, Katherine E. and {Labb{\'e}}, Ivo and {van Dokkum}, Pieter G. and {Brammer}, Gabriel and {Kriek}, Mariska and {Marchesini}, Danilo and {Quadri}, Ryan F. and {Franx}, Marijn and {Muzzin}, Adam and {Williams}, Rik J. and {Bezanson}, Rachel and {Illingworth}, Garth D. and {Lee}, Kyoung-Soo and {Lundgren}, Britt and {Nelson}, Erica J. and {Rudnick}, Gregory and {Tal}, Tomer and {Wake}, David A.},
        title = "{The NEWFIRM Medium-band Survey: Photometric Catalogs, Redshifts, and the Bimodal Color Distribution of Galaxies out to z \raisebox{-0.5ex}\textasciitilde 3}",
      journal = {\apj},
         year = 2011,
        month = jul,
       volume = {735},
       number = {2},
          eid = {86},
        pages = {86},
          doi = {10.1088/0004-637X/735/2/86},
archivePrefix = {arXiv},
       eprint = {1105.4609},
 primaryClass = {astro-ph.CO},
       adsurl = {https://ui.adsabs.harvard.edu/abs/2011ApJ...735...86W}
}

@ARTICLE{Wild2009,
       author = {{Wild}, Vivienne and {Walcher}, C. Jakob and {Johansson}, Peter H. and {Tresse}, Laurence and {Charlot}, St{\'e}phane and {Pollo}, Agnieszka and {Le F{\`e}vre}, Olivier and {de Ravel}, Loic},
        title = "{Post-starburst galaxies: more than just an interesting curiosity}",
      journal = {\mnras},
         year = 2009,
        month = may,
       volume = {395},
       number = {1},
        pages = {144-159},
          doi = {10.1111/j.1365-2966.2009.14537.x},
archivePrefix = {arXiv},
       eprint = {0810.5122},
 primaryClass = {astro-ph},
       adsurl = {https://ui.adsabs.harvard.edu/abs/2009MNRAS.395..144W}
}

@ARTICLE{Wild2010,
       author = {{Wild}, Vivienne and {Heckman}, Timothy and {Charlot}, St{\'e}phane},
        title = "{Timing the starburst-AGN connection}",
      journal = {\mnras},
         year = 2010,
        month = jun,
       volume = {405},
       number = {2},
        pages = {933-947},
          doi = {10.1111/j.1365-2966.2010.16536.x},
archivePrefix = {arXiv},
       eprint = {1002.3156},
 primaryClass = {astro-ph.CO},
       adsurl = {https://ui.adsabs.harvard.edu/abs/2010MNRAS.405..933W}
}

@ARTICLE{Wilkinson2022,
       author = {{Wilkinson}, Scott and {Ellison}, Sara L. and {Bottrell}, Connor and {Bickley}, Robert W. and {Gwyn}, Stephen and {Cuillandre}, Jean-Charles and {Wild}, Vivienne},
        title = "{The merger fraction of post-starburst galaxies in UNIONS}",
      journal = {\mnras},
         year = 2022,
        month = nov,
       volume = {516},
       number = {3},
        pages = {4354-4372},
          doi = {10.1093/mnras/stac1962},
archivePrefix = {arXiv},
       eprint = {2207.04152},
 primaryClass = {astro-ph.GA},
       adsurl = {https://ui.adsabs.harvard.edu/abs/2022MNRAS.516.4354W}
}

@ARTICLE{Williams2009,
       author = {{Williams}, Rik J. and {Quadri}, Ryan F. and {Franx}, Marijn and {van Dokkum}, Pieter and {Labb{\'e}}, Ivo},
        title = "{Detection of Quiescent Galaxies in a Bicolor Sequence from Z = 0-2}",
      journal = {\apj},
         year = 2009,
        month = feb,
       volume = {691},
       number = {2},
        pages = {1879-1895},
          doi = {10.1088/0004-637X/691/2/1879},
archivePrefix = {arXiv},
       eprint = {0806.0625},
 primaryClass = {astro-ph},
       adsurl = {https://ui.adsabs.harvard.edu/abs/2009ApJ...691.1879W}
}

@ARTICLE{Zinger2020,
       author = {{Zinger}, Elad and {Pillepich}, Annalisa and {Nelson}, Dylan and {Weinberger}, Rainer and {Pakmor}, R{\"u}diger and {Springel}, Volker and {Hernquist}, Lars and {Marinacci}, Federico and {Vogelsberger}, Mark},
        title = "{Ejective and preventative: the IllustrisTNG black hole feedback and its effects on the thermodynamics of the gas within and around galaxies}",
      journal = {\mnras},
         year = 2020,
        month = nov,
       volume = {499},
       number = {1},
        pages = {768-792},
          doi = {10.1093/mnras/staa2607},
archivePrefix = {arXiv},
       eprint = {2004.06132},
 primaryClass = {astro-ph.GA},
       adsurl = {https://ui.adsabs.harvard.edu/abs/2020MNRAS.499..768Z}
}

@MISC{Bluck2024b,
       author = {{Bluck}, Asa F.~L. and {Conselice}, Christopher J. and {Ormerod}, Katherine and {Piotrowska}, Joanna M. and {Adams}, Nathan and {Austin}, Duncan and {Caruana}, Joseph and {Duncan}, K.~J. and {Ferreira}, Leonardo and {Goubert}, Paul and {Harvey}, Thomas and {Trussler}, James and {Maiolino}, Roberto},
        title = "{Galaxy Quenching at the High Redshift Frontier: A Fundamental Test of Cosmological Models in the Early Universe with JWST-CEERS}",
      journal = {\apj},
         year = 2024,
        month = feb,
       volume = {961},
       number = {2},
          eid = {163},
        pages = {163},
          doi = {10.3847/1538-4357/ad0a98},
archivePrefix = {arXiv},
       eprint = {2311.02526},
 primaryClass = {astro-ph.GA},
       adsurl = {https://ui.adsabs.harvard.edu/abs/2024ApJ...961..163B}
}

@ARTICLE{Hyndman1996,
       author = {{Hyndman}, Rob J.},
        title = "{Computing and Graphing Highest Density Regions}",
      journal = {The American Statistician},
         year = 1996,
        month = may,
       volume = {50},
       number = {2},
        pages = {120-126},
          doi = {10.1080/00031305.1996.10474359},
       adsurl = {https://www.jstor.org/stable/2684423}
}

@ARTICLE{SamworthWand2010,
       author = {{Samworth}, Richard J. and {Wand}, Matt P.},
        title = "{Asymptotics and Optimal Bandwidth Selection for Highest Density Region Estimation}",
      journal = {The Annals of Statistics},
         year = 2010,
        month = jun,
       volume = {38},
       number = {3},
        pages = {1767-1792},
          doi = {10.1214/09-AOS766},
       adsurl = {https://doi.org/10.48550/arXiv.1010.0591}
}


\appendix
\section{Stellar Mass-Ratio Determination}
\label{sec:App_AA}

\begin{figure*}
    \centering
    \includegraphics[width=0.85\textwidth]{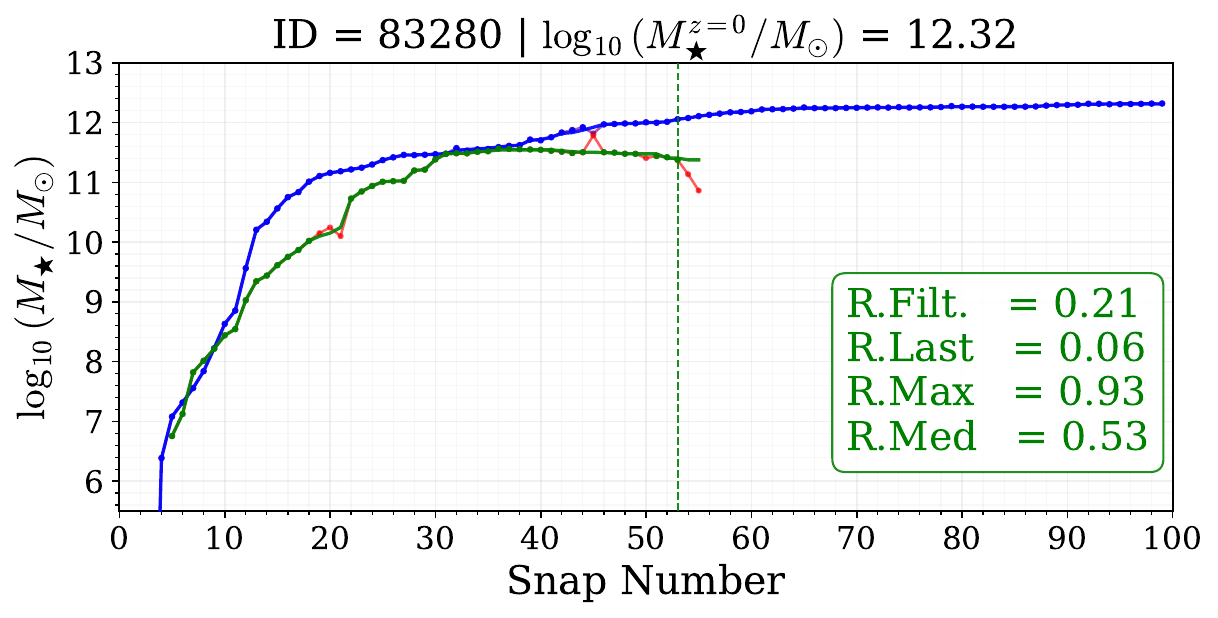}
    \caption{
    Example of the stellar-mass evolution of a merging pair. The raw stellar-mass histories of the host and satellite are shown (purple and red, respectively), together with the filtered histories (blue and green, respectively) obtained using the procedure described in this section. The vertical green dashed line marks the last snapshot prior to coalescence at which the raw stellar-mass histories remain reliable, before significant deviations due to particle misclassification occur. The annotations indicate the stellar mass--ratio predicted for this system by each of the four methods (filtered, maximum-satellite-mass, median, and pre-coalescence), with the filtered estimate highlighted by the dashed vertical line.}
    \label{fig:Example_Ratio}
\end{figure*}

\begin{figure*}
    \centering
    \includegraphics[width=0.85\textwidth]{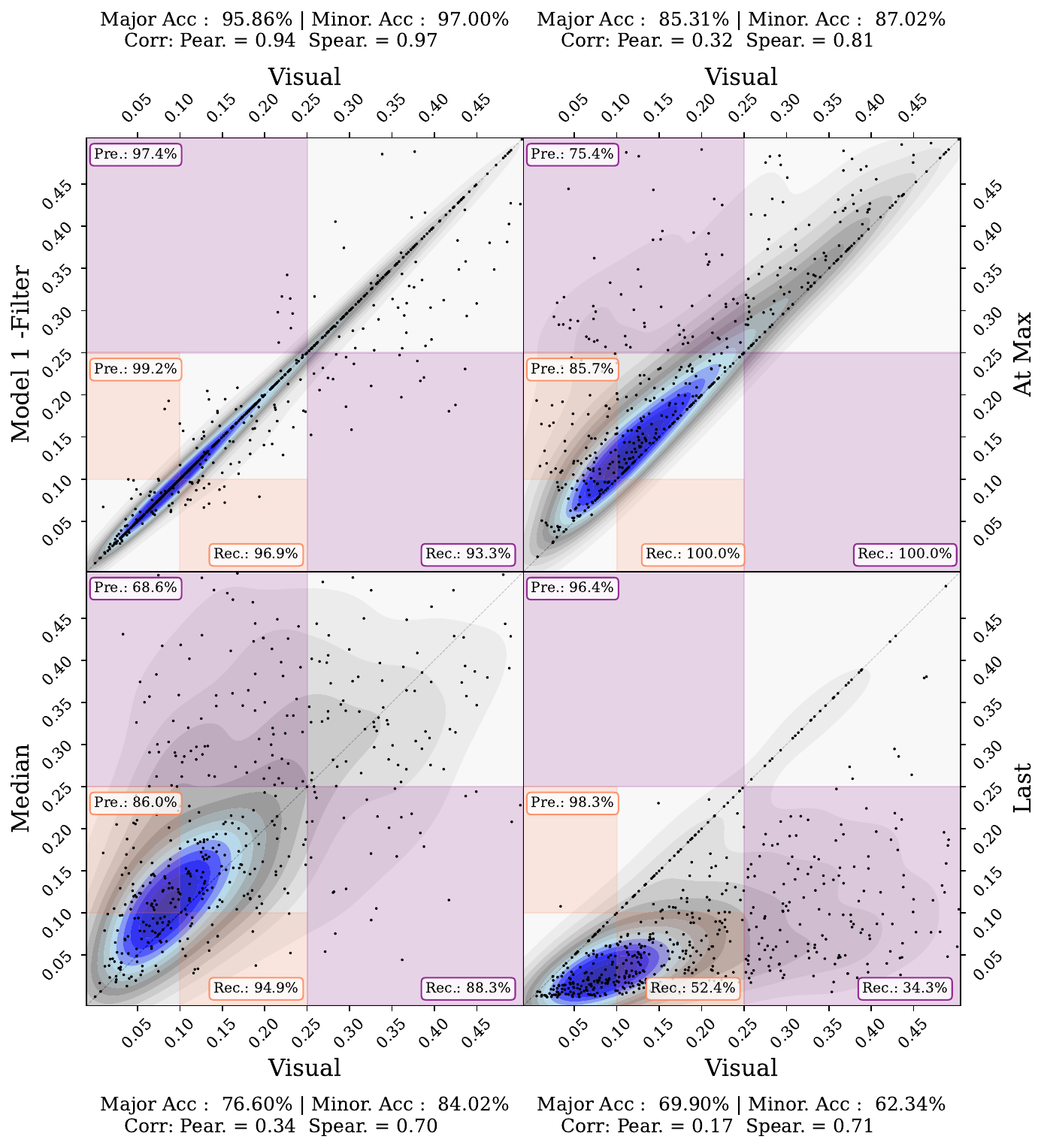}
    \caption{
    Comparison between the stellar mass--ratio obtained from visual inspection and the four methods considered in this work: the filtered method, the maximum-satellite-mass method, the median-mass method, and the pre-coalescence snapshot method. The filtered method shows the strongest agreement with the visual classification.}
    \label{fig:Models_Comp}
\end{figure*}

\begin{figure*}
    \centering
    \includegraphics[width=0.85\textwidth]{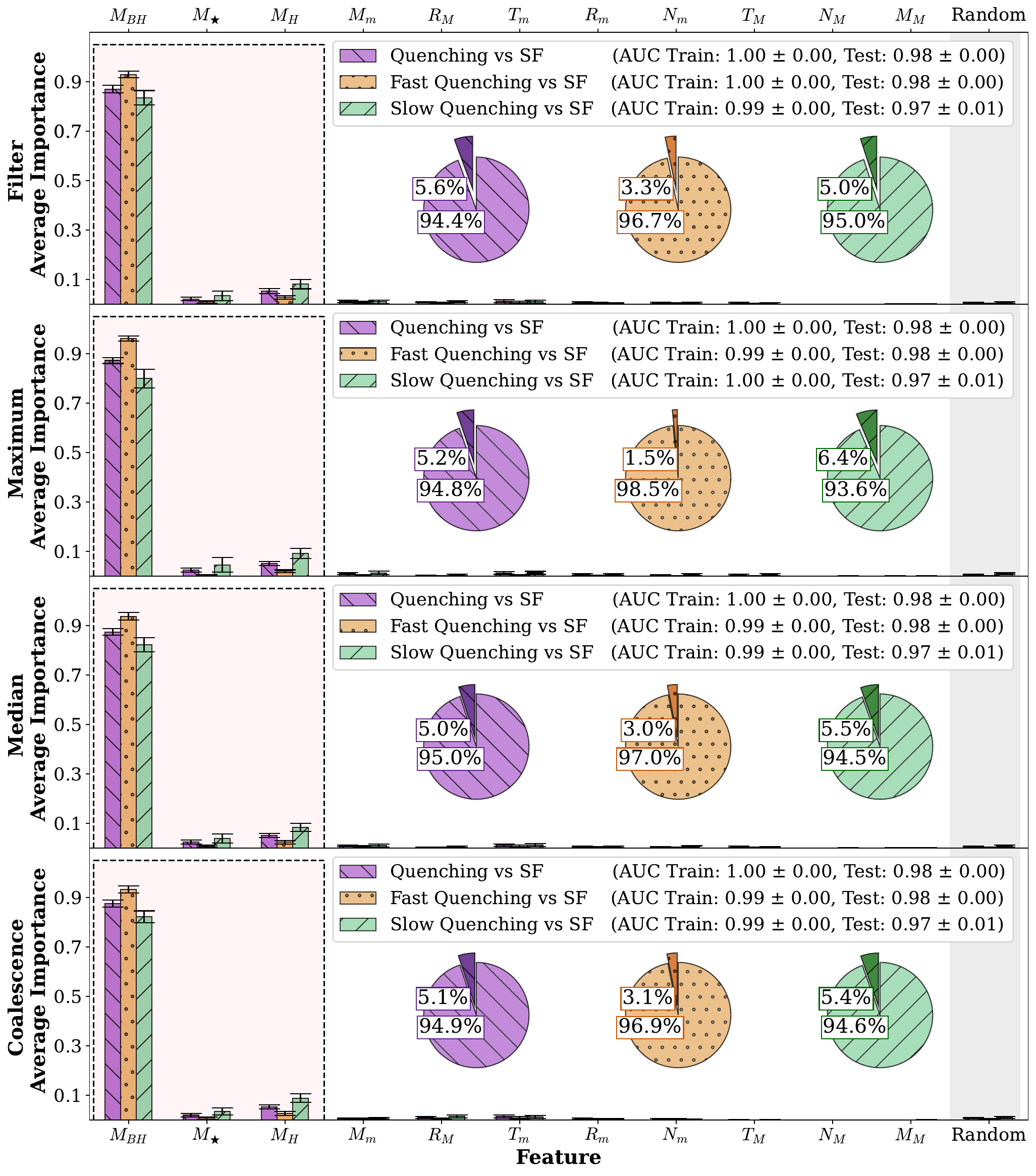}
    \caption{
    The impact of the stellar mass--ratio determination method on the Random Forest classification analysis. The structure of this figure is identical to Fig. 13 (top panel), but shows the stability of this result to different methods to identify the merger ratio. The relative feature importances and overall classification performance remain consistent across all four definitions, demonstrating that the main scientific conclusions are robust to the specific choice of stellar mass--ratio estimator.}
    \label{fig:Model_Comparison_RF}
\end{figure*}

For each merger event, we follow the stellar mass histories of the host and satellite across all snapshots and identify a subset of snapshots that reflect physically meaningful evolution, while excluding numerical artifacts and mass transfer. To achieve this we implement a novel approach which we outline and test here. We also compare against alternative methods and demonstrate that our results and conclusions are stable across various reasonable approaches for dealing with this well-known issue.

First, we suppress short-timescale reversible fluctuations caused by particle misclassification by applying a rolling-median filter with a window scaled to the number of resolved snapshots. Second, we enforce physically motivated monotonicity by applying a recursive constraint that only permits sustained stellar-mass decreases below a fixed threshold (absolute decrease) per snapshot, preventing spurious drops while retaining genuine stripping prior to coalescence. Third, we flag anomalous snapshots where the difference between the raw and filtered stellar-mass histories exceeds a fixed tolerance, identifying times where the raw evolution deviates from the expected physical behavior.

A snapshot is considered valid only if it is unflagged for both the host and satellite, ensuring that the stellar-mass ratio is measured only when both galaxies have reliable mass estimates. The merger stellar mass ratio is then computed from the subset of jointly valid snapshots using the selection criterion illustrated in Fig.~\ref{fig:Example_Ratio}.

For comparison, we also compute three other stellar mass-ratio estimates from the same stellar-mass histories. First, we measure the ratio at the snapshot immediately preceding coalescence, using the last timestep where both galaxies are still identified as separate objects. Second, we measure the ratio at the snapshot where the satellite reaches its maximum stellar mass (as in \citealt{RodriguezGomez2015} and \citealt{Patton2020}), in principle corresponding to the time of peak stellar content prior to any significant stripping. Third, we compute the ratio of the median stellar masses of the host and satellite over the full time interval during which both galaxies coexist as distinct objects.

These three alternative definitions are shown alongside our filtered-measurement method in Fig.~\ref{fig:Example_Ratio}, allowing direct visual comparison between the raw stellar-mass histories and the filtered histories, and between the stellar mass--ratio values returned by each method.

To validate the stellar mass--ratio measurements, we visually inspect a sample of 700 merging galaxy pairs and manually selected, for each system, the last snapshot at which both galaxies appeared undisturbed prior to coalescence, defining a visual reference mass ratio. We then compare this visual estimate to the predictions of the four models described above (our filtered method, the maximum-satellite-mass method, the median-mass method, and the pre-coalescence snapshot method). 

As shown in Fig.~\ref{fig:Models_Comp}, the filtered method exhibits the highest agreement with the visual selection, with strong correlations (Pearson $r \approx 0.94$, Spearman $r \approx 0.97$) and the highest classification accuracy for both major and minor mergers, while the alternative methods show progressively weaker agreement and larger systematic scatter. It is important to note that choosing the last snapshot prior to coalescence systematically underestimates the true merger ratio. Alternatively, choosing the highest mass value of the secondary galaxy systematically overestimates the true merger ratio. No bias is found by using the median of the history, however, the scatter against the true merger ratio is very large in this case. Consequently, by far the most effective method is our novel, filter-based approach.

Finally, we recompute the main Random Forest analysis presented in the top panel of Fig.~\ref{fig:RF_All} using each of the four stellar mass--ratio definitions (filtered, maximum-mass, median, and pre-coalescence). As shown in Fig.~\ref{fig:Model_Comparison_RF}, the relative feature importances, classification performance, and overall trends remain qualitatively unchanged across all definitions. In particular, the same set of physical quantities dominates the predictions in all cases, and the model performance (AUC) varies only at the percent level. This demonstrates that, although our filtering procedure provides a more physically robust estimate of the stellar mass ratio, the main scientific conclusions of this work are not sensitive to the specific definition adopted. 

Explicitly, the complete lack of importance of parameters connected to merging for predicting quenching remains completely stable to all reasonable approaches to deal with the assignment of merger ratios. Hence, the main results and conclusions of this work are completely robust to issues connected with mass transfer and numerical fluctuations during galaxy interactions.

\section{Additional Examples of Star Formation Histories}
\label{sec:App_B}

In this part of the appendix we present additional examples of individual star formation histories (SFHs) for galaxies in each quenching category: star-forming (Fig. B1), fast-quenched (Fig. B2), and slow-quenched (Fig. B3), based on their specific star formation rate (sSFR) evolution. These examples complement the classification approach described in Section~\ref{sec:Q-SF_Class} by illustrating the diversity of SFHs within each class. The galaxies shown span the same stellar mass range and redshift cuts as in the main analysis. 

\begin{figure*}
    \centering
    \includegraphics[width=0.96\textwidth]{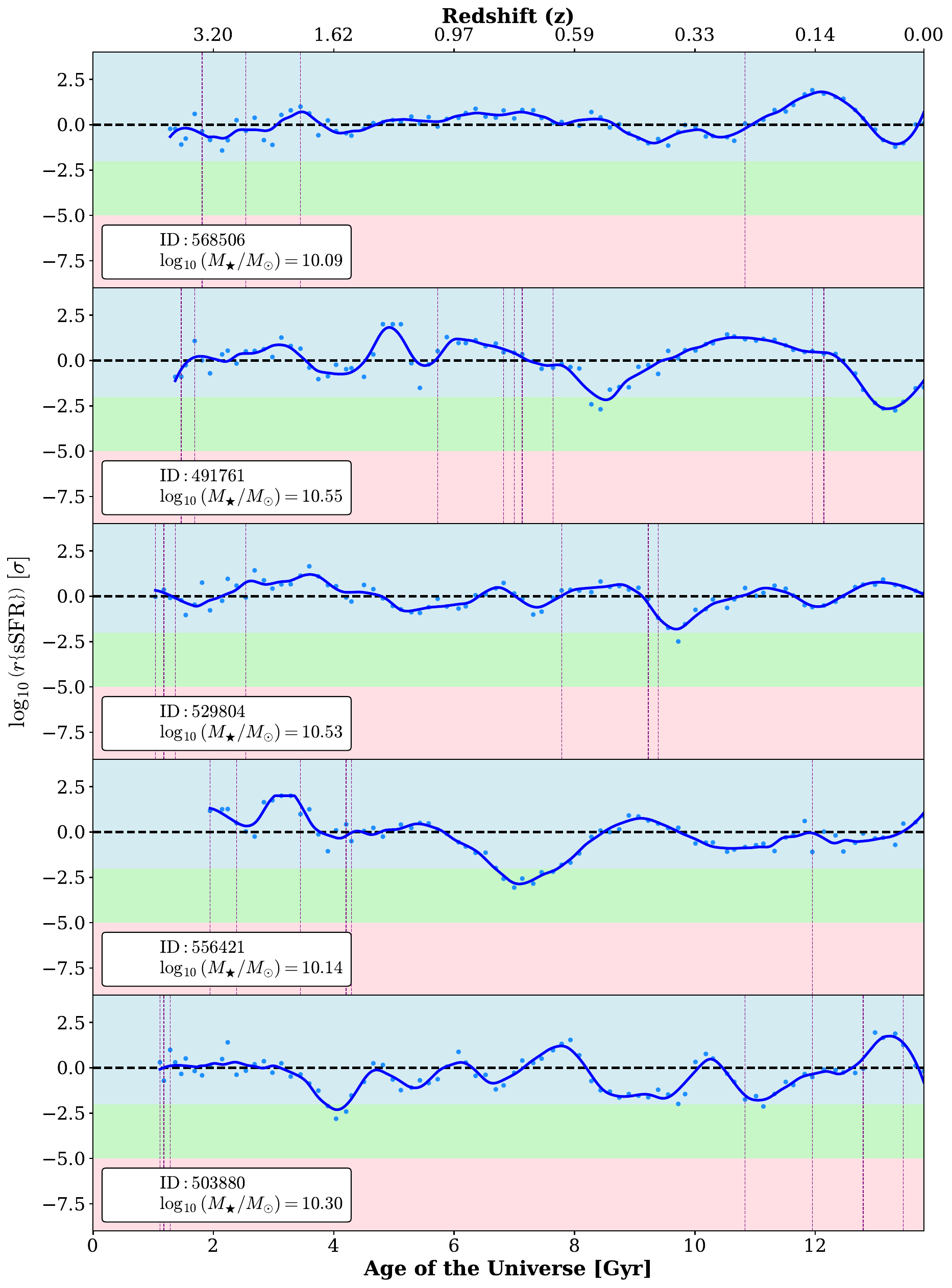}
    \caption{Additional examples of star-forming galaxies, shown in the same format as Figure~\ref{RsSFR_Hist} for consistency. These cases further illustrate the typical features of the star-forming population.}
    \label{fig:extra_SF}
\end{figure*}

\begin{figure*}
    \centering
    \includegraphics[width=0.96\textwidth]{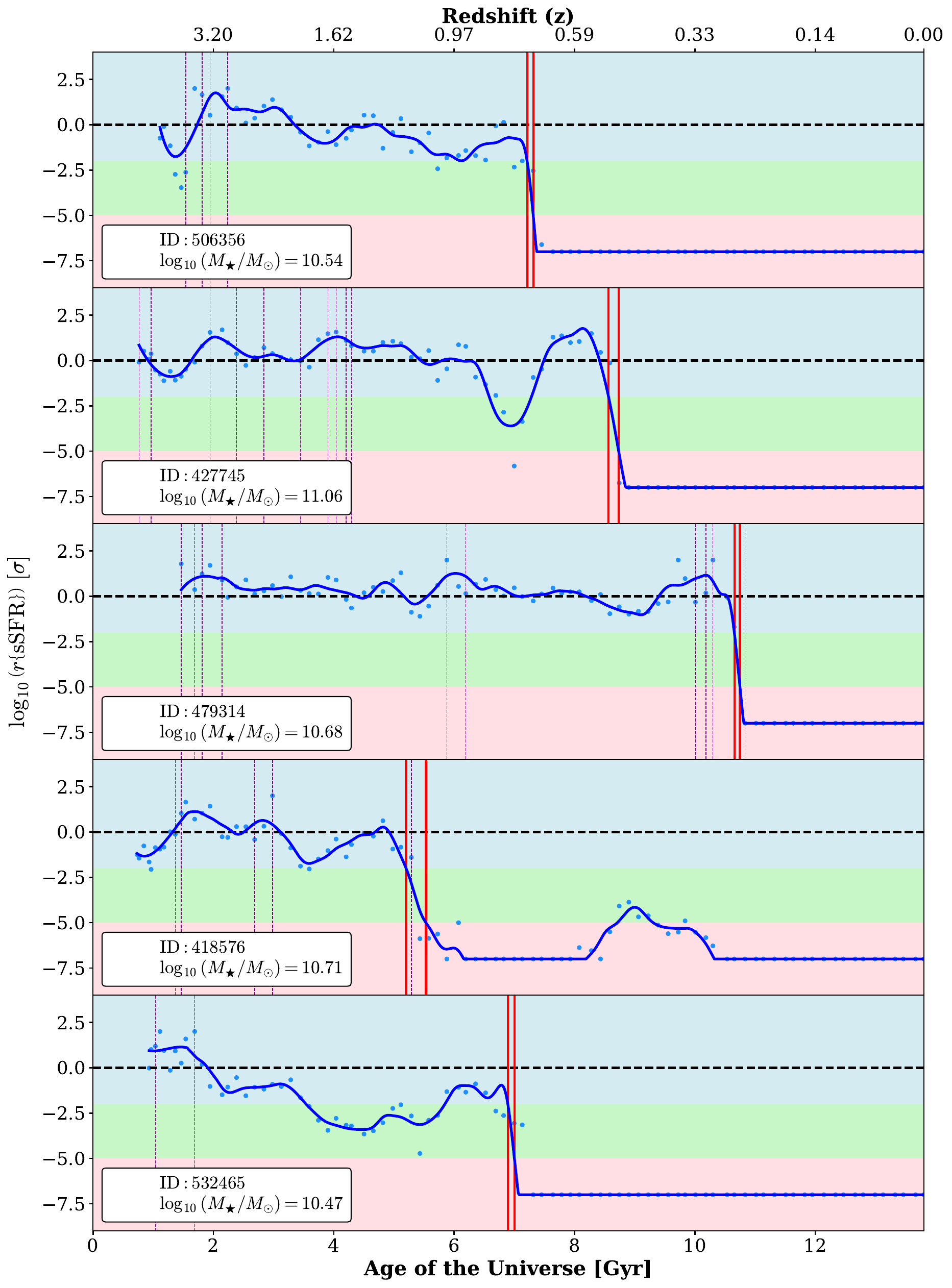}
    \caption{Additional examples of fast quenching galaxies, shown in the same format as Figure~\ref{RsSFR_Hist} for consistency. These cases further illustrate the typical features of the fast quenching population.}
    \label{fig:extra_fast}
\end{figure*}

\begin{figure*}
    \centering
    \includegraphics[width=0.96\textwidth]{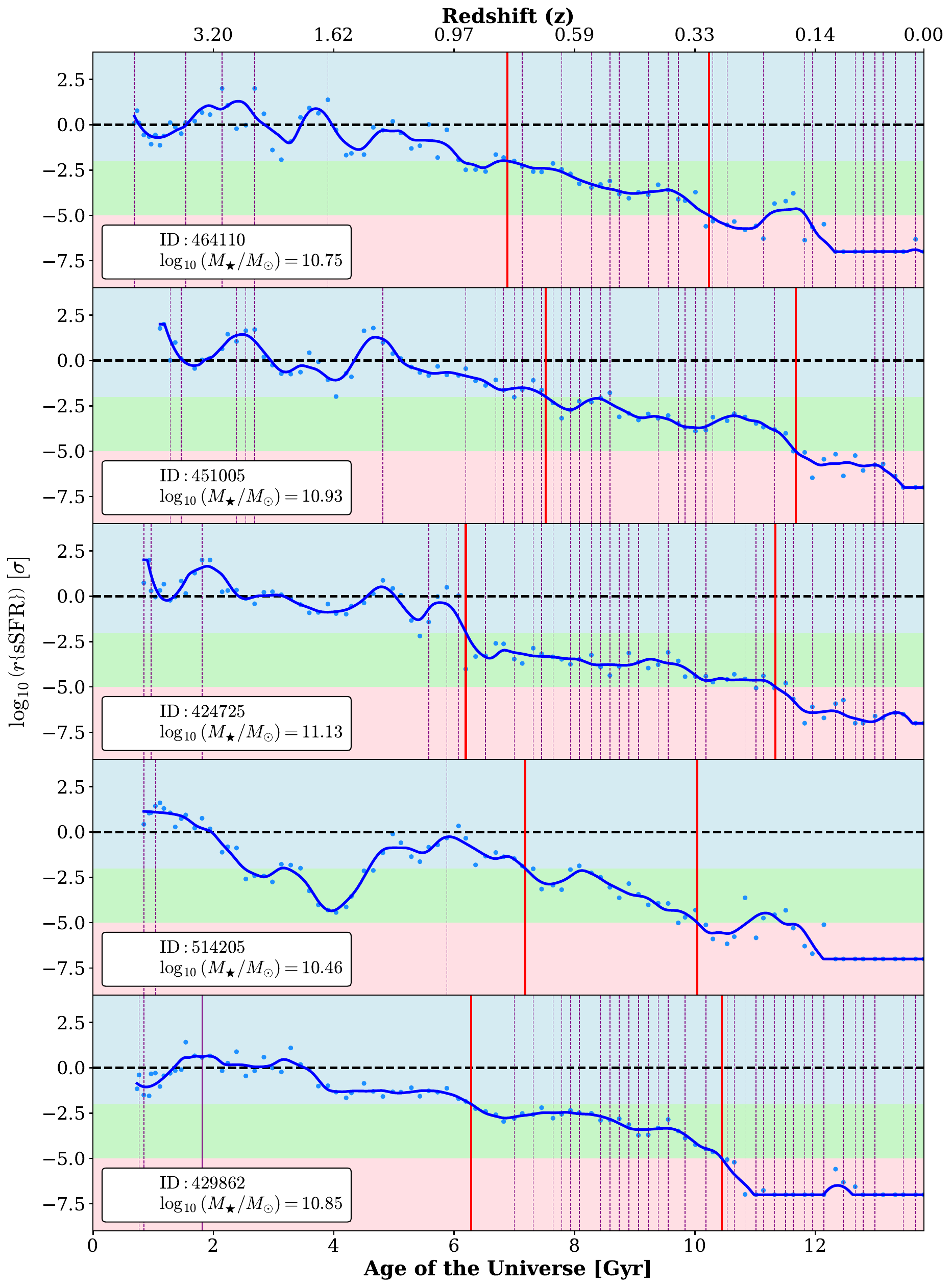}
    \caption{Additional examples of slow quenching galaxies, shown in the same format as Figure~\ref{RsSFR_Hist} for consistency. These cases further illustrate the typical features of the slow quenching population.}
    \label{fig:extra_slow}
\end{figure*}

\section{Machine Learning Reproducibility}
\label{sec:App_C}

In this work we employ Random Forest (RF) models, using the \texttt{scikit‑learn} implementations, for both classification and regression tasks.  Classification models (\texttt{RF\_Class}) are used to distinguish between various galaxy populations, including star-forming (SF) versus quenched (Q), SF versus fast‑quenched, and SF versus slow‑quenched systems, while regression models (\texttt{RF\_Regre}) predict black hole masses based on galaxy and halo properties. This appendix provides the key hyper‑parameters used and explains their role in our experiments.

\subsection*{Hyper‑parameter settings:}

\emph{Method:} We use the RandomForestClassifier and RandomForestRegressor classes from \texttt{scikit‑learn} throughout.  Classification tasks are abbreviated as \texttt{RF\_Class} and regression tasks as \texttt{RF\_Regre} in  Table~\ref{fig:Opt_Table_2}.\\

\emph{Normalized:} All input features are normalized by subtracting the median and dividing by the inter-quartile range.  This scaling mitigates differences in feature distributions and prevents variables with large numerical ranges from dominating the training process.\\

\emph{Number of iterations ($N_{\mathrm{int}}$):} Each RF model is trained 100 times using independent random splits of the data (\texttt{N\_int = 100}).  For classification, the majority class is randomly down‑sampled at each iteration so that the training sample contains equal numbers of galaxies in each class.  Feature importances reported in the main text are the mean values across these 100 iterations, and the associated errors are their standard deviations.\\

\emph{Balanced:} Classification runs use balanced samples (as described above) to counteract class imbalance, especially important when quenched systems are rare at high redshift.  Balanced sampling is not applied to regression tasks.\\

\emph{Train–test ratio:} We adopt a 50:50 split between training and testing sets, which is usually recommended for large datasets.\\

\emph{Number of trees (\texttt{n\_estimators}):} We use 1000 decision trees in each RF model.  A large ensemble of trees reduces variance and stabilizes the feature importance rankings.\\

\emph{Maximum depth (\texttt{max\_depth}):} The maximum depth of trees is left unconstrained (\texttt{max\_depth = None}), allowing the trees to grow until further splits are prohibited by the minimum samples thresholds below.  This lets the model capture complex, non‑linear relationships.\\

\emph{Minimum samples per split and leaf:} The number of samples required to attempt a split is set to 2 (\texttt{min\_samples\_split = 2}), while the minimum number of samples per terminal leaf (\texttt{min\_samples\_leaf}  or MSL) is optimized individually for each model. This approach minimizes constraints on tree growth and allows the model complexity to be controlled primarily through the tuned MSL parameter, ensuring statistically robust leaves while avoiding overfitting.\\

\emph{Number of features per split (\texttt{max\_features}):} All features are considered at each split (\texttt{max\_features = None}), which maximizes the model’s ability to find informative splits and best deals with correlated variables as mentioned before.\\

\emph{Maximum leaf nodes (\texttt{max\_leaf\_nodes}):} We do not restrict the number of leaf nodes (\texttt{max\_leaf\_nodes = None}); the tree structure is controlled by the minimum samples thresholds described above.\\

\emph{Bootstrap:} We enable bootstrap resampling (\texttt{bootstrap = True}) so that each tree is trained on a random subset of the data.  This increases diversity among trees and lowers variance in the ensemble.\\

\emph{Random state:} A fixed random seed (\texttt{random\_state = 42}) is set for all runs to enable exact reproducibility of results.\\

\subsection*{MSL optimization:}

We optimize the MSL separately for each classification and regression model.  Values from 2 up to 200 are explored.  For classification, we compute the test AUC across 100 realizations and select the MSL that yields the maximum test AUC (mean value) while ensuring $\Delta$AUC\,$\equiv$\,AUC$_\mathrm{train}$–AUC$_\mathrm{test} \leq 0.035$.  For regression, we compute the test RMSE over 100 realizations and choose the MSL that minimizes the test RMSE (mean value) with $|\Delta$RMSE|\,$\equiv|$RMSE$_\mathrm{train}$–RMSE$_\mathrm{test}| \leq 0.035$.  

\begin{figure}
    \centering
    \includegraphics[width=0.5\textwidth]{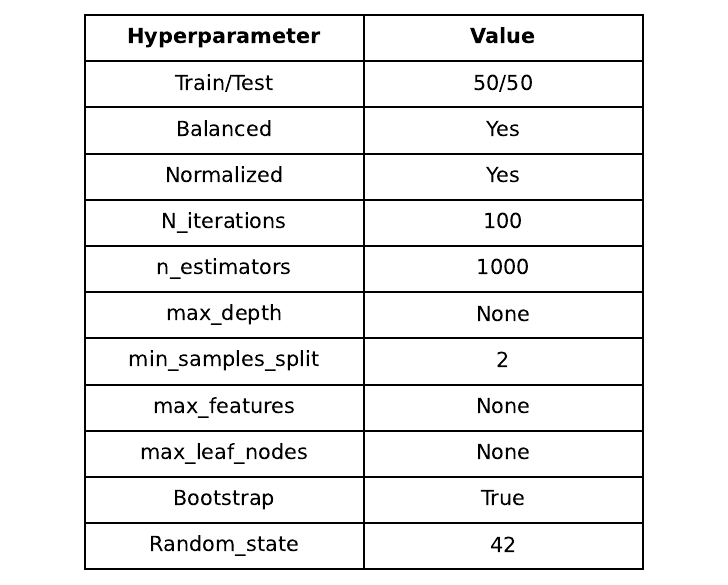}
    \caption{Common Random Forest hyper-parameters adopted in all classification and regression models. A 50:50 train–test split and balanced, normalized data were used; models contain 1,000 trees with unrestricted depth, bootstrap sampling, and random seed = 42}
    \label{fig:Opt_Table_1}
\end{figure}

\begin{figure*}
    \centering
    \includegraphics[width=0.96\textwidth]{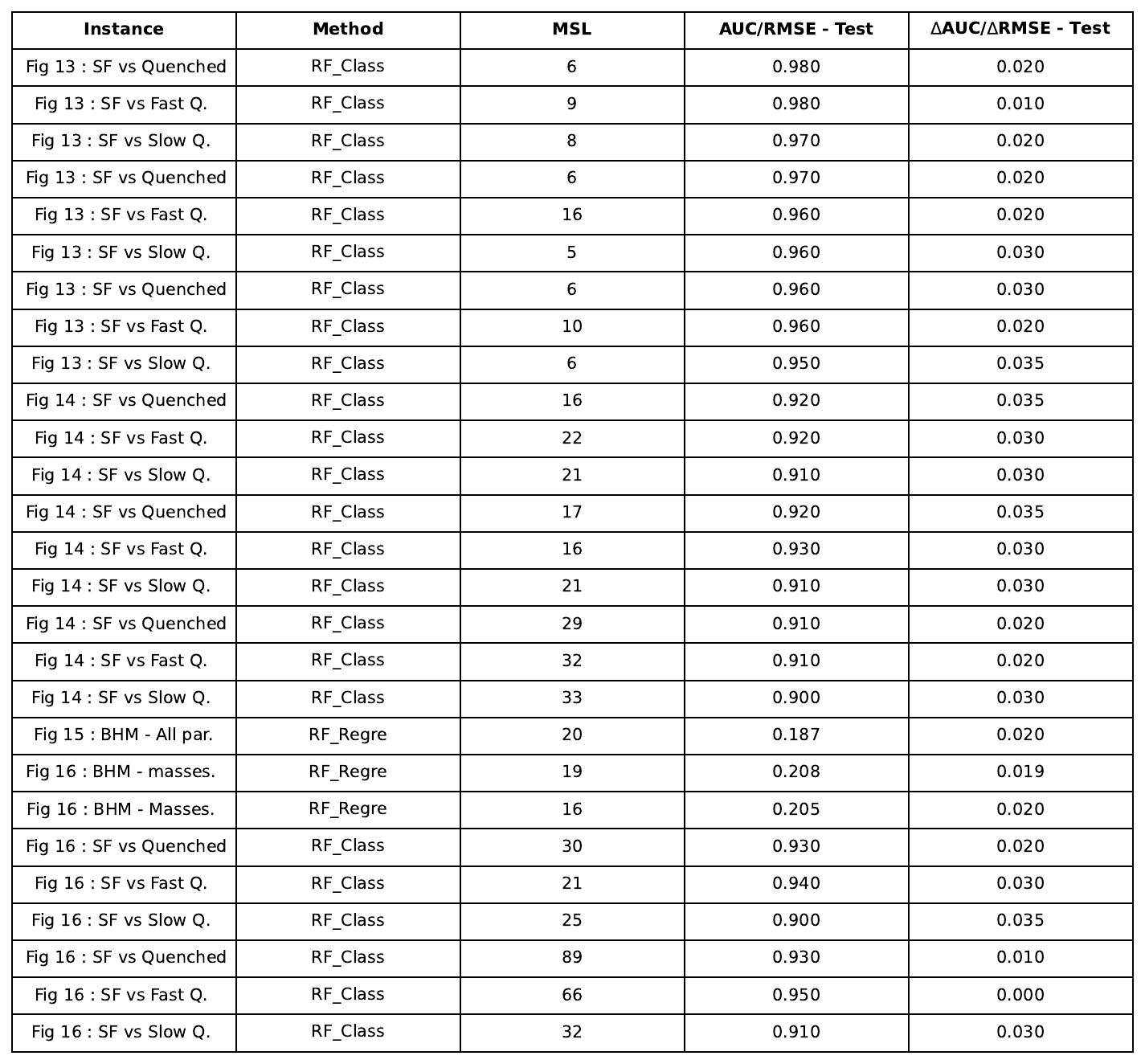}
    \caption{Optimized MSL values and resulting test-set AUC or RMSE for each Random Forest model used in Figures 13–16. Values correspond to the MSL giving maximum AUC (for classifiers) or minimum RMSE (for regressors) with $\Delta$AUC $\leq$ 0.035 or |$\Delta$RMSE| $\leq$ 0.035.}
    \label{fig:Opt_Table_2}
\end{figure*}

The resulting optimal MSLs for all models used in the main paper (Figs. 13–16) are compiled in Table~\ref{fig:Opt_Table_2}.  This table also lists the corresponding test AUC (for classifiers) or test RMSE (for regressors) and the difference between training and test performance ($\Delta$AUC or $|\Delta$RMSE|).  Once the optimal MSL was determined for each model, we retrained the Random Forest using the common hyper‑parameters of Table~\ref{fig:Opt_Table_1} and 100 realizations to produce the final results reported in the main text.

In addition to the optimization described above, we verify that the main scientific conclusions reported in this paper are insensitive to the precise choice of the minimum samples per leaf (MSL) parameter. Exploring a wide range of MSL values, we find that all Random Forest classifiers and regressors produce consistent feature importance rankings and stable performance metrics as long as MSL remains within a reasonable range. In particular, selecting any value between 2 and 90 yields results that are statistically indistinguishable from those obtained with the optimal MSLs listed in Table \ref{fig:Opt_Table_2}. This confirms that our findings are robust to variations in model complexity and that the reported trends do not depend on fine tuning of the Random Forest hyper parameters.

All models are implemented with \texttt{scikit‑learn} \citep{Pedregosa2011}.  This approach ensures reproducibility and facilitates comparison across different Random Forest experiments.

\label{lastpage}
\end{document}